\documentclass[preprint2]{proto}
\usepackage{times}
\usepackage{xcolor}
\usepackage{graphicx}
\usepackage{wasysym}
\usepackage{float}
\usepackage{dblfloatfix}
\usepackage{enumitem}
\usepackage{longtable} % RK for tables
\usepackage{multicol} % RK for table
\usepackage{afterpage} % RK for table
\usepackage{tabularx} % MK for table
\usepackage{booktabs} % MK for table

\graphicspath{{./}{Figures/}}

\usepackage{outlines} % easy indented \item
\usepackage{soul}

\newcommand\edit[1]{\textcolor{cyan}{#1}}

\begin{document}

\title{\textbf{\LARGE PHYSICAL AND SURFACE PROPERTIES OF COMET NUCLEI FROM REMOTE OBSERVATIONS}}

\author {\textbf{\large Matthew M. Knight}}
\affil{\small\em United States Naval Academy}

\author {\textbf{\large Rosita Kokotanekova}}
\affil {\small\em Institute of Astronomy and National Astronomical Observatory, Bulgarian Academy of Sciences}
\affil{\small\em European Southern Observatory}

\author {\textbf{\large Nalin H. Samarasinha}}
\affil{\small\em Planetary Science Institute}

\begin{abstract}

\begin{list}{ } {\rightmargin 1in}
\baselineskip = 11pt
\parindent=1pc
{\small 
We summarize the collective knowledge of physical and surface properties of comet nuclei, focusing on those that are obtained from remote observations. We now have measurements or constraints on effective radius for over 200 comets, rotation periods for over 60, axial ratios and color indices for over 50, geometric albedos for over 25, and nucleus phase coefficients for over 20. The sample has approximately tripled since the publication of {\it Comets~II}, with IR surveys using Spitzer and NEOWISE responsible for the bulk of the increase in effective radii measurements. Advances in coma morphology studies and long-term studies of a few prominent comets have resulted in meaningful constraints on rotation period changes in nearly a dozen comets, allowing this to be added to the range of nucleus properties studied. The first delay-Doppler radar and visible light polarimetric measurements of comet nuclei have been made since {\it Comets~II} and are considered alongside the traditional methods of studying nuclei remotely. We use the results from recent in situ missions, notably Rosetta, to put the collective properties obtained by remote observations into context, emphasizing the insights gained into surface properties and the prevalence of highly elongated and/or bilobate shapes. We also explore how nucleus properties evolve, focusing on fragmentation and the likely related phenomena of outbursts and disintegration. Knowledge of these behaviors has been shaped in recent years by diverse sources: high resolution images of nucleus fragmentation and disruption events, the detection of thousands of small comets near the Sun, regular photometric monitoring of large numbers of comets throughout the solar system, and detailed imaging of the surfaces of mission targets. Finally, we explore what advances in the knowledge of the bulk nucleus properties may be enabled in coming years.
\\~\\~
%\\~
\vspace{-2mm}
}%leave this in to get the correct vertical space after the abstract
\end{list}
\end{abstract}  

%-----------------------------------------------------------------------------------------------------------
\section{\textbf{INTRODUCTION}}
\label{sec:intro}
\vspace{-1mm}
Comets are among the best known and most accessible of astronomical phenomena. They have undoubtedly been noticed -- and quite often feared -- as long as there have been humans to do so. Records of comet observations are extant from well over 2000 years ago \citep[cf.][]{Xi1984,Kronk1999}. And yet, an accurate understanding of the physical properties of cometary nuclei only came about in the last few decades of the 20th century, far later than one might have assumed given their prominent places in the sky. 

The modern concept of a comet as a small, consolidated nucleus was established by \citet{Whipple1950}, though holdouts in support of the earlier ``sandbank'' model \citep[e.g.,][]{Lyttleton1953} forced continued debate in the literature for over a decade \citep[e.g.,][]{Whipple1963}. Whipple's ``dirty snowball'' model was primarily motivated by the desire to explain the non-gravitational forces evident on comet orbits as well as observed gas production rates, but a natural consequence was that comet nuclei were quite small, of order a few km in radius rather than the 10s of km that had been inferred observationally from the apparent sizes of central condensation of comets.

The first conclusive measurements of nucleus properties came in the 1980s due to the proliferation of modern instrumentation over the previous two decades. Key advances included the ability to observe at wavelengths beyond the near-UV and visible, bigger telescopes at higher altitude sites around the world, the introduction of CCD cameras, and the ability to observe from space. Notably, \citet{Millis1985,Millis1988} made the first convincing measurement of nucleus size and albedo (of 49P/Arend-Rigaux), while \citet{Cruikshank1985} concluded that 1P/Halley's albedo was $<$10\%. When the Vega 1, Vega 2, and Giotto spacecraft reached Halley and unequivocally measured its size and albedo \citep[e.g.,][]{Keller1986,Sagdeev1986}, they confirmed that observations from Earth could successfully determine nucleus properties.

Despite technological advances, knowledge of nucleus properties has remained elusive due to the often quixotic nature of their study. When active, comets' gas and dust comae typically obscure the nucleus, but when inactive, their small and dark nuclei are often too faint to detect. What is more, comets' elliptical orbits bring them into the inner solar system only infrequently, of order 10 years for the short period, low inclination Jupiter family comets (JFCs), a few decades for the higher inclination Halley-type comets (HTCs), and hundreds to thousands of years or more for long period comets from the Oort cloud (LPCs). Even during these infrequent passages, an individual comet can only be studied well if it happens to pass close to Earth. The result is that our knowledge of comets often comes in bursts, via either predictable and long-anticipated favorable apparitions of individual JFCs and HTCs, or unexpected arrivals of LPCs.

Comets are renowned as ``fossils'' left over from the formation of the solar system, with their study motivated by the insight they provide into the conditions in the Sun's protoplanetary disk. However, they are not static, and their properties, individually or as a population, must be properly contextualized. An individual comet will have formed between approximately 15--30 au \citep[e.g.,][]{Gomes2004}, been scattered to a more distant orbit during the solar system's early evolution where it remained in relative stasis, though not completely unmodified, before being eventually perturbed into the inner solar system (see Kaib and Volk in this volume; Fraser et al.\ in this volume). Some comets will settle into relatively stable orbits in the inner solar system while others will experience large orbital changes due to gravitational perturbations.  Comets are subject to a host of fates, ranging from complete disappearance from sublimation of volatile ices, quiescence due to loss or burial of accessible volatiles near the surface, spontaneous disruption for a variety of reasons, impact with the Sun or planets, or ejection from the solar system. Furthermore, each comet is at a different place in its evolution, and its specific history is unknown. The current comet population is thought to be in a quasi-steady state, with roughly equal numbers of comets newly arriving from the outer solar system reservoirs as being lost via the various mechanisms just discussed.  While the studies conducted today merely give a snapshot in time of an individual comet, they sample numerous members of various populations at different times in their evolution. By combining observations and dynamical modeling, researchers seek to assemble all of these snapshots into coherent stories.

\vspace{-0.3mm}
\subsection{Nomenclature}
What is a ``comet'' has become difficult to precisely define as more objects with unusual or ambiguous properties are discovered. For this chapter we will concentrate on objects that are traditionally cometary; that is, they contain volatile ices which sublimate to produce a gas and/or dust coma. We further consider primarily those objects whose dynamics bring them into the inner solar system where sublimation can readily occur. Except as noted, we are generally excluding active asteroids  \citep[e.g.,][]{Jewitt2015_Asteroids4}, main belt comets \citep[e.g.,][]{Hsieh2006}, asteroidal objects on comet-like orbits \citep[ACOs, e.g.,][]{Weissman2002}, and outer solar system objects like centaurs and Kuiper belt objects.

A common means of assigning taxonomic classifications to orbits is the Tisserand parameter with respect to Jupiter
\vspace{-1mm}
\begin{equation}
    T_\mathrm{J}=\frac{a_\mathrm{J}}{a}+2~\sqrt{\frac{a}{a_\mathrm{J}}(1-e^2)}~\mathrm{cos}~i
\vspace{-1mm}
\end{equation}
where $a$, $e$, and $i$ are the orbit's semi-major axis, eccentricity, and inclination, respectively, and $a_\mathrm{J}$ is the semi-major axis of Jupiter's orbit  \citep{Kresak1972,Carusi1987}. From a dynamical standpoint, comets are objects that have T$_\mathrm{J}~{\lesssim}~3.0$. We follow traditional definitions \citep[e.g.,][]{Levison1996} to distinguish between JFCs having $2<\mathrm{T_J}<3$, and HTCs and LPCs having $\mathrm{T_J}<2$.  A stricter method of orbit classification to distinguish between comets and asteroids was presented by \citet{Tancredi2014}, but is not needed in this chapter.

At times, we will distinguish between returning Oort cloud comets and dynamically new comets (DNC) that are statistically likely to be on their first passage through the inner solar system. DNCs are identified dynamically by their reciprocal original semi-major axes ($1/a_{0}$). 
We use $1/a_{0}~<~5{\times}10^{-5}$~au$^{-1}$ to identify comets that are unlikely to have previously passed close enough to the Sun for substantial outgassing to occur (e.g., \citealt[][]{AHearn1995}; see also \citealt{Oort1950}). 
Note, however, that more stringent requirements are needed to ensure a high likelihood that an object is genuinely new \citep[cf.][]{Dybczynski2015}. Unless explicitly stated that we are discussing DNCs, we include DNCs in the term LPCs.

Other populations to which we will refer on occasion are centaurs, damocloids, and Manx comets. Centaurs are small bodies on orbits that are intermediate between JFCs and objects residing entirely in the trans-Neptunian region. Note that a strict classification of the centaur population on orbital grounds causes some overlap with other populations \citep[e.g.,][]{Gladman2008}. Damocloids, a term coined by \citet{Jewitt2005},  are objects on HTC or LPC orbits that do not exhibit cometary activity, and are presumed to be inert (or nearly so) nuclei. Introduced by \citet{Meech2016}, Manx comets are objects on Oort cloud and DNC orbits with no activity – suggesting (especially for DNCs) that they may never have had much (or any) ice. It is thought that centaurs and damocloids, along with other minor bodies in the outer solar system such as Kuiper belt objects (aka trans-Neptunian objects, or TNOs), Jupiter trojans (asteroids that share Jupiter's orbit, but librate around its L4 or L5 Lagrange points), Hildas (asteroids on a 3:2 resonance with Jupiter, and residing within its orbit), and irregular satellites formed in the same region as traditional comets \citep[][and references therein]{Dones2015}. Manx comets may have formed with little or no water (e.g., asteroidal material); some may represent early solar system building blocks that formed near the water-ice line in our solar system. See the chapters by Fraser et al.\ and Jewitt and Hsieh in this volume for additional discussion of these objects.

\subsection{Overview and Related Chapters}
Our understanding of the nucleus properties of comets has grown substantially since the publication of {\it Comets~II}, driven in large part by an unprecedented string of successful space missions (see review by Snodgrass et al.\ in this volume).  As a result, several chapters in {\it Comets~III} deal with insights into comet nuclei gained from these missions including interior and global structure and density (Guilbert-Lepoutre et al.\ in this volume), surface properties (Pajola et al.\ in this volume), and nucleus activity and surface evolution (Filacchione et al.\ in this volume). Although we will touch on many of the following topics, readers should consult other chapters for more detailed discussion of planetesimal/comet formation (Simon et al.\ in this volume), dynamical population of comet reservoirs (Kaib and Volk in this volume), the journey from the Kuiper belt and transneptunian objects to comets (Fraser et al.\ in this volume), asteroid/comet transition objects (Jewitt and Hsieh\ in this volume), and comet science with astrophysical assets (Bauer et al.\ in this volume).

The current chapter will largely focus on the knowledge gained from remote observations -- made using ground- and space-based telescopes in the vicinity of Earth, as opposed to in situ studies by dedicated spacecraft -- of a larger number of comets. As such, it builds on a number of comprehensive papers published in the last two decades, notably including \citet{Boehnhardt2004}, \citet{Lamy2004}, \citet{Samarasinha2004}, \citet{Weissman2004}, \citet{Snodgrass2006}, \citet{Lowry2008}, \citet{Fernandez2009}, \citet{Fernandez2013}, \citet{Bauer2017}, and \citet{Kokotanekova2017}. As will be discussed later, remote studies must contend with a variety of limitations in order to ascertain the properties of the nuclei under study, and a broad understanding of an individual comet typically requires the synthesis of many different investigations. Due to page limitations, this chapter cannot provide a comprehensive list of citations for all nucleus properties. We will cite individual papers whenever possible and will frequently refer to earlier review papers, but readers are encouraged to seek out the original sources when citing in future work. 

This chapter is laid out as follows. Section~\ref{sec:how_we_know} gives a brief summary of the observational techniques by which nucleus properties are measured or constrained and Section~\ref{sec:what_we_know} discusses the ensemble properties of comet nuclei. For each property, we first describe how observations are translated to the relevant measurements, then review the known properties and discuss what insights were learned from them. We conclude with a discussion of future advances that are likely to shape our understanding in Section~\ref{sec:future} and a brief summary in Section~\ref{sec:summary}.

\section{\textbf{OBSERVATIONAL TECHNIQUES}}
\label{sec:how_we_know}

\vspace{-0.5mm}
\subsection{Optical Studies}
By far the most common method for determining comet nucleus properties is optical studies, which we define to broadly include the near-UV to mid-IR wavelengths that are available to ground-based observers. We introduce these techniques first, before moving on to radar studies and space missions in the following subsections.

\vspace{-0.5mm}
\subsubsection{The Difficulty with Studying Comet Nuclei}
Comet nuclei are small -- as will be discussed below, typically a few km -- so the vast majority of remote observations are not capable of resolving the nucleus. At a distance of 1 au, 1{\arcsec} corresponds to $\sim$725~km. Thus, a 5~km diameter nucleus would need to pass within 0.07 au of Earth to extend 0.1{\arcsec} arcsec and appear two pixels wide with 0.05~{\arcsec}/pix resolution typical of Hubble Space Telescope (HST) or adaptive optics images. More realistically it would need to be several times closer in order for it to be resolved sufficiently for a meaningful constraint on the size to be made. Approaches to Earth of even 0.07 au are extremely rare. Since HST's launch in 1990, just four comets have been observed passing this close; of these only 252P/LINEAR was observed with high resolution imaging \citep{Li2017}. Thus, comet nucleus sizes and other physical properties are generally not directly measurable, but must be deduced via other means.

\vspace{-0.5mm}
\subsubsection{(Mostly) Direct Detections of the Nucleus}
\label{sec:direct_detection}

Cometary activity tends to obscure the nucleus, and much of this section deals with how this obscuration can be mitigated. However, in certain cases, it is possible to detect the nucleus directly. Direct detection occurs when the nucleus is inactive or so weakly active that its signal dominates that from the coma and tail. When the nucleus signal can be assumed to dominate, its physical properties can be investigated.

The techniques were first successfully applied to the largest comets, since they could be detected when much further from the Sun, and thus less active. Prominent early examples include 28P/Neujmin~1 \citep{Campins1987}, 10P/Tempel~2 \citep{Jewitt1988,Ahearn1989}, and the aforementioned 49P, all of which were initially characterized by the late 1980s. Comets with very low dust-to-gas ratios in their coma (cf. \citealt{AHearn1995}) can also have their nuclei directly detected even when active, e.g, 2P/Encke (see \citealt{Fernandez2000} and references therein) and 162P/Siding Spring \citep{Fernandez2006}. Technological advancements have allowed smaller comets to be reliably observed at larger heliocentric distances, enabling subsequent studies to include far more objects. Notable surveys have been published by \citet{Lowry1999,Lowry2003b}, \citet{Licandro2000},  \citet{Lowry2001,Lowry2005}, \cite{Lowry2003a}, \citet{Meech2004}, \citet{Snodgrass2006}, and \citet{Kokotanekova2017}.

\subsubsection{Nucleus Detection Via Coma Fitting}
Since most comets have coma contribution too prominent for a direct detection, techniques have been developed that can successfully remove the coma under certain circumstances. When there is sufficiently high spatial resolution that the coma can reliably be extrapolated all the way to the nucleus, the coma signal at the center can be removed and the excess signal attributed to the nucleus. The technique has been used most effectively for space-based observations, but can be utilized in ground-based observations of comets coming extremely close to Earth or for facilities having exceptional angular resolution. Lamy and collaborators have utilized HST extensively in such studies \citep[e.g.,][]{Lamy1995,Lamy2009_hst,Lamy2011}, and the technique is described in more detail in \citet{Lamy2004}.

A similar procedure can be applied to mid-IR observations of comets despite lower spatial resolution when there is high enough contrast between the nucleus and the dust coma, typically at heliocentric distances beyond $\sim$3~au. A tweak to the approach removes a range of scaled PSFs, allowing the  nucleus signal to be extracted when the coma cannot be fit with a single power-law. \citet{Lamy2004} and \citet{Fernandez2013} describe the IR nucleus extraction processes in detail; the latter processed their sample independently using both techniques and found good agreement. The approach is limited to the few facilities capable of making thermal IR (roughly $5-25$ ${\mu}$m) observations such as the ground-based IRTF \citep{Lisse1999} and space-based facilities including Infrared Space Observatory \citep[ISO;][]{Jorda2000,Lamy2002,Groussin2004}, Spitzer Space Telescope \citep[e.g.,][]{Lisse2005,Groussin2009}, and Wide-field Infrared Survey Explorer, called WISE during its cryogenic phase in 2009--2010 and NEOWISE henceforth \citep[e.g.,][]{Bauer2011,Bauer2012}. Two major surveys have been conducted in the thermal IR: SEPPCoN, a targeted survey of JFCs with  Spitzer \citep{Fernandez2013}, and a compilation of all comets observed during the cryogenic phase of WISE/NEOWISE \citep{Bauer2017}.

\vspace{-0.5mm}
\subsubsection{Coma Morphology}
\label{sec:coma_morphology}
A method that has been utilized more frequently since {\it Comets~II} is studying varying structures in the coma to infer nucleus properties. If repeating features can be identified, their temporal spacing and/or rate of motion can allow a rotation period to be determined or constrained. The observed morphology will vary as the viewing geometry changes, so 3-D modeling of activity can yield the pole orientation and, frequently, estimates of the location and extent of the active regions on the surface producing the observed coma features. Variations in the observed morphology can indicate seasonal changes and/or be diagnostic of the extent of any non-principal axis (NPA) rotation (see Section~\ref{sec:spin_state} for additional details).

This method is best applied to bright comets making reasonably close approaches to Earth ($\lesssim$1.0~au) since the spatial resolution and signal-to-noise are highest. The ability to resolve features in the coma is dependent on the nature of those structures, including their contrast relative to the ambient coma, the number and projected velocity of the features, and the rotation period and spin state of the nucleus. See reviews by \citet{Schleicher2004} and \citet{Farnham2009} for detailed discussion. Asymmetries in the coma can often be identified by eye in raw images, but a variety of image enhancement techniques \citep[see reviews by][]{Schleicher2004,Samarasinha2014} are used to accentuate the small differences, often just a few percent in brightness, between the structures of interest and the ambient coma.

Coma morphology studies require much brighter comets than the direct detections described above, but they have several advantages. Most critically, coma morphology can be used to infer the rotation period and spin state of active comets when the nucleus is heavily obscured. Many comet nuclei are only accessible in this manner, and the technique is particularly helpful for expanding our knowledge of LPC nucleus properties since these nuclei often remain active until the nucleus is too faint to detect. Another advantage is that morphology studies are much less sensitive to observing conditions than photometric studies, and often can be conducted without the need for absolute calibrations. Meaningful constraints can often be set with sparse sampling of just 1--2 visits per night over several nights.

Coma structures can be due to gas or dust. Although easier to detect, dust is generally harder to interpret because it is slower moving than gas (requiring better spatial resolution to resolve), has large velocity dispersion (thus smearing out features), and its trajectory is altered by solar radiation pressure. Gas is harder to detect than dust, but travels at a much higher velocity ($\sim$1~km/s vs $\sim$0.1 km/s) and is much less affected by solar radiation pressure. The gases seen at optical wavelengths are so called ``fragment species,'' daughter or grand-daughter gases descended from the parent ices that left the nucleus \citep[e.g.,][]{Feldman2004}. Although these fragment species gain excess velocities in random directions during their production, the excess velocities are essentially randomized and the bulk outward motion of the parents leaving the nucleus is approximately preserved. As a result, gas species have proven to be far more useful for rotational studies. 

In order to study the gas coma, the gas must be isolated from the dust. This is most commonly accomplished using specialized narrowband filters \citep{Farnham2000,Schleicher2004}, with the CN filter being by far the most heavily utilized due to CN's bright emission band and high contrast relative to reflected solar continuum at the same wavelengths. In principle, the OH filter is also effective for such studies, but it suffers from severe atmospheric extinction, and many telescopes have very low throughput at the relevant wavelengths. With recent improvements in integral field unit (IFU) spectroscopy, it is becoming possible to study coma morphology in gas ``images'' constructed from spectra \citep{Vaughan2017,Opitom2019}, with the added possibility of studying lines which are too spread out for conventional narrowband filters, like NH$_2$. Coma morphology techniques are not limited to near-UV and visible wavelengths. However, few other methods result in images with sufficient signal-to-noise, spatial resolution, and  temporal coverage to conduct such studies. CO, CO$_2$, and/or dust features are detectable in the comae of some comets imaged by Spitzer's 4.5 $\mu$m channel \citep[e.g.,][]{Reach2013}, so similar studies should be possible for some comets observed by NEOWISE, JWST, and future space-based IR telescopes if image duration and cadence permit. 

Attempts to interpret coma morphology in order to infer properties of the nucleus date back many decades \citep[e.g.,][]{Whipple1978,Sekanina1979}. These images were generally dominated by dust, and more recent work has shown that the assumptions in these early modeling efforts yielded results that are incompatible with modern solutions \citep[cf.][]{Sekanina1991,Knight2012}. 

The first results using gas filters were achieved by studying CN in 1P \citep{AHearn1986,Samarasinha1986,Hoban1988}. With the improvement in CCDs and the production of the ESA and HB filters in the 1990s \citep{Farnham2000,Schleicher2004}, coma morphology studies have now been applied to many comets. Work has largely been concentrated among a few groups with access to narrowband filters and sufficient telescope time to constrain rotation periods: Schleicher and collaborators at Lowell Observatory \citep[e.g.,][]{Schleicher1998,Farnham2005,Knight2011,Bair2018}, Jehin and collaborators using the TRAPPIST telescopes \citep[e.g.,][]{Jehin2010,Opitom2015,Moulane2018}, Samarasinha and colleagues using Kitt Peak National Observatory \citep{Mueller1997,Farnham2007,Samarasinha2011}, and Waniak and collaborators at Rozhen National Observatory in Bulgaria \citep{Waniak2009,Waniak2012}.
The technique has matured to the point that off-the-shelf filters with a quasi-CN bandpass have been successfully employed \citep{Ryske2019}.

\subsubsection{Other Methods of Constraining Nucleus Properties}
\label{sec:other_methods}
Coma lightcurves have been used for decades to constrain rotation periods \citep[e.g.,][]{Millis1986,Feldman1992}, with the technique becoming much more prevalent in the modern era \citep[e.g.,][]{Anderson2010,SantosSanz2015,Manzini2016}. Aperture photometry is typically employed, with a signal that is assumed to be dominated by dust and/or gas in the coma as opposed to the nucleus. Except in rare cases, such lightcurves tend to have very small amplitudes (often just a few 0.01 mag from peak-to-trough) that can be easily affected by seeing variations, background contamination, calibration systematics, etc. When phased to a ``best'' period, interpretation is dependent on an assumption about the source of the variations (frequently assumed to be a single source), but unless this can be constrained, e.g., by coma morphology, it can lead to aliasing problems. Even in comets for which much information is known a priori, conclusive interpretation of a coma lightcurve can be challenging \citep[cf.][]{Schleicher2015}, so we generally consider periods from coma lightcurves to be less reliable than those obtained by the means discussed above.

Various efforts have been made to extract nucleus information from observations not necessarily designed for that purpose. Compilations of multi-wavelength constraints on individual comets to determine their nucleus radius include C/1995 O1 (Hale-Bopp) \citep{Weaver1997,Fernandez2002} and C/1983 H1 (IRAS-Araki-Alcock) \citep{Groussin2010}. Non-gravitational accelerations (caused by momentum transfer to the nucleus from outgassing materials) from orbit calculations and overall brightness have been used to infer mass \cite[e.g.,][]{Rickman1986, Rickman1989, Sosa2009,Sosa2011}. When sizes are also known, this can also yield estimates of the bulk density of the nucleus \citep[e.g.,][]{Farnham2002,Davidsson2006}. For a detailed discussion of nucleus densities and how they are estimated, see the chapter by Guilbert-Lepoutre et al., this volume.

Secular lightcurves (brightness/activity as a function of heliocentric distance) imply seasonal variations in activity of some comets; these can provide constraints on pole orientations and spin state \citep[e.g.,][]{Schleicher2003,Farnham2005}. \citet{Boe2019} combined a cometary activity model with a survey simulator to statistically characterize the size distribution of LPCs. \citet{Tancredi2000,Tancredi2006}, \citet[e.g.,][]{Ferrin2010}, and \citet{Weiler2011} have assembled data from a variety of published observations to attempt to extract nucleus magnitudes from observations at large distances. The comets studied are primarily JFCs near aphelion, but are occasionally LPCs when they are assumed to be inactive. As acknowledged by these authors, such observations can be problematic since many JFCs still show distant activity near aphelion \citep[e.g.,][]{MazzottaEpifani2007,MazzottaEpifani2008,Kelley2013}. \citet{Hui2018} showed that the nucleus must account for more than 10\% of the total signal for a reliable extraction from typical ground-based observations, effectively ruling out nucleus extraction for more active JFCs and LPCs.

\vspace{-0.5mm}
\subsection{Radar Studies}
\label{sec:radar}
Radar contributes unique insight into comet nuclei, being capable of imaging nuclei at spatial scales only achieved otherwise by space missions and of measuring the rotation rate of the nucleus directly. Since radar observations are conducted by sending a burst of microwaves towards a target and measuring the power of the returned echo, the received signal varies as ${\Delta}^{-4}$, where $\Delta$ is the geocentric distance.  This effectively limits the detectable population to only those approaching within 0.1 au of Earth unless their nuclei are especially large \citep{Harmon1999}. A detailed review of radar observations was provided by \citet{Harmon2004}; we discuss the technique briefly and provide updates of key observations since then in Section~\ref{sec:nucleus_shape}.

The highest quality radar observations are delay-Doppler imaging which measure both the echo Doppler spectrum as well as the time delay of the echo, resulting in spatial and rotational rate information about different positions on the nucleus. Delay-Doppler data can be inverted to form a 3-D model of the nucleus. While common for asteroids, this has only been achieved for a handful of comets. Doppler-only detections yield a radar cross section, which has a degeneracy between nucleus size and radar albedo unless additional information is considered. The Doppler signal can be interpreted to give constraints on the rotation period, polarization by the surface, surface roughness, and density of the surface layer. Interestingly, radar albedos have been found to be similar to optical albedos despite probing meters into the surface \citep{Harmon2004}. Although beyond the scope of this chapter, radar observations can also provide information about the properties of large grains in the inner coma and even their position relative to the nucleus.

All radar detections of comet nuclei to date have been achieved using Arecibo and/or Goldstone. Bistatic observations, where the emitter and receiver are located at different telescopes, with Greenbank Observatory as the receiver have been made occasionally. With the loss of Arecibo in 2020 and few known upcoming close approaching comets, there are no foreseeable opportunities for radar comet observations until the mid-2030s. New comet radar detections in the next decade will require as yet undiscovered comets passing close to Earth (E.~Howell, pers.\ comm., 2021).

%\vspace{-0.5mm}
\subsection{Space Missions}
Dedicated space missions to comets provide definitive measurements of comet nucleus properties since they fully resolve the nucleus and are so close that the dust and gas along the line of sight are negligible. Six comet nuclei have been imaged in situ: 1P by Vega 1 and 2 and Giotto, 19P/Borrelly by Deep Space~1, 81P/Wild~2 by Stardust, 9P/Tempel 1 by Deep Impact and Stardust, 103P/Hartley 2 by Deep Impact/EPOXI, and 67P/Churyumov-Gerasimenko by Rosetta. These are discussed extensively in Snodgrass et al.\ in this volume, but we note that some properties of these nuclei were constrained by remote observations before the mission encounters and were in good agreement with mission findings. Successful examples include the surprisingly small size of 103P for its activity level \citep{Lisse2009} and the rotation period of 67P \citep{Lowry2012}. The consistency between the remote observations and spacecraft results is a validation that the remote techniques are reliable. The spacecraft results provide a ground-truth for the selected objects, allowing us to trust that, under the proper conditions, the same techniques are giving good results for other objects as well.

%\vspace{-0.5mm}
\section{\textbf{THE RANGES OF PHYSICAL AND SURFACE PROPERTIES}}
\label{sec:what_we_know}
The catalog of measured nucleus properties is now approximately three times as large as when they were reviewed in {\it Comets~II}. We also now have detailed information from several new space missions which influence our interpretation of these measurements. We tabulate the physical (Table~\ref{tab:physical}) and surface (Table~\ref{tab:surface}) properties for all comets having reliable determinations of any of the following: rotation period, axial ratio, albedo, nucleus phase function, or nucleus color. The distributions of these properties are plotted in Figure~\ref{fig:nucleus_histograms}. For any comet with one of the other properties measured, we tabulate what we conclude to be the most reliable effective radius in Table~\ref{tab:physical}. This is not a complete review of all measured sizes; it aims to enable searching for the relation between nucleus size and the other parameters. Due to the limited space, it is not possible to list all measured effective radii and provide the necessary context to evaluate the accuracy of the measurements. Instead, we plot in Figure~\ref{fig:nucleus_histograms} all effective radii determined from the two major thermal IR surveys \citep{Fernandez2013, Bauer2017}. Although this is not a complete set of effective nucleus measurements, it encompasses 213 comets and the methodology was the same throughout, allowing more meaningful comparisons. Owing to the still sparse collection of properties for HTCs and LPCs, these are grouped together in the figure for comparison with the JFCs, but are discussed individually below as warranted.

\renewcommand{\thetable}{1A}
\begin{table*}[p!]
\footnotesize
\caption{Physical properties of comet nuclei. Comets are grouped by dynamical classification, and not all properties are known for a given comet. Column 1 gives the comet name. Columns 2, 4, and 6 give the effective radius ($r_\mathrm{n}$), rotation period ($P$), and axial ratio ($a/b$), respectively. Columns 3, 5, and 7 give the corresponding reference code. Reference codes are expanded in the footnotes.} 
  \label{tab:physical}
%\begin{tabularx}{0.935\textwidth}{lcccccc}
\begin{tabularx}{\textwidth}{lp{1.8cm}p{1cm}p{3.4 cm}p{1.3cm}p{2cm}p{0.8cm}}
\noalign{\vskip 3 pt}
\hline
\noalign{\vskip 3 pt}
Comet & $r_n$\textsuperscript{\dag} [km]&  Ref\textsuperscript{*} & $P$ [h] &  Ref\textsuperscript{*} & $a/b$\textsuperscript{\ddag} & Ref\textsuperscript{*} \\
\cmidrule(lr){2-3}\cmidrule(lr){4-5} \cmidrule(lr){6-7}
\noalign{\vskip 2 pt}
Jupiter-family comets & &&&&& \\
\hline
\noalign{\vskip 2 pt}
                     2P/Encke &                    2.4 $\pm$ 0.3 &                       F00 &             11.0830 $\pm$ 0.0030\textsuperscript{a,b?} &       L07, B05 &                           $\geq$ 1.44 (0.06) &       L07 \\
                      4P/Faye &                  1.77 $\pm$ 0.04 &                       L09 &                                                     &           &                                  $\geq$ 1.25 &       L04 \\
                  6P/d’Arrest &  $\mathrm{2.23^{+0.13}_{-0.15}}$ &                       F13 &                                     6.67 $\pm$ 0.03\textsuperscript{b?} &       G03 &                                  $\geq$ 1.08 &       G03 \\
             7P/Pons-Winnecke &                  2.64 $\pm$ 0.17 &                       F13 &                        $\mathrm{7.9^{+1.6}_{-1.1}}$ &       S05 &                             $\geq$ 1.3 (0.1) &       S05 \\
                  9P/Tempel 1 &                   2.83 $\pm$ 0.10 &                      T13a &               41.335 $\pm$ 0.005\textsuperscript{a} &       B11 &                      1.28\textsuperscript{c} &      T13a \\
                 10P/Tempel 2 &                  5.98 $\pm$ 0.04 &                       L09 &                8.948 $\pm$ 0.001\textsuperscript{a} &       S13 &                                   $\geq$ 1.9 &       J89 \\
                     14P/Wolf &                  2.95 $\pm$ 0.19 &                       F13 &                                     9.07 $\pm$ 0.01 &       K18 &                           $\geq$ 1.41 (0.06) &       K17 \\
                   17P/Holmes &                  2.41 $\pm$ 0.53 &                      BG17 &                                   7.2/8.6/10.3/12.8 &       S06 &                             $\geq$ 1.3 (0.1) &       S06 \\
                 19P/Borrelly &                   2.5 $\pm$ 0.1 &                       B04 &                   26.0 $\pm$ 1.0\textsuperscript{a} &       M02 &              2.5 (0.07)\textsuperscript{c} &       B04 \\
         21P/Giacobini-Zinner &                                1 &                       T00 &    7.39 $\pm$ 0.01; 10.66 $\pm$ 0.01\textsuperscript{a} &       G23 &                                   $\geq$ 1.5 &       M92 \\
                    22P/Kopff &                  2.15 $\pm$ 0.17 &                       F13 &                                     12.30 $\pm$ 0.8 &      L03a &                           $\geq$ 1.66 (0.11) &      L03a \\
         26P/Grigg-Skjellerup &                              1.3 &                       L04 &                                                     &           &                                   $\geq$ 1.1 &       B99 \\
                28P/Neujmin 1 &                  $\sim$10.7  &                       L04 &                                    12.75 $\pm$ 0.03 &       D01 &                           $\geq$ 1.51 (0.07) &       D01 \\
   31P/Schwassmann-Wachmann 2 &  $\mathrm{1.65^{+0.11}_{-0.12}}$ &                       F13 &                                     5.58 $\pm$ 0.03 &       L92 &                            $\geq$ 1.6 (0.15) &       L92 \\
                  36P/Whipple &                  2.31 $\pm$ 0.29 &                      BG17 &                                           $\sim$ 40 &       S08 &                             $\geq$ 1.9 (0.1) &       S08 \\
37P/Forbes & $\mathrm{1.23^{+0.08}_{-0.09}}$ & F13	&  &  &  & \\
  41P/Tuttle-Giacobini-Kres{\'a}k &                              0.7 &                       T00 &                      19.75 $-$ 20.05\textsuperscript{a,b?} &      BF18, H18 &                                          $\geq$ 2.1 &      BR20 \\
44P/Reinmuth & 2.55 $\pm$ 0.15 & BG17	&  &  &  & \\
                    45P/Honda-Mrkos-Pajdu{\v{s}}{\'a}kov{\'a} &                         0.6 $-$ 0.65 &                      LH22 &                                       7.6 $\pm$ 0.5 &       S22 &                                   $\geq$ 1.3 &      LT99 \\
                 46P/Wirtanen &                  0.56 $\pm$ 0.04 &                       B02 &  $\mathrm{8.94^{+0.02}_{-0.01}}$\textsuperscript{a} &       F21 &                             $\geq$ 1.4 (0.1) &       B02 \\
         47P/Ashbrook-Jackson &  $\mathrm{3.11^{+0.20}_{-0.21}}$ &                       F13 &                                     15.6 $\pm$ 0.1  &       K17 &                           $\geq$ 1.36 (0.07) &       K17 \\
                  48P/Johnson &  $\mathrm{2.97^{+0.19}_{-0.20}}$ &                       F13 &                                   29.00 $\pm$ 0.04 &       J04 &                           $\geq$ 1.34 (0.06) &       J04 \\
             49P/Arend-Rigaux &                  4.57 $\pm$ 0.06 &                      KW17 &                13.450 $\pm$ 0.005 &       E17 &                           $\geq$ 1.63 (0.07) &       M88 \\
50P/Arend & 1.49 $\pm$ 0.13  & F13	&  &  &  & \\
53P/Van Biesbroeck & 3.33 $-$ 3.37 & M04	&  &  &  & \\
59P/Kearns-Kwee & 0.79 $\pm$ 0.03 & L09	&  &  &  & \\
          61P/Shajn-Schaldach &                  2.28 $\pm$ 0.64 &                      BG17 &                                       4.9 $\pm$ 0.2 &       L11 &                                   $\geq$ 1.3 &       L11 \\
                   63P/Wild 1 &                 1.46  $\pm$ 0.03 &                       L09 &                                          14 $\pm$ 2 &     BdS20 &                            $\geq$ 2.19 (0.02) &     BdS20 \\
                      67P/Churyumov-Gerasimenko &                1.649 $\pm$ 0.007 &                       J16 &               12.055 $\pm$ 0.001\textsuperscript{a} &   Rosetta &              1.67 (0.01)\textsuperscript{c} &       J16 \\
                   70P/Kojima &                  1.84 $\pm$ 0.09 &                       L11 &                                                     &           &                                   $\geq$ 1.1 &       L11 \\
71P/Clark & 0.79 $\pm$ 0.03 & L09	&  &  &  & \\
73P-C/Schwassmann-Wachmann 3\textsuperscript{f} &                    0.4 $\pm$ 0.1 &                      GJ19 &                                    20.76 $\pm$ 0.08 &      GJ19 &                             $\geq$ 1.8 (0.3) &       T06 \\
         74P/Smirnova-Chernykh &  $\mathrm{3.31^{+0.60}_{-0.69}}$ &                       F13 &                                                     &           &                                  $\geq$ 1.14 &       L11 \\
    76P/West-Kohoutek-Ikemura &                 0.31 $\pm$ 0.01 &                       L11 &                                       6.6 $\pm$ 1.0 &       L11 &                                  $\geq$ 1.45 &       L11 \\
                   81P/Wild 2 &                  1.98 $\pm$ 0.05 &                      SB04 &                                      13.5 $\pm$ 0.1 &       M10 &                                  1.38 (0.04) &       D04 \\
                82P/Gehrels 3 &                  0.59 $\pm$ 0.04 &                       L11 &                                   $\geq$ 24 $\pm$ 5 &       L11 &                                  $\geq$ 1.59 &       L11 \\
84P/Giclas & 0.90 $\pm$ 0.05 & L09	&  &  &  & \\
                   86P/Wild 3 &                  0.42 $\pm$ 0.02 &                       L11 &                                                     &           &                                  $\geq$ 1.35 &       L11 \\
                      87P/Bus &                  0.26 $\pm$ 0.01 &                       L11 &                                          32 $\pm$ 9 &       L11 &                                   $\geq$ 2.2 &       L11 \\
                  92P/Sanguin &                  2.08 $\pm$ 0.01 &                       S05 &                                     6.22 $\pm$ 0.05 &       S05 &                             $\geq$ 1.7 (0.1) &       S05 \\
                  93P/Lovas 1 &                  2.59 $\pm$ 0.26 &                       F13 &                        $\mathrm{18.2^{+1.5}_{-15}}$ &       K17 &                            $\geq$ 1.21(0.06) &       K17 \\
                94P/Russell 4 &  $\mathrm{2.27^{+0.13}_{-0.15}}$ &                       F13 &                                    20.70 $\pm$ 0.07 &       K17 &                             $\geq$ 2.8 (0.2) &       K17 \\
               96P/Machholz 1\textsuperscript{f} &                  3.40 $\pm$ 0.20 &                       E19 &                4.096 $\pm$ 0.002 &       E19 &                             $\geq$ 1.6 (0.1) &       E19 \\
               103P/Hartley 2 &                 0.580 $\pm$ 0.018 &                      T13b &                   16.4 $\pm$ 0.1\textsuperscript{a,b} &       M11 &                      3.11\textsuperscript{c} &      T13b \\
106P/Schuster & 0.94 $\pm$ 0.03 & L09	&  &  &  & \\
               110P/Hartley 3 &                  2.20 $\pm$ 0.10 &                       L11 &                                 10.153 $\pm$ 0.001  &       K17 &                           $\geq$ 1.20 (0.03) &       K17 \\
112P/Urata-Niijima & 0.90 $\pm$ 0.05 & L09	&  &  &  & \\
113P/Spitaler & 1.70 $\pm$ 0.10 & F13	&  &  &  & \\
114P/Wiseman-Skiff & 0.78 $\pm$ 0.05 & L09	&  &  &  & \\
                  121P/Shoemaker-Holt 2  &  $\mathrm{3.87^{+0.26}_{-0.27}}$ &                       F13 &                             $\mathrm{10^{+8}_{-2}}$ &       S08 &                           $\geq$ 1.15 (0.03) &       S08 \\
            123P/West-Hartley &                  2.18 $\pm$ 0.23 &                       F13 &                                                     &           &                                    $\geq$ 1.6 (0.1) &       K17 \\
131P/Mueller 2 & $\mathrm{1.11^{+0.09}_{-0.07}}$ & F13	&  &  &  & \\
                  137P/Shoemaker-Levy 2  &  $\mathrm{4.04^{+0.31}_{-0.32}}$ &                       F13 &                                                     &           &                                  $\geq$ 1.18 (0.05) &       K17 \\
             143P/Kowal-Mrkos &  $\mathrm{4.79^{+0.32}_{-0.33}}$ &                       F13 &                                    17.20 $\pm$ 0.02 &       K18 &                           $\geq$ 1.49 (0.05) &       L04 \\
       147P/Kushida-Muramatsu &                  0.21 $\pm$ 0.02 &                       L11 &                        10.5 $\pm$ 1.0 / 4.8 $\pm$ 0.2 &       L11 &                                  $\geq$ 1.53 &       L11 \\
149P/Mueller 4 & $\mathrm{1.42^{+0.09}_{-0.10}}$ & F13	&  &  &  & \\
           162P/Siding Spring &  $\mathrm{7.03^{+0.47}_{-0.48}}$ &                       F13 &                                     32.864  $\pm$ 0.001 &       D23 &                           1.56 ($_{-0.16}^{+0.44}$)\textsuperscript{c} &       D23 \\
                    169P/NEAT &  $\mathrm{2.48^{+0.13}_{-0.14}}$ &                       F13 &                                 8.4096 $\pm$ 0.0012 &       K10 &        $\geq$ 1.74 (0.03)\textsuperscript{d} &       W06 \\
                  209P/LINEAR &                      $\sim$ 1.53 &                       H14 &                                   10.93 $\pm$ 0.020 &  H14, S16 &                                  $\geq$ 1.55 &       H14 \\
249P/LINEAR & 1.0 $-$ 1.3 & FL17	&  &  &  & \\
                  252P/LINEAR &                   0.3 $\pm$ 0.03 &                       L17 &   5.41 $\pm$ 0.07/7.24$\pm$ 0.07\textsuperscript{b?} &       L17 &                                              &           \\
\end{tabularx}
\end{table*}

\renewcommand{\thetable}{1A}
\begin{table*}
\footnotesize
\caption{Physical properties (continued)}
\begin{tabularx}{0.998\textwidth}{lcccccc}
\noalign{\vskip 5 pt}
\hline
\noalign{\vskip 3 pt}
Comet & $r_n$\textsuperscript{\textdagger} [km]&  Ref\textsuperscript{*} & $P$ [h] &  Ref\textsuperscript{*} & $a/b$\textsuperscript{\ddag} & Ref\textsuperscript{*} \\
\cmidrule(lr){2-3}\cmidrule(lr){4-5} \cmidrule(lr){6-7}
Jupiter-family comets (continued) & &&&&& \\
\hline
\noalign{\vskip 2 pt}
260P/McNaught &  $\mathrm{1.54^{+0.09}_{-0.08}}$ &                       F13 &                                     8.16 $\pm$ 0.24 &       M14 &               $\geq$ 1.07\textsuperscript{d} &       M14 \\
280P/Larsen & $\mathrm{1.23^{+0.15}_{-0.17}}$ & F13	&  &  &  & \\
                    322P/SOHO\textsuperscript{x} &                    0.075 $-$ 0.16 &                       K16 &                                       2.8 $\pm$ 0.3 &       K16 &                                   $\geq$ 1.3 &       K16 \\
                    323P/SOHO\textsuperscript{f,x} &                0.086 $\pm$ 0.003 &                      H22 &                                0.522024 $\pm$ 0.000002 &      H22 &                $\sim$ 1.25\textsuperscript{c} &      H22 \\
333P/LINEAR & 3.04 & S21	&  &  &  & \\
      P/2016 BA14 (PANSTARRS) &    0.55 $-$ 0.8                              &           K22             &                                            $\sim$40 &       N16 &                                              &           \\
\hline
\noalign{\vskip 1 pt}
Halley-type comets & &&&&& \\
\hline
\noalign{\vskip 2 pt}
                    1P/Halley &                              5.5 &                       L04 &                        $\sim$ 68\textsuperscript{b} &       S04 &  2.0 (0.1)\textsuperscript{c} &       K87 \\
                    8P/Tuttle &                              2.9 &                       G19 &                                                11.4 &       H10 &                 1.41 (0.07)\textsuperscript{cb} &       G19 \\
            55P/Tempel-Tuttle &                              1.8 &                       L04 &                                                     &           &                                   $\geq$ 1.5 &       H98 \\
            109P/Swift-Tuttle &                   15 $\pm$ 3 &                       F95 &                                                69.4 &       L04 &                                              &           \\
        C/2001 OG108 (LONEOS) &                    7.6 $\pm$ 1.0 &                       A05 &                                      57.2 $\pm$ 0.5 &       A05 &                                   $\geq$ 1.3 &       A05 \\
         C/2002 CE10 (LINEAR) &                  8.95 $\pm$ 0.45 &                       S18 &                                     8.19 $\pm$ 0.05 &       S18 &                             $\geq$ 1.2 (0.1) &       S18 \\
            P/1991 L3 (Levy) &     5.8 $\pm$ 0.1                  &      F94          &                                                8.34 &       F94 &                                   $\geq$ 1.3 &       F94 \\
P/2006 HR30 (Siding Spring) & 11.95 $-$ 13.55 & BI17	&  &  &  & \\
\hline
\noalign{\vskip 1 pt}
Long period comets & &&&&& \\
\hline
\noalign{\vskip 2 pt}
            C/1983 H1 (IRAS-Araki-Alcock) &                    3.4 $\pm$ 0.5 &                       G10 &                                     51.3 $\pm$ 0.3  &       S88 &                                              &           \\
             C/1990 K1 (Levy) &                                  &                           &                   17.0 $\pm$ 0.1\textsuperscript{a} &       F92 &                                              &           \\
        C/1995 O1 (Hale-Bopp) &                               37 &                      SK12 &                                    11.35 $\pm$ 0.04 &       J02 &                           $\geq$ 1.72 (0.07) &      SK12 \\
        C/1996 B2 (Hyakutake) &                    2.4 $\pm$ 0.5 &                      LF99 &                                   6.273 $\pm$ 0.007 &       S02 &                                              &           \\
C/2002 VQ94 (LINEAR) & 40.7 & J05	&  &  &  & \\
         C/2004 Q2 (Machholz) &                                  &                           &                                    17.60 $\pm$ 0.05 &       F07 &                                              &           \\
            C/2007 N3 (Lulin) &                  6.10 $\pm$ 0.25 &                      BG17 &                                    41.45 $\pm$ 0.05 &      BS18 &                                              &           \\
            C/2009 P1 (Garradd) & 13.50 $\pm$ 2.50  & BG17	& 11.1 $\pm$ 0.8 & I17 &  & \\
            C/2012 F6 (Lemmon) &  & & 9.52 $\pm$ 0.05 & O15 &  & \\
       C/2012 K1  (PANSTARRS) &              8.67 $\pm$ 0.08         &    BdS20               &                                       9.4 $\pm$ 0.4 &     BdS20 &                                  $\geq$ 1.44 (0.03) &     BdS20 \\
   C/2013 A1  (Siding Spring) &   $\sim$ 1                               &  F17                         &                                                 8.00 $\pm$ 0.08 &    L16    &                                      &           \\
          C/2014 Q2 (Lovejoy) &                                  &                           &                                    17.89 $\pm$ 0.17 &      SR15 &                                              &           \\
        C/2014 S2 (PANSTARRS) &                              1.3 &                     BdS18 &                                          68 $\pm$ 2 &     BdS18 &        $\geq$ 1.45 (0.13)\textsuperscript{d} &     BdS18 \\
C/2014 UN271 (B-B) & 137 $\pm$ 15/119 $\pm$ 13 & LM22,HJ22	&  &  &  & \\
          C/2020 F3 (NEOWISE) &                              2.5 &  J. Bauer (unpubl.\ data) &                                       7.8 $\pm$ 0.2 &       M21 &                                              &           \\
          
\hline
\hline
\vspace{-6 mm}
\end{tabularx}
\end{table*}

\renewcommand{\thetable}{1B}
\begin{table*}[p!]
\footnotesize
\caption{Surface properties of comet nuclei. Comets are grouped by dynamical classification, and not all properties are known for a given comet. Column 1 gives the comet name. Columns 2, 4, and 6 give the $V$-band albedo ($p_\mathrm{V}$), phase coefficient ($\beta$), and $(V-R)$ color, respectively. Columns 3, 5, and 7 give the corresponding reference code. Reference codes are expanded in the footnotes.}
  \label{tab:surface}
%\begin{tabularx}{0.828\textwidth}{lcccccc}
\begin{tabularx}{\textwidth}{lp{2.1cm}p{1.2cm}p{2.4cm}p{1.3cm}p{3cm}p{0.8cm}}
\noalign{\vskip 5 pt}
\hline
\noalign{\vskip 3 pt}
Comet & $p_{\mathrm{V}}$ &  Ref\textsuperscript{*} & $\beta$ [mag/$^{\circ}$] &  Ref\textsuperscript{*} &              $(V-R)$ & Ref\textsuperscript{*} \\
\cmidrule(lr){2-3}\cmidrule(lr){4-5} \cmidrule(lr){6-7}
\noalign{\vskip 2 pt}
Jupiter-family comets & &&&&& \\
\hline
\noalign{\vskip 3 pt}
              2P/Encke &                  0.04 $\pm$ 0.03\textsuperscript{R} &         F00 &  0.053 $\pm$ 0.003\textsuperscript{WM} &  F00, B08 &                                      0.44 $\pm$ 0.06 &      LT09 \\
               4P/Faye &                                                       &             &                                        &           &                                      0.45 $\pm$ 0.04 &       L09 \\
           6P/d’Arrest &                                                       &             &                                        &           &                                       0.51 $\pm$ 0.10 &      LT09 \\
      7P/Pons-Winnecke &                                                       &             &                                        &           &                                      0.49 $\pm$ 0.03 &       S05 \\
           9P/Tempel 1 &                                     0.056 $\pm$ 0.007 &        L07a &                      0.046 $\pm$ 0.007 &      L07a &                                       0.50 $\pm$ 0.01 &      L07a \\
          10P/Tempel 2 &                    $\mathrm{0.022^{+0.004}_{-0.006}}$ &        AH89 &                      0.037 $\pm$ 0.004 &       S91 &                                      0.54 $\pm$ 0.03 &      LT09 \\
              14P/Wolf &  0.046 $\pm$ 0.007\textsuperscript{$\mathrm{r_{PS}}$} &         K17 &                       0.060 $\pm$ 0.005 &       K17 &                                      0.57 $\pm$ 0.07 &       S05 \\
            17P/Holmes &                                                       &             &                                        &           &                                      0.53 $\pm$ 0.07 &       S06 \\
          19P/Borrelly &                  0.06 $\pm$ 0.02\textsuperscript{R} &        L07b &                      0.043 $\pm$ 0.009 &      L07b &                                      0.25 $\pm$ 0.78 &      L03b \\
  21P/Giacobini-Zinner &                                                       &             &                                        &           &                                       0.50 $\pm$ 0.02 &       L93 \\
             22P/Kopff &                                     0.042 $\pm$ 0.006 &         L02 &                                        &           &                                      0.50 $\pm$ 0.05 &      LT09 \\
  26P/Grigg-Skjellerup &                                                       &             &                                        &           &                                        0.3 $\pm$ 0.1 &       B99 \\
         28P/Neujmin 1 &                  0.03 $\pm$ 0.01\textsuperscript{R} &         J88 &                      0.025 $\pm$ 0.006 &       D01 &                                      0.47 $\pm$ 0.05 &      LT09 \\
           36P/Whipple &                                                       &             &                       0.060 $\pm$ 0.019 &       S08 &                                      0.47 $\pm$ 0.02 &       S08 \\
            37P/Forbes &                                                       &             &                                        &           &                                      0.29 $\pm$ 0.03 &       L09 \\
44P/Reinmuth &                                                       &             &                                        &           &                                      0.62 $\pm$ 0.08 &       L09 \\
45P/Honda-Mrkos-Pajdu{\v{s}}{\'a}kov{\'a} &                                                       &             &                                   0.06 &       L04 &                                      0.44 $\pm$ 0.05 &      LT09 \\
46P/Wirtanen &                                                       &             &                                        &           &                                      0.45 $\pm$ 0.06 &      LT09 \\
47P/Ashbrook-Jackson &  0.054 $\pm$ 0.008\textsuperscript{$\mathrm{r_{PS}}$} &         K17 &                                        &           &                                      0.43 $\pm$ 0.09 &      LT09 \\
48P/Johnson &                                                       &             &                      0.059 $\pm$ 0.002 &       J04 &                                        0.5 $\pm$ 0.3 &       L00 \\
49P/Arend-Rigaux &                                     0.04 $\pm$ 0.01 &         C95 &                                        &           &                                      0.47 $\pm$ 0.05 &      LT09 \\
             50P/Arend &                                                       &             &                                        &           &                                      0.81 $\pm$ 0.10 &       L09 \\
53P/Van Biesbroeck &                                                       &             &                                        &           &                                      0.34 $\pm$ 0.08 &       M04 \\
59P/Kearns-Kwee &                                                       &             &                                        &           &                                      0.62 $\pm$ 0.07 &       L09 \\
\vspace{-10mm}
\end{tabularx}
\end{table*}
\begin{table*}[ht!]
\footnotesize
\setlength{\tabcolsep}{0.05in}
\caption{Surface properties (continued)}
%\begin{tabularx}{0.91\textwidth}{lcccccc}
%\begin{tabularx}{\textwidth}{lp{2.5cm}p{1.3cm}p{2 cm}p{1.3cm}p{3cm}p{0.8cm}}
\begin{tabularx}{\textwidth}{lp{2.5cm}p{1.3cm}p{2 cm}p{1.3cm}p{3cm}p{0.8cm}}
\noalign{\vskip 5 pt}
\hline
\noalign{\vskip 3 pt}
Comet & $p_{\mathrm{V}}$ &  Ref\textsuperscript{*} & $\beta$ [mag/$^{\circ}$] &  Ref\textsuperscript{*} &              $(V-R)$ & Ref\textsuperscript{*} \\
\cmidrule(lr){2-3}\cmidrule(lr){4-5} \cmidrule(lr){6-7}
 Jupiter-family comets (continued) & &&&&& \\
\hline
\noalign{\vskip 2 pt}
            63P/Wild 1 &                                                       &             &                                        &           &                                      0.50 $\pm$ 0.05 &       L09 \\
    67P/Churyumov-Gerasimenko &                                     0.059 $\pm$ 0.002 &         S15 &                      0.047 $\pm$ 0.002 &       F15 &                                      0.57 $\pm$ 0.03 &       C15 \\
70P/Kojima &                                                       &             &                                        &           &                                       0.60 $\pm$ 0.09 &       L11 \\
             71P/Clark &                                                       &             &                                        &           &                                      0.64 $\pm$ 0.07 &       L09 \\
               73P/Schwassmann-Wachmann 3-C &                                                       &             &                                        &           &                                       0.45 $\pm$ 0.20 &      LT09 \\
            81P/Wild 2 &                                     0.059 $\pm$ 0.004 &       LAH09 &                    0.0513 $\pm$ 0.0002 &     LAH09 &                                                      &           \\
            84P/Giclas &                                                       &             &                                        &           &                                      0.32 $\pm$ 0.03 &       L09 \\
            86P/Wild 3 &                                                       &             &                                        &           &                                       0.86 $\pm$ 0.10 &       L11 \\
           92P/Sanguin &                                                       &             &                                        &           &                                      0.54 $\pm$ 0.04 &       S05 \\
         94P/Russell 4 &  0.043 $\pm$ 0.007\textsuperscript{$\mathrm{r_{PS}}$} &         K17 &                      0.039 $\pm$ 0.002 &       K17 &                                      0.62 $\pm$ 0.05 &       S08 \\
        96P/Machholz 1 &                                                       &             &                                        &           &                                      0.41 $\pm$ 0.04 &       E19 \\
        103P/Hartley 2 &                                     0.045 $\pm$ 0.009 &         L13 &                      0.046 $\pm$ 0.002 &       L13 &                                      0.43 $\pm$ 0.04 &       L13 \\
         106P/Schuster &                                                       &             &                                        &           &                                      0.52 $\pm$ 0.06 &       L09 \\
        110P/Hartley 3 &                                                       &             &                      0.069 $\pm$ 0.002 &       K17 &                                      0.67 $\pm$ 0.09 &       L11 \\
    112P/Urata-Niijima &                                                       &             &                                        &           &                                      0.53 $\pm$ 0.04 &       L09 \\
         113P/Spitaler &                                                       &             &                                        &           &                   0.58 $\pm$ 0.08\textsuperscript{C} &      SI12 \\
    114P/Wiseman-Skiff &                                                       &             &                                        &           &                                      0.46 $\pm$ 0.02 &       L09 \\
           121P/Shoemaker-Holt 2  &                                                       &             &                                        &           &                                      0.53 $\pm$ 0.03 &       S08 \\
        131P/Mueller 2 &                                                       &             &                                        &           &                                      0.45 $\pm$ 0.12 &       S08 \\
           137P/Shoemaker-Levy 2  &  0.030 $\pm$ 0.005\textsuperscript{$\mathrm{r_{PS}}$} &         K17 &                      0.035 $\pm$ 0.004 &       K17 &                                      0.71 $\pm$ 0.18 &       S06 \\
     143P/Kowal-Mrkos &  0.044 $\pm$ 0.008\textsuperscript{$\mathrm{r_{PS}}$} &         K18 &                      0.043 $\pm$ 0.014 &       J03 &                                      0.58 $\pm$ 0.02 &      LT09 \\
        149P/Mueller 4 &  0.030 $\pm$ 0.005\textsuperscript{$\mathrm{r_{PS}}$} &         K17 &                        0.03 $\pm$ 0.02 &       K17 &                                                      &           \\
    162P/Siding Spring &  0.021  $\pm$ 0.002 &         D23 &                      0.051 $\pm$ 0.002  &       D23 &                                      0.45 $\pm$ 0.01 &       C06 \\
             169P/NEAT &                                     0.03 $\pm$ 0.01 &        DM08 &                                        &           &                                      0.43 $\pm$ 0.02 &       K10 \\
            249P/LINEAR &                                                       &             &                                        &           &                                      0.37 $\pm$ 0.01\textsuperscript{S'} &       K21 \\
           280P/Larsen &                                                       &             &                                        &           &                                      0.49 $\pm$ 0.03 &       S08 \\
             322P/SOHO\textsuperscript{x} &                                                       &             &    0.031 $\pm$ 0.004                                    &      K16     &                                      0.41 $\pm$ 0.04 &       K16 \\
             323P/SOHO\textsuperscript{x} &                                                       &             &      0.0326 $\pm$ 0.0004                                  &       H22    &  0.05 $\pm$ 0.06; 0.13 $\pm$ 0.09\textsuperscript{D} &      H22 \\
             333P/LINEAR &                                                       &             &                                        &           &                                      0.44 $\pm$ 0.01\textsuperscript{S'} &       S21 \\
%             323P/SOHO\textsuperscript{x} &                                                       &             &      0.0326 $\pm$ 0.0004                                  &       H22b    &  0.05 $\pm$ 0.06; 0.13 $\pm$ 0.09\textsuperscript{D} &      H22b \\
      P/2016 BA14 (PANSTARRS) &                                           0.01 $-$ 0.03 &         K22 &                                        &           &                                                      &           \\
\hline
\noalign{\vskip 2 pt}
Halley-type comets & &&&&& \\
\hline
\noalign{\vskip 2 pt}
            1P/Halley &                       $\mathrm{0.04^{+0.02}_{-0.01}}$ &         S86 &                                        &           &                                      0.41 $\pm$ 0.03 &       T89 \\
             8P/Tuttle &                  0.04 $\pm$ 0.01\textsuperscript{R} &         G19 &                           0.033 $-$ 0.04 &       L12 &                                      0.53 $\pm$ 0.04 &      LT09 \\
     55P/Tempel-Tuttle &                  0.05 $\pm$ 0.02\textsuperscript{R} &         C02 &                                  0.041 &       L04 &                                      0.51 $\pm$ 0.05 &      LT09 \\
     109P/Swift-Tuttle &                          0.02–0.04\textsuperscript{R} &         L04 &                                        &           &                                                      &           \\
 C/2001 OG108 (LONEOS) &                                     0.04 $\pm$ 0.01 &         A05 &              0.034\textsuperscript{TW} &       A05 &                                      0.46 $\pm$ 0.02 &       A05 \\
  C/2002 CE10 (LINEAR) &                                     0.03 $\pm$ 0.01 &         S18 &                                        &           &                                    0.568 $\pm$ 0.039 &       S18 \\
     P/2006 HR30 (Siding Spring) &                                           0.035-0.045 &        BI17 &                                        &           &                                      0.45 $\pm$ 0.01 &       H07 \\
\hline
\noalign{\vskip 1 pt}
Long period comets & &&&&& \\
\hline
 \noalign{\vskip 2 pt}    
     C/1983 H1 (IRAS-Araki-Alcock) &                                     0.04 $\pm$ 0.01 &         G10 &                                   0.04 &       G10 &                                                      &           \\
 C/1995 O1 (Hale-Bopp) &              0.04 $\pm$ 0.03\textsuperscript{var} &         C02 &                                        &           &                                                      &           \\
C/2002 VQ94 (LINEAR) &                                                       &             &                                        &           &                                      0.50 $\pm$ 0.02 &  J05 \\

    C/2014 UN271 (Bernardinelli-Bernstein) &                  0.034 $\pm$ 0.008 &  LM22 &                                        &           &                   0.46 $\pm$ 0.04\textsuperscript{C} &       B21 \\
%    C/2014 UN271 (Bernardinelli-Bernstein) &                  0.034 $\pm$ 0.008; 0.044 $\pm$ 0.011 &  LM22; H22a &                                        &           &                   0.46 $\pm$ 0.04\textsuperscript{C} &       B21 \\
\hline
\hline
\end{tabularx}

%\footnotesize
%\caption{Footnotes and references}
%  \label{tab:surface2}
%\tablecomments{Footnotes to Table 1A:}
%\noalign{\vskip 3 pt}
\vspace{2 mm}
 \textsuperscript{*}Additional references are available for some comets. The table lists the most recent and/or most precise measurements. All other literature values have been omitted due to space limitations. Published precision has been preserved, resulting in differing numbers of significant digits;
\textsuperscript{x}Excluded from bulk calculations and Figures~\ref{fig:nucleus_histograms}-\ref{fig:P_vs_r} for the following reasons: 322P: suspected asteroid only active near extremely small perihelion of 0.071 au \citep{Knight2016}; 323P: disintegrating, only active near extremely small perihelion of 0.039 au \citep{Hui2022b}. 

\vspace{2 mm}
{\bf Footnotes to 1A: }
\textsuperscript{\dag}Nucleus radius measurements are given for any comet with $P$, $a/b$, $\beta$, $p_V$, or $(V-R)$ measurements (see text for additional details);
\textsuperscript{$\ddag$}The uncertainty of the axial ratio value is listed in brackets whenever available;
\textsuperscript{a}Period changes observed; This is the minimum known value measured with sufficient precision; \textsuperscript{b}NPA rotation (suspected NPA rotators indicated with \textsuperscript{b?}); 
\textsuperscript{c}A shape model was ``quoted'' in the cited paper; The provided axial ratio is obtained by dividing the largest shape model axis by the second largest axis; 
\textsuperscript{cb}A shape model consisting of two spheres with radii $R_1 > R_2$ was presented in the cited paper; The axial ratio was approximated by $(R_1+R_2)/R_1$;
\textsuperscript{d}Calculated from the brightness variation in the cited paper;
\textsuperscript{f}Ongoing fragmentation observed.

\vspace{2 mm}
{\bf Footnotes to 1B: }
\textsuperscript{C}Converted to $(V-R)$ using \cite{Jester2005}; 
\textsuperscript{D}Different colors before and after perihelion; Converted to $(V-R)$ using \cite{Jester2005};
\textsuperscript{R}Original geometric albedo in $R$-band converted to $V$-band using the $(V-R)$ color of the nucleus in this table. Due to the large uncertainty in the color index of 19P, the average $(V-R)$ = 0.50 $\pm$ 0.03 for JFC nuclei \citep{Lamy2009,Jewitt2015} was used instead; 
\textsuperscript{S'} Using the expression in eq. (2) in \cite{Luu1990b} can be approximated to a color index $(V-R)$, using a $(V-R)$ = 0.354 $\pm$ 0.010 mag for the Sun \citep{Holmberg2006};
\textsuperscript{$\mathrm{r_{PS}}$}Original geometric albedo in $r_\mathrm{PS}$-band converted to $V$-band using $p_\mathrm{V}$ = $p_\mathrm{rPS}$ $\times$ 0.919 for the mean color index $(B-V)$ = 0.87 $\pm$ 0.05 mag \citep{Lamy2009}; 
\textsuperscript{TW}Linear fit derived from data in \cite{Abell2005}; 
\textsuperscript{var}Albedo variations observed \cite[][]{Szabo2012}; 
\textsuperscript{WM}Weighted~mean.

\vspace{2 mm}
{\bf References: }A05: \cite{Abell2005}; AH89: \cite{Ahearn1989}; B02: \cite{Boehnhardt2002}; B04: \cite{Buratti2004}; B05: \cite{Belton2005}; B08: \cite{Boehnhardt2008}; B11: \cite{Belton2011}; B21: \cite{Bernardinelli2021}; B99: \cite{Boehnhardt1999}; BdS18: \cite{Betzler2018}; BdS20: \cite{Betzler2020}; BF18: \cite{Bodewits2018}; BG17: \cite{Bauer2017}; BI17: \cite{Bach2017}; BR20: \cite{Boehnhardt2020}; 
\end{table*}

\begin{table*}[ht!]
\footnotesize

\vspace{2 mm}
{\bf References (continued): }  BS18: \cite{Bair2018}; C02: \cite{Campins2002}; C06: \cite{Campins2006}; C15: \cite{Ciarniello2015}; C95: \cite{Campins1995}; D01: \cite{Delahodde2001}; D04: \cite{Duxbury2004}; D23 \cite{Donaldson2023}; DM08: \cite{DeMeo2008}; E17: \cite{Eisner2017}; E19: \cite{Eisner2019}; F00: \cite{Fernandez2000}; F07: \cite{Farnham2007}; F13: \cite{Fernandez2013}; F15: \cite{Fornasier2015}; F17: \cite{Farnham2017}; F21: \cite{Farnham2021}; F92: \cite{Feldman1992}; F94: \cite{Fitzsimmons1994a}; F95: \cite{Fomenkova1995}; G03: \cite{Gutierrez2003a}; G10: \cite{Groussin2010}; G19: \cite{Groussin2019}; G23: \cite{Goldberg2023}; GJ19: \cite{Graykowski2019}; H07: \cite{Hicks2007}; H10: \cite{Harmon2010}; H14: \cite{Howell2014}; H18: \cite{Howell2018}; H22: \cite{Hui2022b}; H98: \cite{Hainaut1998}; I17: \cite{Ivanova2017}; J02: \cite{Jorda2002}; J03: \cite{Jewitt2003}; J04: \cite{Jewitt2004b}; J05: \cite{Jewitt2005}; J16: \cite{Jorda2016}; J88: \cite{Jewitt1988}; J89: \cite{Jewitt1989}; K10: \cite{Kasuga2010}; K16: \cite{Knight2016}; K17: \cite{Kokotanekova2017}; K18: \cite{Kokotanekova2018}; K21: \cite{Kareta2021}; K22: \cite{Kareta2022}; K87: \cite{Keller1987}; KW17: \cite{Kelley2017}; L00: \cite{Licandro2000}; L02: \cite{Lamy2002}; L03a: \cite{Lowry2003a}; L03b: \cite{Lowry2003b}; L04: \cite{Lamy2004}; L07: \cite{Lowry2007}; L07a: \cite{Li2007_Tempel}; L07b: \cite{Li2007_Borrelly}; L09: \cite{Lamy2009_hst}; L11: \cite{Lamy2011}; L12: \cite{Lamy2012}; L13: \cite{Li2013}; L16 \cite{Li2016}; L17: \cite{Li2017}; L92: \cite{Luu1992}; L93: \cite{Luu1993}; LAH09: \cite{Li2009}; LF99: \cite{Lisse1999}; LH22: \cite{Lejoly2022}; LM22: \cite{Lellouch2022}; LT09: \cite{Lamy2009}; LT99: \cite{Lamy1999}; M02: \cite{Mueller2002}; M04: \cite{Meech2004}; M10: \cite{Mueller2010}; M11: \cite{Meech2011}; M14: \cite{Manzini2014}; M21: \cite{Manzini2021}; M88: \cite{Millis1988}; M92: \cite{Mueller1992}; N16: \cite{Naidu2016}; O15: \cite{Opitom2015}; Rosetta: Rosetta; S02: \cite{Schleicher2002}; S04: \cite{Samarasinha2004}; S05: \cite{Snodgrass2005}; S06: \cite{Snodgrass2006}; S08: \cite{Snodgrass2008}; S13: \cite{Schleicher2013}; S15: \cite{Sierks2015}; S16: \cite{Schleicher2016}; S18: \cite{Sekiguchi2018}; S21: \cite{Simion2021}, private communication; S22: \cite{Springmann2022}; S86: \cite{Sagdeev1986}; S88: \cite{Sekanina1988}; S91: \cite{Sekanina1991}; SB04: \cite{Sekanina2004_81P}; SI12: \cite{Solontoi2012}; SK12: \cite{Szabo2012}; SR15: \cite{Serra-Ricart2015}; T00: \cite{Tancredi2000}; T06: \cite{Toth2006}; T13a: \cite{Thomas2013Tempel1}; T13b: \cite{Thomas2013}; T89: \cite{Thomas1989}; W06: \cite{Warner2006}.
\end{table*}

In the following subsections we discuss these cumulative properties and provide updated mean, standard deviation, and median values when it reasonable to do so. Section~\ref{sec:size_shape} discusses nucleus sizes, the nucleus size-frequency distribution, and nucleus shapes. Section~\ref{sec:surface_prop} discusses the cumulative surface properties, with separate subsections for albedo (\ref{sec:albedo}), phase function and phase reddening (\ref{sec:phase_function}), and nucleus colors/spectroscopy (\ref{sec:color_and_spectra}). Rotation periods and changes in rotation period are discussed in Section~\ref{sec:spin_state}.

This extensive compilation of nucleus properties provides an opportunity to explore correlations between properties. We conducted Spearman rank correlations for all comets and for only the JFC population (the HTC plus LPC data are too sparse for their own test) for all combinations of the data compiled in Tables~\ref{tab:physical} and \ref{tab:surface}. The pairings resulting in the highest Spearman rank correlations ($\rho$) were albedo and phase coefficient (${\rho}=0.52$ for all comets and ${\rho}=0.43$ for JFCs; previously identified by \citealt{Kokotanekova2017}), period and axial ratio (${\rho}=0.33$), and phase coefficient with radius (${\rho}=-0.31$). The first can be considered to have moderate correlation, while the remainder have low correlation. We caution that for some properties there are still few enough measurements that Spearman ranks can change substantially with the addition of a single new measurement. The lack of correlation between some quantities that might reasonably be assumed to correlate such as color and albedo (${\rho}=-0.09$), or color and phase function (${\rho}=-0.07$) may be informative. The scatter in some properties (albedo, phase coefficient, color) is larger for smaller radii. While this may be an indication of a greater diversity of surface properties in small nuclei, more data are needed to rule out observational bias since most have large uncertainties. We discuss these and other interesting comparisons between properties in the relevant subsections. We also performed Spearman rank correlations to check for any possible correlations between the nucleus properties in Tables~\ref{tab:physical} and \ref{tab:surface} and orbital parameters. The intriguing possibility for a correlation between $(V-R)$ and perihelion distance is discussed in Section \ref{sec:color_and_spectra}.

\subsection{Nucleus Sizes and Shapes}
\label{sec:size_shape}

\subsubsection{Effective Radius}
\label{sec:nucleus_size}
Except in rare instances of radar observations or space missions, the actual width of the nucleus cannot be measured directly. Instead, the effective radius ($r_\mathrm{n}$; the radius of a sphere having the same cross-section as the comet nucleus), is used to quantify its size. More nuanced measurements of the radius are sometimes used ($r_{n,a}$ and $r_{n,v}$; see \citealt{Lamy2004}), but since many authors do not specify which they have measured, we generically use the term $r_\mathrm{n}$ in this chapter. As shown by \citet{Lamy2004}, $r_{n,a}$ and $r_{n,v}$ agree to within 10\% for axial ratios up to 3, so any discrepancies between nomenclature are minimal.

The apparent magnitude, $m$, can be translated into $r_\mathrm{n}$ (in meters) from reflected-light observations by the following equation, which was originally derived by \citet{Russell1916}:
\begin{equation}
    p{\Phi}({\alpha})r_\mathrm{n}^{2}=2.238{\times}10^{22}r_\mathrm{H}^{2}{\Delta}^{2}10^{0.4(m_{\odot}-m)}.
    \label{eq:r_nucleus}
\end{equation}
Here, $p$ is the geometric albedo, ${\Phi}({\alpha})$ is the phase function of the nucleus at solar phase angle $\alpha$ (solar phase angle, henceforth simply ``phase angle,'' is the Sun-comet-observer angle), $r_\mathrm{H}$ is the heliocentric distance in au, $\Delta$ is the distance from the observer to the nucleus in au, and $m_\odot$ is the apparent magnitude of the Sun. The quantities $p$, ${\Phi}({\alpha})$, $m_\odot$, and $m$ need to be taken in the same spectral band. The phase function is typically assumed to be linear with phase angle (see Section~\ref{sec:phase_function}), and is given by
\begin{equation}
    -2.5\mathrm{log}[{\Phi}({\alpha})]={\alpha}{\beta}
    \label{eq:phase_function}
\end{equation}
where $\beta$ is the linear phase coefficient in mag/$^\circ$. Traditionally, this phase coefficient is taken to be independent of wavelength; however, this should be investigated with additional photometric data taken at multi-bandpasses in the future.

Although not discussed further in this chapter, a common method of comparing comet nucleus sizes and in searching for unresolved activity contaminating the photometric aperture is via the absolute magnitude, $H$, given by \begin{equation}
    H = m - 5~\mathrm{log}(r_\mathrm{H}{\Delta}) - {\alpha}{\beta}.
    \label{eq:absolute_mag}
\end{equation}
This is the magnitude the nucleus would have at $r_\mathrm{H}={\Delta}=1$~au and ${\alpha}=0^{\circ}$, a physically impossible scenario, but one that is convenient for normalizations.

Thermal IR observations are usually dominated by thermal re-radiation from the nucleus and, when paired with thermophysical modeling, can be used to determine nucleus sizes. The process utilizes the size, shape, rotation period, spin axis orientation, and surface properties (thermal inertia, surface roughness) to predict the observed thermal flux. This is more mathematically complex than the reflected light procedure just discussed. Detailed explanations are given in \citet{Lamy2004} and \citet{Fernandez2005}; the key result of relevance here is that it yields an effective nucleus radius. This is best constrained with concurrent visible wavelength observations, but single-wavelength thermal IR observations yield small enough uncertainties \citep{Fernandez2013,Bauer2017} that they are generally considered reliable.

\begin{figure*}[t!]
\begin{center}
\includegraphics[width=6.5in]{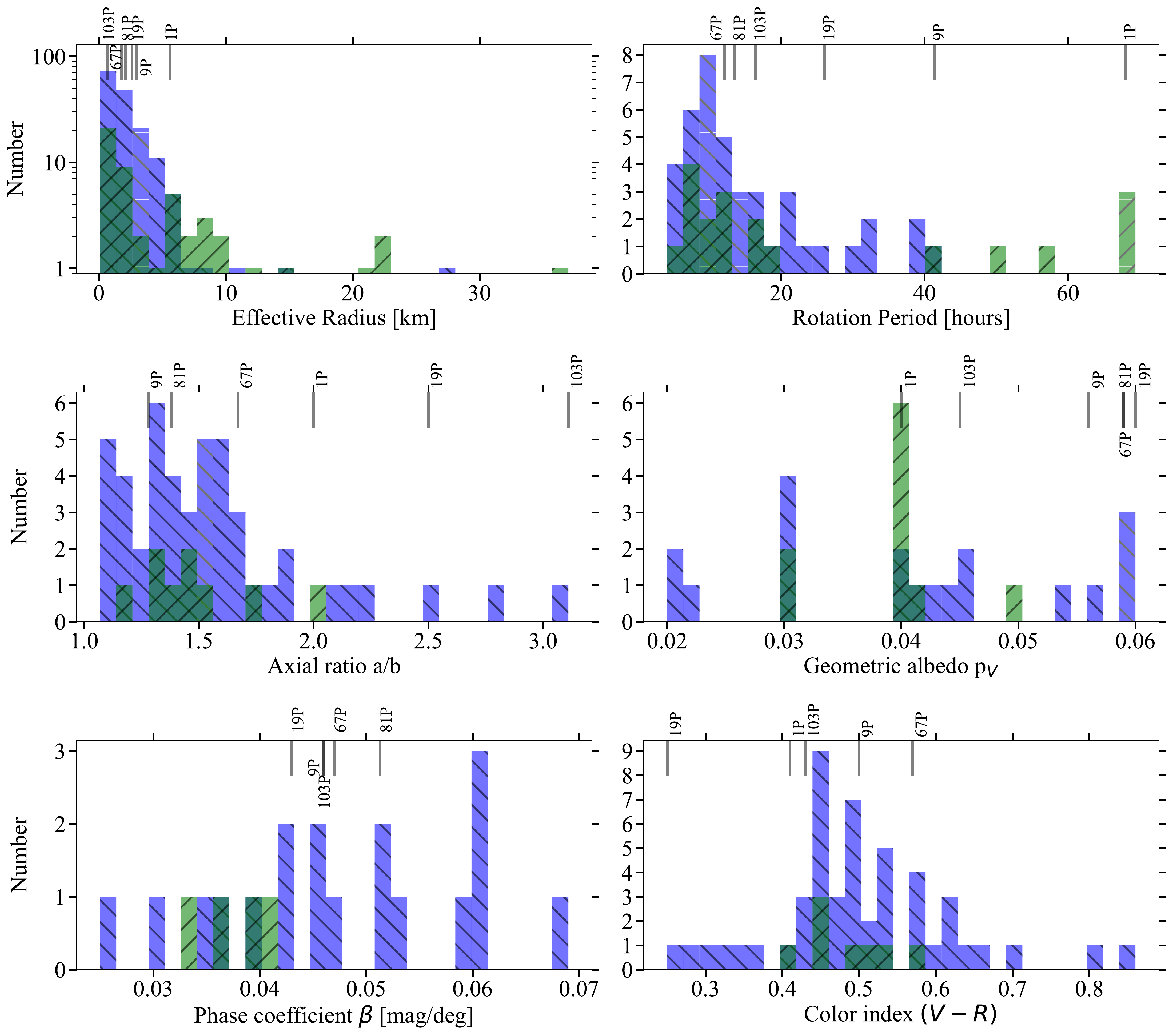}
\caption{Histogram of the measured nucleus properties for JFCs (blue backward ($\backslash$) hashing) and HTCs plus LPCs (green forward (/) hashing). The values for comets with in-situ spacecraft measurements are indicated by vertical lines at the top axis. This figure aims to provide an overview of each parameter’s range and the sample size for which it has been measured. The property being plotted is given in the abscissa label for each plot. There are different numbers of comets in each sample because only a few comets have had all properties measured. As discussed in the text, the effective radius values are taken only from the thermal-IR studies of \citet{Fernandez2013} and \cite{Bauer2017}, while all other quantities represent our best effort to include all published values. In the cases when a comet had multiple measurements of a certain property, we have displayed the most recent sufficiently precise measurement, therefore only plotting a given comet's property once. Note that most axial ratios plotted are lower limits. See the text for additional details. }
\label{fig:nucleus_histograms}
\end{center}
\end{figure*}

Together, these methods have revealed a large diversity of comet sizes, ranging from effective radii of a few hundred meters up to tens of kilometers (see Table~\ref{tab:physical} and Figure~\ref{fig:nucleus_histograms}). The largest known nucleus is C/2014 UN271 (Bernardinelli-Bernstein) \citep[${r_\mathrm{n}}~{\sim}~60$~km,][]{Hui2022,Lellouch2022}, with C/2002 VQ94 and Hale-Bopp also having ${r_\mathrm{n}}~{\gtrsim}~30$~km. As noted earlier, we exclude active centaurs like Chiron from consideration. Based on the existence of objects with radius of $\sim$100s of km in the centaur \citep{Stansberry2008} and SDO populations \citep{Mueller2020}, it is conceivable that comparably large comet nuclei are present in the Oort cloud but have not yet been discovered due to their lower abundance. Large ($r_\mathrm{n}~{\gtrsim}~10$~km) ACOs also exist, including (3552) Don Quixote, which has shown recurrent activity \citep{Mommert2020}. It is noteworthy that the debiased average size of LPCs has been found to be 1.6$\times$ larger than the debiased average JFC nucleus size (\citealt{Bauer2017}; see next subsection).

\subsubsection{Size-Frequency Distribution}
\label{sec:sfd}
Collecting the $r_\mathrm{n}$ of a large sample of objects and studying their cumulative size-frequency distribution (SFD) provides an invaluable tool to probe the formation and subsequent evolution of comets. The SFDs of minor-planet populations in the solar system are expressed as a power law of the form:
\begin{equation}
N(> r_\mathrm{n}) \propto r_\mathrm{n}^{-a},
\end{equation}
\noindent where $N$ is the number of objects with radius larger than $r_\mathrm{n}$. According to analytical models, collisionally relaxed populations of self-similar bodies with identical physical parameters have a power-law SFD with $a = 2.5$. \citep{Dohnanyi1969}. In contrast, a shallower slope, $a = 2.04$ is predicted for collisionally relaxed populations of strengthless bodies \citep{OBrien2003}. Observations of asteroids reveal that their SFD shows characteristic “breaks”, or changes in the slope of the power-law distribution, which can be used to probe the material strength and the population evolutionary processes \citep[e.g.,][]{OBrien2005,Bottke2005}. 

Traditionally comets are presumed to have had a very different evolutionary history than the asteroid belt (e.g., they are not expected to have reached a collisionally steady state). Moreover they undergo sublimation-driven mass loss which is expected to result in significant differences between the SFDs of cometary populations and their source populations. However, recent work reviewed in \cite{Weissman2020} and \cite{Morbidelli2021} suggests that the source population of comets and TNOs in the primordial trans-Neptunian disk may have evolved to reach collisional equilibrium prior to being dispersed. While the debate about this possibility remains open, some of the deciding evidence may come from a better understanding of the small-end of the SFD of today's remnants from the primordial transplanetary disk and specifically from comets. It is therefore informative to examine the observational evidence on the SFDs of short- and long-period comets in an attempt to distinguish the signatures of recent activity-driven evolution from those of planetesimal formation and/or early collisional history.

Earlier attempts to derive the SFD of JFCs from optical observations resulted in slightly different slopes, $a$ \citep[e.g.,][]{Lowry2003b,Lamy2004,Meech2004,Tancredi2006,Weiler2011}. This highlighted the need to assess the uncertainty of the power-law slope determination by assessing the contribution of the various assumptions of the size estimates (e.g., on the albedo, phase function and shape of the nucleus, as well as photometric uncertainty). This was addressed by \cite{Snodgrass2011} and yielded a SFD with $a = 1.92~{\pm}~0.20$ for \edit{JFCs} with $r_\mathrm{n}~{\geq}~1.25$~km. This result is comparable to the SFD slopes determined for JFCs from thermal-IR observations \citep[$1.92~{\pm}~0.23$;][]{Fernandez2013}; both are consistent with the expected slope for a collisionally relaxed population of strengthless bodies \citep{OBrien2003}, though as just discussed, the implications of this are not yet settled.

These studies, however, did not take into account the observational biases influencing the SFD, the most prominent being the bias against detecting small comet nuclei. \citet{Bauer2017}, therefore, performed careful debiasing of the NEOWISE comet size estimates and determined a slope $a = 1.0~{\pm}~0.1$ for LPCs and a steeper slope, $a = 2.3~{\pm}~0.2$, for JFCs. This analysis, however, has left a few prominent questions unresolved. Theory suggests that there is an under-abundance of sub-km comets (e.g., \citealt{Samarasinha2007, Jewitt2021}); it is important for future observations to determine whether there is a paucity of small JFCs and what it reveals about comet disruption and the primordial SFD of outer-solar system planetesimals \citep{Fernandez2013,Bauer2017}. Other features, such as the small bump in the SFD of JFCs between 3 and 6 km \citep{Fernandez2013}, also remain to be confirmed and explained in the context of planetesimal formation or sublimation evolution \citep{Kokotanekova2018}.  For more details about these debates, we refer the reader to Bauer et al.\ in this volume where the details of the telescope surveys used to derive the comet SFD are covered and to Fraser et al.\ in this volume where the comet size distribution is discussed in the context of other outer solar system populations.

\subsubsection {Nucleus Shapes}
\label{sec:nucleus_shape}
Three main observational techniques can be employed to study the shapes of comet nuclei: rotational lightcurves, radar observations, and spacecraft data. The least complex and most easily available are rotational lightcurves in which the nucleus signal dominates the flux in the photometric aperture. Lightcurves in which coma flux dominates yield  fundamentally different information and are discussed in Section~\ref{sec:spin_state}.

Observations that are frequent enough, ideally on the same night or over several consecutive nights, can be combined to create a lightcurve, in which the magnitude is plotted as a function of time or rotational phase (if known). Lower limits to the projected nucleus axial ratio can be determined from the the peak-to-trough amplitude (${\Delta}m = m_\mathrm{min}-m_\mathrm{max}$) as
\begin{equation}
    \frac{a}{b}~{\geq}~10^{0.4(m_\mathrm{min}-m_\mathrm{max})} 
    \label{eq:axis_ratio}
\end{equation}
where $a$ and $b$ are the semi-long and and semi-intermediate axes in a triaxial ellipsoid in simple rotation, and $m_\mathrm{min}$ and $m_\mathrm{max}$ are the magnitudes at lightcurve minimum and maximum, respectively.

Broadly available for a large number of comet nuclei, $\Delta m$ allows the study of a large sample of comets \citep[e.g.,][]{Lamy2004} under the assumption that they are approximately triaxial ellipsoids.  \citet{Kokotanekova2017} updated the sample of well-constrained JFC lightcurves and estimated a median axial ratio of $a/b = 1.5$, in agreement with the previous estimate from \citet{Lamy2004}. We show in Figure~\ref{fig:nucleus_histograms} an updated version that includes all known comet axial ratios; these are also tabulated in Table~\ref{tab:physical}. The known range extends from 1.07 to 3.11 for the elongated nucleus of comet 103P. The axial ratios tabulated in Table~\ref{tab:physical} have a mean of $1.55~{\pm}~0.40$ and a median of 1.45. Similar results are obtained when considering only JFCs versus HTCs plus LPCs.

It is important to keep in mind that the projected axial ratio derived from rotational lightcurves is just a lower limit unless the spin axis is normal to the observer's line-of-sight. If the lightcurve is observed at an unfavorable geometry or when the nucleus is surrounded by an undetected coma, the actual nucleus elongation can be significantly underestimated. This limitation becomes evident when the axial ratios of comets observed in-situ by spacecraft are compared to the total population average. Most spacecraft targets have axial ratios ($\geq$1.5), with comets 103P and 19P reaching some of the largest $a/b$ of 3.1 and 2.5 (Figure~\ref{fig:nucleus_histograms}).

Rotational lightcurves taken at a wide variety of different observing geometries can also be analysed using the convex lightcurve inversion (CLI) technique \citep{Kaasalainen2001a, Kaasalainen2001b}. This technique has been successfully applied to produce shape models for thousands of asteroids \citep{Durech2010}; a few comets have thus far been modeled, including 67P \citep{Lowry2012}, 162P \citep{Donaldson2023}, and 323P/SOHO \citep{Hui2022b}, as well as Don Quixote \citep{Mommert2020}. This method is limited to producing convex shape models and cannot recreate the concavities now known to be characteristic for comet nuclei (see below). Despite this limitation, CLI can accurately determine the object’s pole orientation and axial ratio with great precision. Another challenge posed by this method is that it requires a lot of observing time on comparatively large ($>$2-m) telescopes, given the faintness of bare comet nuclei. Additionally, only a small number of comets are inactive at large portions of their orbits, limiting the possibility to probe different observing geometries. However, an increasing number of bare nuclei have well-observed rotational lightcurves collected mainly to study changes in their rotation rates (see Section~\ref{sec:spin_state} below). The addition of data from future all sky surveys (see Section~\ref{sec:future}) may enable the shapes of additional comets to be constrained. Moreover, large flat surfaces on the asteroid convex shape models can be used to infer the existence of concavities \citep{Devogele2015}. Combined with large elongations, this could reveal more contact binaries among the known JFC population and has the potential to improve our understanding of the binary fraction among comet nuclei (see below).

In the rare occasions when a comet passes sufficiently close to the Earth, radar delay-Doppler imaging can be performed to constrain the shapes of comet nuclei (see Section \ref{sec:radar}). At the time of \citet{Harmon2004}'s writing, nine comets had Doppler-only detection with radar and none had delay-Doppler imaging. Thanks to technological improvements and a confluence of close approaching comets, at least eight now have delay-Doppler imaging of their nucleus: 300P/Catalina \citep{Harmon2006}, 73P/Schwassmann-Wachmann 3 fragments B \& C \citep{Nolan2006}, 8P/Tuttle \citep{Harmon2010}, 103P \citep{Harmon2011}, 209P/LINEAR \citep{Howell2014}, P/2016 (BA14 PANSTARRS) \citep{Naidu2016}, and 45P/Honda-Mrkos-Pajdu{\v{s}}{\'a}kov{\'a} \citep{Lejoly2017}. Doppler only detections since \citet{Harmon2004} include 252P, 289P/Blanpain, 41P/Tuttle-Giacobini-Kres{\'a}k \citep{Howell2017}, and 46P/Wirtanen \citep{Lejoly2019}. As evidenced by these studies, high-resolution radar imaging is occasionally possible. In such cases, precise shape modeling can be achieved by combining radar observations with optical light-curve modeling \citep[see][]{Ostro2002}.

Finally, owing to the space missions equipped with on-board cameras (see Snodgrass et al.\ in this volume) the shapes of six comets have been studied in great detail (1P, 9P, 19P, 67P, 81P, and 103P). Four were found to be highly elongated or potentially bi-lobed: 1P \citep{Keller1986}, 19P \citep{Britt2004,Oberst2004}, 103P \citep{Thomas2013}, and 67P \citep{Sierks2015}. Additionally, radar observations of comet 8P were consistent with a contact binary shape \citep{Harmon2010}.

The progress in characterizing comet nucleus shapes in the last decade revealed a striking overabundance of highly-elongated/bilobate objects in comparison to other minor planets. The comparison with other populations is somewhat complicated by the different definitions used by the different communities. Works focusing on NEAs often use a strict contact binary definition which sets a limit on the components' mass ratio and implies that the objects might have been separate in the past \citep[see][]{Benner2015}. TNO studies, on the other hand, are now making the first steps in understanding the shapes of individual objects and use a less restrictive definition \citep[e.g.,][]{Thirouin2019}. It is nevertheless informative to outline the contrasting findings for the different populations. Out of the six comets visited by spacecraft, four are highly-elongated/bi-lobed. If 8P is also considered, this sample of well-constrained comet shapes contains more than two-thirds highly-elongated/bi-lobed shapes. In comparison, only 14\% of the almost 200 radar-imaged NEAs are bilobate \citep{Taylor2011,Benner2015}. The large abundance of bilobate shapes cannot be traced to the centaur region where no contact binary has been identified \citep{Peixinho2020}, while the contact-binary fraction among TNOs is estimated as 10\%–25\% for cold classicals or up to 50\% for Plutinos \citep{Thirouin2018,Thirouin2019}. However, recently \cite{Showalter2021} presented evidence that the contact-binary fraction in the Kuiper Belt can even be higher if the shapes and the directional distribution of the objects’ rotation poles are accounted for.

The unusually large abundance of highly-elongated/bi-lobate objects among comets, prompted a number of works to investigate which combination of the comet's physical properties and/or evolutionary processes have led to the formation of a large number of contact binaries. This motivation was further enhanced by the finding that New Horizons' target in the cold classical Kuiper Belt, Arrokoth, is also a contact binary \citep{Stern2019}. Arrokoth's shape is consistent with  formation by merger of a collapsed binary system \citep{McKinnon2020}. However, unlike Cold Classical KBOs, the progenitors of JFCs have undergone giant-planet encounters and possibly a significant collisional evolution \citep{Morbidelli2020} which have most likely destroyed any distant binary systems early on. Instead, the formation of bilobate comet nuclei is better explained by the re-accretion of material ejected from catastrophic collisions in the early solar system \citep{Jutzi2017,Schwartz2018} or even from multiple fission and reconfiguration cycles \citep{Hirabayashi2016}. Alternatively, modeling work by \cite{Safrit2021} shows that bilobate shapes can also form after comet nuclei experience rotational disruption caused by sublimation-driven torques soon after the onset of activity (e.g. during the centaur phase). As shown by \citet{Zhao2021}, Arrokoth and by extension other icy bodies too could have evolved their shapes through sublimation to produce objects with enhanced elongated shapes if the conditions were right. It is remarkable that these scenarios not only reproduce the highly-elongated/bilobate shapes of comet nuclei but can also preserve the cometary physical properties and volatile content \citep{Schwartz2018}.

\subsection{Cumulative Surface Properties of the Comet Population}
\label{sec:surface_prop}

The merit in exploring the cumulative surface properties lies in its potential to reveal dependencies between the comets’ surface properties and the physical properties or orbital characteristics which could, in turn, shed light on cometary evolution. On the other hand, comparing the bulk properties of comet nuclei with other minor planet populations could be used to establish the dynamical and evolutionary links among the diverse small-body populations in the solar system.

\subsubsection{Albedo}
\label{sec:albedo}
The energy balance on the surface of a comet can be described using several photometric properties. The most frequently constrained is the geometric albedo $p_\lambda$ for a given wavelength $\lambda$, defined as the ratio between the disk-integrated reflectance at opposition and that of a perfectly reflective flat disk with the same size. If the phase function of the object at wavelength $\lambda$ is denoted by $\Phi$\edit{$_{\lambda}$}$(\alpha)$, where $\alpha$ is the phase angle, and observations cover a large phase-angle range (typically covering a phase angle range of 70$^\circ$ and above, \citealt{Verbiscer1988}), its phase integral $q_\lambda$ can be derived from: 
   \begin{equation}
   q_\lambda = 2\int{\Phi_\lambda(\alpha) \sin(\alpha) d\alpha}.
   \end{equation}
\noindent In such cases, the spherical albedo at wavelength $\lambda$, $A_\lambda$ (sometimes referred to as the Bond albedo), can be calculated as 
\begin{equation}
A_{\lambda}=p_{\lambda}q_{\lambda}. 
\end{equation}
\noindent Physically, this is the fraction of power of the incident radiation that is scattered back into space over all angles and wavelengths. See \citet{Hanner1981} for further discussion of terminology.

The phase integral has been constrained by in-situ spacecraft observations of five JFCs: 19P \citep{Li2007_Borrelly}, 9P \citep{Li2007_Tempel}, 81P \citep{Li2009}, 103P  \citep{Li2013} and 67P \citep{Ciarniello2015}. Besides 81P which has an exceptionally low phase integral of 0.16, the other JFCs have small phase integrals in the range 0.2-0.3, similar to small low-albedo asteroids \citep{Verbiscer2019}.

As indicated by Equation~\ref{eq:r_nucleus},
determination of reliable nucleus sizes requires knowledge of the geometric albedo. Albedo can be determined by coupling 
simultaneous observations of sunlight reflected by the nucleus with observations that depend on the nucleus size but {\it not} its ability to reflect sunlight (most commonly via thermal re-radiation in the mid-IR). The geometric albedo can be determined using ground- and space-based telescopes (see \citealt{Lamy2004} for details) and therefore, the sample of comet nuclei with well-constrained albedo in the visible range is comparatively large. The geometric albedos of JFCs were most recently reviewed by \cite{Snodgrass2011} and \cite{Kokotanekova2017}, while the HTCs and LPCs were last summarized in \cite{Lamy2004}.

As seen in Table~\ref{tab:surface} and Figure~\ref{fig:nucleus_histograms}, this sample of 29 comet nuclei (19 JFCs and 10 HTCs plus LPCs) contains $V$-band albedos ($p_V$) between 0.02 and 0.06. Even though the absolute range is small, albedos vary by a factor $\sim$3 from darkest to brightest. This distribution clearly identifies comets as some of the darkest objects in the solar system. The unweighted average albedo of the whole sample of comets is $0.040~\pm~0.011$ (median 0.040). Subdividing into JFCs or HTC/LPC yield virtually identical values. Thus, the common practice of assuming $p=0.04$ when albedo is unknown \citep[cf.][]{Lamy2004} continues to be reasonable. Since $p_{\lambda}~{\sim}~0.04$ and $q_{\lambda}~{\sim}~0.2-0.3$, this means that $A_{\lambda}~{\sim}~0.01$, and thus $\sim$99\% of the incident solar energy is absorbed by the comet.

Interestingly, as pointed out by \citet{Kokotanekova2018}, the objects with the largest geometric albedos are all comets with albedo estimates from spacecraft observations (9P, 19P, 67P, 81P). On the other hand, the darkest surfaces belong mostly to the largest \citep{Fernandez2016,Kokotanekova2017} and to the least active comets whose nuclei are easiest to characterize with telescope observations \citep{Kokotanekova2018}. The latter result may be an observational selection effect since small and dark objects will be more difficult to discover, but the former almost certainly is not -- spacecraft targets were primarily selected for orbital accessibility.

As compared to other populations, comet nuclei span a narrow range of geometric albedos. The average albedo of comet nuclei is comparable to that of C, D and P-type asteroids \citep{Mainzer2011} but unlike the asteroids of the respective classes, the comet population lacks objects with larger albedos.  The consistently low albedo of comet nuclei has been utilized as a criterion to distinguish extinct comets from asteroids in the ACO populations identified by using dynamical criteria \citep{Fernandez2001,Licandro2016}. Centaurs and Scattered Disk objects, thought to be the progenitors of current short-period comets, contain objects with albedos as low as those of comet nuclei. However, the mean geometric albedo of these populations is somewhat higher (0.056 for centaurs and 0.057 for SDOs) and they contain objects with albedos of up to 0.25. The known high-albedo surfaces of centaurs and TNOs are also associated with redder colors (larger spectral slopes in the visible) and form the ``bright red'' surface type with $18 < S' < 58$ \%/100 nm ($S'$ is the normalized reflectivity gradient; see Section~\ref{sec:color_and_spectra}), and albedo $>$0.06 that is thought to disappear with the onset of centaur activity (see \citealt{Jewitt2015}). On the other hand the ``dark grey'' centaurs have albedos similar to these of Jupiter trojans and Hildas \citep{Romanishin2018} which has been interpreted as evidence of the common origins of these populations and consequently between comet nuclei, Jupiter trojans, and Hildas.

An interesting question that has been difficult to answer broadly from telescope observations is whether comet surfaces have large-scale albedo variations. Synchronous visible/IR lightcurves \citep[e.g., 10P by ][]{Ahearn1989} suggest there are not, but can only be accomplished for a small number of low activity nuclei. It was, however, possible to search for surface areas with different albedos on the comets visited by spacecraft. While the brightness variations on the surface of 19P have been found to be up to a factor of 2 \citep{Buratti2004,Li2007_Borrelly}, they are thought to be caused by surface roughness variations rather than differences in the albedo. All other comets that have been measured have smaller albedo variations, ${\lesssim}16\%$; 81P \citep{Li2009}, 9P \citep{Li2007_Tempel}, 103P \citep{Li2013}, and 67P \citep{Fornasier2015}.

\vspace{-0.5mm}
\subsubsection{Phase Function and Phase Reddening}
\label{sec:phase_function}
The observed spectrophotometric properties of a reflecting surface are known to change depending on the phase angle, $\alpha$, of the observations. The spectral slope (as a function of wavelength) increase with $\alpha$ is referred to as phase reddening while the phase darkening, or phase function, describes the decrease of an object’s brightness with increasing $\alpha$. Moreover, at small phase angles $\alpha \leq 10^{\circ}$, the phase function can undergo a sharp nonlinear increase, known as the opposition effect \citep[e.g.,][]{Gehrels1956}.

These patterns are characteristic for airless regolith surfaces of solar system bodies. They can be analysed using physically motivated photometric models which relate the reflectance changes with the varying geometry and the properties of the object's surface layer. The currently most widely used models follow the Hapke formalism \citep[e.g.,][]{Hapke1981} and provide an opportunity to characterize the surface properties on small scales below the resolution of the observing instruments available on telescopes and most spacecraft. For example, careful modeling of the opposition effect or of the phase function beyond 90$^\circ$ can be used to derive properties of the surface regolith at small scales. The moderate angle phase function, on the other hand, can be used to study the surface roughness and topography \citep[see][]{Verbiscer2013}. In many cases, however, remote telescope observations of minor planets provide insufficient coverage to constrain the analytic photometric models. Instead, the phase function is approximated by a simpler parametric formalism, such as the IAU HG phase function \citep{Bowell1989} which generally works well for of asteroids.

The phase functions of comet nuclei present an observational challenge because they require relatively large telescopes that are able to detect the bare nuclei and a substantial amount of observing time in order to characterize the rotation rate and correct 
for rotational variability. Despite the diversity of comet orbits, in practice, all nucleus phase functions observed from the ground span a narrow phase angle range of less than 20$^\circ$ \citep[see][]{Kokotanekova2018} with the exception of 2P ($2.5-177^\circ$; \citealt{Fernandez2000}). None of the nucleus phase functions (reviewed in \citealt{Lamy2004}, \citealt{Snodgrass2011}, and \citealt{Kokotanekova2017}) provide evidence for the presence of an opposition effect and are in excellent agreement with linear phase function fits, hence its use in Equation~\ref{eq:phase_function}. Note that a different, non-linear phase function is used to characterize coma dust (e.g., Schleicher-Marcus model; \citealt{Schleicher2011})..

In situ observations during the fly-by missions covered a broad phase-angle range, but were insufficient to model the phase function close to opposition \citep{ Li2007_Tempel,Li2007_Borrelly,Li2009,Li2013}.  Rosetta’s rendezvous with 67P was the first opportunity to observe a comet nucleus close to opposition. Even though the phase function between 0.5$^\circ$ and 12.5$^\circ$ of 67P from ground-based observations was well described by a linear function with a coefficient  ${\beta} = 0.059~{\pm}~0.006$ \citep{Lowry2012}, the disc integrated phase function of 67P observed by OSIRIS between $1.3-54^\circ$ exhibited a strong opposition effect and was better characterized by an IAU $HG$ model with $G=-0.13$ \citep{Fornasier2015}. The shape of the phase curve during the opposition effect derived from disk-resolved observations of different nucleus areas also enabled studies of the structure and properties of 67P’s surface (e.g., \citealt{Masoumzadeh2017,Hasselmann2017}).

Table~\ref{tab:surface} lists the phase coefficients of 24 comets (20 JFCs and 4 HTCs and LPCs) ranging from 0.025 to 0.07 mag/$^\circ$. From this sample, the average phase coefficient for JFCs is $0.047~\pm~0.012$ mag/$^\circ$ (with median 0.046 mag/$^\circ$). As shown in  Figure~\ref{fig:nucleus_histograms}, the phase function slopes measured for LPCs and HTCs are smaller than the JFC average, though with the caveat that there are only five combined LPC plus HTC measurements. All phase coefficients determined from in-situ measurements, presumed to provide the most reliable estimates, are larger than the commonly used phase coefficient of 0.04 mag/$^\circ$ \citep[e.g.][]{Lamy2004}. We therefore recommend an updated value of 0.047 mag/$^\circ$ to be assumed as an approximation when the phase coefficient of a nucleus is unknown.

Analyzing disk-resolved observations with VIRTIS, on-board Rosetta, \citet{Longobardo2017} found that terrains with larger roughness have steeper phase functions, At the same time a comparison with the phase functions from previous missions indicated that nuclei known to have rougher surfaces (81P and 9P) have steeper phase functions than 103P and 19P, characterized by smoother surfaces. Interestingly, dynamical studies of these five comets by \citet{Ip2016} suggested that the comets with smoother surfaces and shallower phase functions have spent more time in the inner solar system and have experienced more sublimation-driven erosion. 

\begin{figure}[t!]
\begin{center}
\includegraphics[width=3.3in]{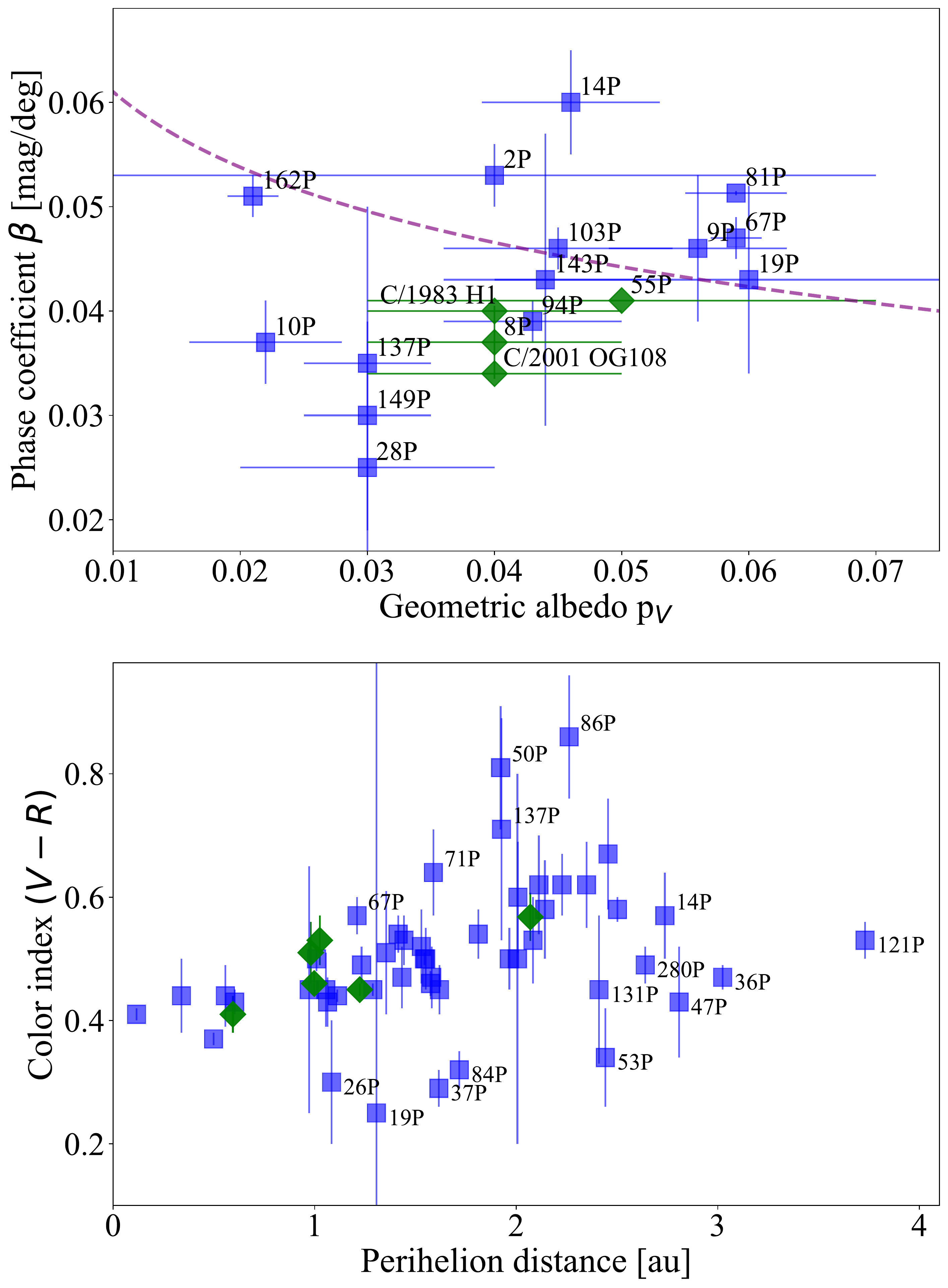}
\caption{Plots of the two highest correlations we found in the datasets compiled here: phase coefficient as a function of geometric albedo (top panel) and $(V-R)$ color versus perihelion distance (bottom panel). The blue squares are JFCs and the green diamonds are HTCs and LPCs. The purple dashed line in the top panel is the correlation between phase coefficient and albedo for asteroids \citep{Belskaya2000}. A recent update of 162P's phase coefficient to a higher value has made the trend with albedo first identified in \citet{Kokotanekova2017} less pronounced. As identified by \citet{Lamy2009} and discussed further in Section~\ref{sec:color_and_spectra}, there is a group of 10 comets that run parallel to, but below the main cluster in the color versus perihelion distance plot.}
\label{fig:trends}
\vspace{-2.5mm}
\end{center}
\end{figure}

The notion that the phase functions of comet nuclei can reveal the extent of their surface evolution was strengthened when ground observations of JFC nuclei were included in the sample. \cite{Kokotanekova2017} identified a possible correlation of increasing phase coefficient with increasing albedo for the 14 JFCs with both of these parameters constrained (updated version shown in Figure~\ref{fig:trends}). This was interpreted as a possible evolutionary path of JFCs where the least evolved comet nuclei have bright surfaces with steep phase possibly associated with volatile-rich surface layers and topography dominated by deep pits and cliffs \citep[see][]{Vincent2017,Kokotanekova2018,Vincent2019}. This hypothesis suggests that sublimation-driven evolution is responsible for gradually eroding their surfaces which, in turn, decreases their albedos and phase coefficients. If confirmed by future observations, this idea provides a compelling possibility to characterize the evolutionary state of comets and to distinguish between extinct comets and objects of asteroidal origin \citep{Kokotanekova2018}. Alternatively, it is also conceivable that better phase-angle coverage and therefore better characterized nucleus phase functions in the future will challenge this hypothesis. For example, the phase coefficient of 162P was updated to a significantly larger value of 0.051 $\pm$ 0.002 mag/$^\circ$ \citep{Donaldson2023} which leads to a weaker Spearman rank correlation ($\rho = 0.43$) as compared to the sample in \cite{Kokotanekova2018}. Additionally, the smallest known JFC phase coefficient, that of 28P, is also likely to be much larger, $\sim$~0.05mag/$^\circ$ \citep{Schleicher2022}.

Nevertheless, it is interesting to note that asteroids follow a reverse correlation of decreasing phase coefficient for increasing geometric albedo \citep{Belskaya2000}. However, their dataset does not allow conclusive results for objects with albedo similar to the darkest comet nuclei (smaller than 0.05). Another comparison between comet nuclei and asteroids worth considering is the presence of an opposition effect. The opposition effect amplitude (defined as the magnitude difference between the phase function at 0.3$^{\circ}$ and the extrapolation of the linear part of the phase curve) is found to decrease for small albedos \citep{Belskaya2000}. Moreover, some low-albedo Hilda asteroids and Jupiter trojans do not show evidence for the presence of an opposition effect down to very small phase angles (0.1$^\circ$, \citealt{Shevchenko2008}). It is conjectured that the strong opposition effect found for 67P can be related to its comparatively large albedo. It is therefore essential to probe the phase functions of more comet nuclei of different albedos in order to verify how good of an approximation the linear phase function is.

Although attempts were made to determine the phase reddening of earlier fly-by mission targets, they were not successful. The long duration of the Rosetta rendezvous allowed the first detections of phase-reddening of a nucleus in the visible and near-infrared range \citep{Ciarniello2015,Fornasier2015}. The spectral slopes at different phase angles were analyzed for both disk-integrated and disk-resolved data and show a significant phase reddening in the $1.3^{\circ}-54^\circ$ phase angle range (from spectral slope $S'$ of 11\%/(100 nm) to 16\%/(100 nm); \citealt{Fornasier2015}). In addition to this, the phase reddening was observed to vary with the changing heliocentric distance, first decreasing toward perihelion and then increasing again in the outbound orbit \citep{Fornasier2016}, indicating a significant seasonal variability.

Although the opposition effect and phase reddening have not yet been detected in ground-based data, both may be feasible to detect in the coming era (Section~\ref{sec:future}), and future investigators should be mindful of the possibility when analyzing their data. As shown by recent findings, the spectrophotometric properties of comet nuclei can be used to probe not only the physical characteristics of the surface layers of comets but also their sublimation-driven evolution. This highlights the need for future photometric and spectroscopic observations of bare nuclei both at different geometries and heliocentric distances.

%\vspace{-0.5mm}
\subsubsection{Color and Spectra}
\label{sec:color_and_spectra}
Provided that sufficiently sensitive instruments are used, spectroscopy is the most desirable means of characterizing the nucleus surface as, in principle, it can reveal both precise color determinations and the presence of any absorption features due to surface ice or minerals. This compelling possibility motivated the early comet nucleus spectroscopy study by \cite{Luu1993} which, however, revealed mostly featureless spectra. The absence of absorption features in the visible and near-IR has been confirmed by all subsequent spectroscopic observations (previously reviewed by, e.g.,  \citealt[][]{Lamy2004, Kelley2017, Licandro2018}). As a result, the normalized reflectivity gradient $S'$ (\% per 100 nm) defined by \citet{AHearn1984} is the most commonly used parameter to characterize nucleus spectra.
For visible wavelengths, reflectance is often normalized to 1 at some reference wavelength, frequently chosen to be 550 nm.

Though less informative than spectroscopy, the color index is determined from photometry by measuring the bare nucleus brightness in two filters and subtracting the magnitude at the longer wavelength from the shorter, e.g., $V-R$. Since photometric techniques are generally more sensitive than spectroscopy, this allows the colors of fainter nuclei to be measured with the caveat that they risk contamination by emission lines if unrecognized coma is present. This technique has been used to obtain the color indices of a wide variety of comet nuclei, building statistically significant samples covering the different classes of comets.  A direct expression provided in \cite{Luu1990c} allows an easy conversion between $S'$ and the color index in the corresponding wavelength range (e.g., $V-R$, $B-R$, $R-I$, $g-r$, etc.) and allows for comparison of the surface colors of comets observed with either spectroscopy or photometry to other solar system populations. An important confirmation of the utility of ground-based spectroscopy was the finding that the spectrum of 67P near aphelion \citep{Tubiana2011} was consistent with the visible and near-IR spectrum from OSIRIS on-board Rosetta \citep{Fornasier2015}.

In the visible to near-IR, comet surface spectra have spectral gradients up to 20\% per 1000~{\AA} (see \citealt{Campins2007,Jewitt2002,Kelley2017}). While some spectra have been classified as closer T- and X-type asteroids \citep{DeMeo2008}, most comet spectra have been identified to be closest to those of the primitive D-type asteroids which have similar spectral slopes (9.1 $\pm$ 1.1 \% per 1000~{\AA}; \citealt{Fitzsimmons1994,Fornasier2007}) and to some moderately red centaurs and TNOs \citep{Fornasier2009}. In the near-IR, comet nuclei exhibit spectral diversity similar to that known for Jupiter trojan asteroids \citep[and references therein]{Emery2003,Campins2007}. The similarity between comets and Jupiter trojans extends to further spectral features such as the 10-$\mu$m emission plateau attributed to a surface layer of fine dust \citep{Kelley2017,Licandro2018} and has been used to investigate the possible dynamical links between short-period comets and Jupiter trojans proposed by dynamical studies \citep[e.g.,][]{Morbidelli2005}. 

Since the existing comet surface spectra probe mainly weakly active or extinct comets (e.g. 28P, 49P, 162P, 249P/LINEAR, 364P/PANSTARRS, 196256 (2003 EH1), P/2006 HR30 Siding Spring), the spectra of these objects have been used to study the transition to dormancy. It is a notoriously difficult problem to distinguish inactive objects with a cometary origin in the outer solar system (ACOs) from asteroids using purely dynamical criteria (see \citealt{Fernandez2005} and \citealt{Tancredi2014}). Probing the spectra of ACOs is therefore considered to provide important clues on understanding the differences between these populations \citep{Licandro2018,Simion2021}. Additionally, increasing numbers of so called ``active asteroids'' \citep[objects on typically asteroidal orbits that exhibit cometary activity; cf.][]{Hsieh2017} and short-period comets whose dynamics suggest an origin in the asteroid belt \citep{Fernandez2015,Hsieh2020} are being discovered. Considering this complexity, attributing any observed spectral signatures to the objects' origin and evolution can be enhanced if combined with dynamical studies. This approach was assumed in the recent work by \citet{Kareta2021} who compared the reflectance spectra of three different  dormant comets and inactive solar system small body 2003 EH1 in an attempt to study the diversity among the spectra of dormant comets and objects identified as meteor shower sources.

Previously, the surface colors of comet nuclei were reviewed by \cite{Lamy2004}, \citet{Lamy2009}, and \citet{Solontoi2012}; we have updated these reviews with subsequently published colors in Table~\ref{tab:surface} and Figure~\ref{fig:nucleus_histograms}. \cite{Jewitt2015} calculated the average color indices $B-V$, $V-R$, $R-I$, and $B-R$, distinguishing between nucleus and coma observations of JFCs and LPCs, and comparing them to the different TNO populations, centaurs, Jupiter trojans, and damocloids. This study reinforced  previous findings that the average colors of the coma of JFCs and LPCs and centaurs are indistinguishable from the nucleus colors within the uncertainties. Additionally, the comparison between the nucleus colors of JFCs, LPCs, damocloids, and trojans are indistinguishable within the uncertainties for each class \citep{Jewitt2015}. 

One of the main motivations behind comparing the surface colors of comet nuclei with centaurs and TNOs is understanding the absence of the so called ``ultrared'' material on comet nuclei \citep{Jewitt2002}. While a significant fraction of centaurs and TNOs contain the very red surfaces of spectral slopes $S'>25$\% \cite[and references therein]{Lacerda2014}  and $B-R > 1.5$ \citep[see][]{Tegler2008,Peixinho2012, Fraser2012,Tegler2016,Wong2017} alongside more neutral surfaces, none of the observed comet nuclei exhibit such extreme surface colors (with an average $B-R$ of 1.22 $\pm$ 0.03 and 1.37 $\pm$ 0.08 for LPC and JFC nuclei, respectively; \citealt{Jewitt2015}). According to \citet{Jewitt2015}, the disappearance of the ultrared matter coincides with a perihelion distance of $\sim$10~au where centaur activity is observed to begin and is hypothesized to be due to blanketing of the primordially red material by fallback material ejected during the onset of activity. 

Despite the effort to link surface color to different orbital and nucleus properties (perihelion distance, inclination, nucleus size, and active fraction) of comets, no statistically significant correlations have been found to date \citep[cf.][]{Lamy2009, Jewitt2015}. 
We investigated correlations between our expanded collection of nucleus properties and various dynamical parameters (perihelion distance, aphelion distance, semi-major axis, eccentricity, inclination, and $T_\mathrm{J}$). The only relationship evincing some correlation is color with perihelion distance (plotted in Figure~\ref{fig:trends}), with bluer nuclei having smaller perihelion distances (and a coupled trend of higher eccentricity JFCs having bluer nuclei), mimicking the behavior seen in the asteroid belt \citep[e.g.,][]{Gradie1982} and among near-Earth asteroids \citep[e.g.,][]{Marchi2006}. Previously noticed by \citet{Lamy2009}, this apparent trend is not statistically significant (Spearman rank 0.47 for JFCs or 0.44 for all comets in our sample). However, similarly to \citet{Lamy2009}, we identify two distinct groups in the $(V-R)$ versus $q$ distribution. The main group consisting of 39 JFCs. Considering only comets from this group, results in a Spearman rank of 0.8, suggestive of a moderate to high correlation. Notably, all HTCs from the table also agree with the observed trend while both LPCs have perihelia much larger than 3~au and do not follow the observed trend. The second group consists of 10 JFCs which are bluer than the other group and follow a similar pattern of increasing redness with increasing heliocentric distance. We hypothesize that the existence of the second group can be connected with recent orbital changes, but interpreting these results is beyond the scope of this chapter.

As discussed briefly at the beginning of Section~\ref{sec:what_we_know}, we found no statistically significant correlations between surface color and the geometric albedo or phase coefficient for the JFCs. Owing to the paucity of polarization data (see next subsection), its correlation with other surface properties remains unexplored. The above links should be explored further with the precise color determinations expected from all-sky surveys such as the Vera Rubin Observatory's Legacy Survey of Space and Time (LSST).

%\vspace{-0.5mm}   
\subsubsection{Polarization}   
\label{sec:polarization}
Polarimetry is the study of the degree of polarization of light. Light becomes polarized -- its electromagnetic wave having a preferred plane of oscillation -- by scattering off of particles. The degree of linear polarization can be easily calculated (see \citealt{Kolokolova2004} for details), and its variation reveals information about the particles from which it scattered. The degree of linear polarization of airless solar system bodies changes with phase angle and wavelength, producing a phase-polarization curve with a negative branch between 0$^\circ$ and 20$^\circ$ and a positive branch for higher phase angles that peaks in the range $90-100^\circ$ \citep[cf.][]{Kolokolova2004}.
Characterizing the phase-polarization curve of comet nuclei is limited by the faintness and the narrow phase angle range when comets are sufficiently far from the Sun to be inactive.  Although polarimetric observations of comets date to the early 1800s (reviewed by \citealt{Kiselev2015}), it has only been applied to comet nuclei since 2004.

The first polarization studies of a comet nucleus were by \cite{Jewitt2004} who studied 2P at phase angles of 22$^\circ$ as well as 93–100$^\circ$. Later, \cite{Boehnhardt2008} observed the coma-free nucleus of 2P between 2.7 and 2.1~au, constraining its linear phase-polarization curve between 4$^\circ$ and 28$^\circ$ in $R$ and $V$-band. In this range, the polarization in both bands increases linearly with increasing phase angle, suggesting an inversion angle $\leq$4$^\circ$. Similar low values of the inversion angle are characteristic for F-type asteroids \citep{Belskaya2005} and have been found for active asteroid 107P/Wilson-Harrington \citep[see][]{Belskaya2019} and main-belt comet 133P/Elst–Pizarro \citep{Bagnulo2010}. 

\citet{Kuroda2015} observed the linear polarization of the highly anemic 209P at large phase angles from 92.2$^\circ$ to 99.5$^\circ$. This was possible because 209P was very weakly active even at heliocentric distance $\sim$1~au and the nucleus polarization could be determined separately from that of the coma. They found no significant difference in 209P's $R$ and $J$-band polarization and estimated that maximum polarization $P_\mathrm{max} = 30.8$\% is reached around the phase angles of the observations. Interestingly, the phase angle at which maximum polarization occurs coincides with that of asteroids, but the value of the maximum polarization of 209P is significantly higher. \cite{Boehnhardt2008} and \cite{Kuroda2015} explored whether existing empirical phase-polarization relationships for asteroids can be applied to comet nuclei. This subject remains to be explored further when the polarization of more comet nuclei is characterized in detail. 
     
%\vspace{-0.5mm}
\subsection{Spin State}
%\vspace{-0.5mm}
\label{sec:spin_state}
\subsubsection{Overview and Explanation of Terminology}
\vspace{-1mm}
The spin state, or the rotational state, of a comet fully describes the rotational properties of the nucleus. In this chapter, we use the terms ``spin state'' and ``rotational state'' interchangeably. A comet could be in (a) the least rotational kinetic energy ``simple rotation'' around its maximum moment of inertia (the latter is physically identified with the short axis, or more precisely the principal short axis defined by the moment of inertia ellipsoid) or (b) in an excited rotational state. An ``excited'' rotational state, by definition, has more rotational kinetic energy for a given rotational angular momentum than the least energy state described earlier.

A simple (i.e., a principal-axis) rotation requires three independent parameters and an initial condition to uniquely define it (the rotation period, two parameters defining the direction of the rotational angular momentum vector such as {{\it RA}} and {\it{Dec}}, and the orientation of a reference longitude of the nucleus at a specific time). In theory, a simple rotation could represent a rotation around any one of the three principal axes. However, a simple rotation around the intermediate principal axis or that around the long principal axis is extremely rare and both can be changed due to mechanical damping of energy. Thus, in this chapter, the term ``simple rotation'' means the least kinetic energy rotation around the short principal axis. The observational techniques for identifying and analyzing properties of simple rotations are discussed in the review by {\citet{Samarasinha2004}} in {\it Comets~II} and thus not repeated here.

Except for the Principal-Axis (PA) rotational states around the intermediate principal axis or the long principal axis, a rotationally excited nucleus undergoes three concurrent component rotations, of which two are dynamically coupled, and such a spin state is known as a Non-Principal-Axis (NPA) rotation. In general, these component rotations occur at rates that vary with time but are periodic and therefore they can be expressed by time-averaged periods (e.g., \citealt{Landau1976}).

To uniquely define an NPA spin state, one requires six independent parameters and two initial conditions. These specific parameters and other details of the NPA rotation are described in {\it Comets~II}. The NPA rotational states are the most general spin states and PA spin states can be considered limiting cases. The vast majority of cometary nuclei are observationally determined to be PA rotators in the least energy state; however, it is likely that many of them might be in slightly excited NPA states and groundbased observations are not sensitive enough to detect them (such as 67P; \citealt{Gutierrez2016}).

\subsubsection{Detection of NPA Rotation}
As remarked earlier, many cometary nuclei appear to be simple rotators. However, delving deeper into the characteristics of NPA rotation helps us comprehend the observations of cometary spin and spin evolution of comets coherently. As described in the previous subsection, an NPA spin state is extremely complicated. Now, consider the possible NPA states of a nucleus that is near, but not exactly prolate. This nuclear shape is considered since different space mission images show many cometary nuclei are like that; lightcurves of nuclei, too, provide additional supporting evidence for this prevalent shape. Such a nucleus, if only slightly excited, is likely to be in a Short Axis Mode (SAM) whereas a highly excited nucleus will be in a Long Axis Mode (LAM) NPA state. The SAM and LAM NPA states are described in detail in {\citet{Samarasinha2004}} of {\it Comets~II} and see their Figure 1 for visualizations of the NPA rotations described here.

The near-prolate nucleus undergoing a SAM NPA rotation has its long axis (a) rotating around the rotational angular momentum vector at a time-averaged angle of 90\textdegree, (b) a roll back-and-forth motion around the long axis itself with a back-and-forth half-rolling angle between 0{\textdegree} and 90\textdegree, and (c) a low-amplitude (i.e., a few degree) nodding motion. For a distant observer, the resultant lightcurve would be similar to that of a simple rotator within the observational uncertainties. Furthermore, if coma morphology is due to source region(s) near the end of the long axis, the coma morphology too would mimic a simple rotation. Only if the source region is located on the waist of the nucleus and the half-rolling angle is $\gg$0\textdegree, might the coma morphology hint an NPA rotation. 

Similarly, if the nucleus is in a LAM NPA state, the long axis has (a) a precessional motion around the rotational angular momentum vector and making an angle $<$90\textdegree, (b) a full rolling motion around itself, and (c) a low-amplitude (i.e., a few degree) nodding motion. Again, the lightcurve may not be distinguishable from a simple rotator nor would the coma morphology caused by a source region(s) near the end of the long axis indicate an NPA rotation. However, a source region on the waist of the nucleus could point to an NPA state. 
 
 These examples point to the difficulty that observers face in conclusively identifying NPA rotators, especially if the signal-to-noise of the data is not high or there are not extensive data from multiple observing geometries. It would not be surprising if future observations indicate that some comets which we now consider to be simple rotators are in NPA spin states, or if discrepancies are found between simple/NPA spin states for different observational techniques.

\subsubsection{Observational Evidence for NPA Spin and Excitation into and Damping of NPA Spin}
\label{sec:NPA_spin_changes}

So far, only two comets -- 1P (e.g., \citealt{Schleicher1990, Belton1991, Samarasinha1991}) and 103P (e.g., \citealt{AHearn2011,Samarasinha2011, Knight2011,Drahus2011}) -- have been definitively shown to be in NPA rotation from remote observations. In both cases, extensive observations were required, a necessity that makes NPA rotation difficult to conclusively establish. Rosetta observations of 67P demonstrated that it is also excited, albeit at a level too low to detect from Earth \citep{Gutierrez2016,Kramer2019a}. Several other suspected NPA rotators are indicated by \textsuperscript{b?} in Table~\ref{tab:physical}. Possible NPA spin states were suggested to explain conflicting observations or observations which were deemed incompatible with simple rotations. Clearly additional observations and analyses are needed to confirm NPA states for these comets.

An NPA spin state has been proposed for active centaur 29P/Schwassmann-Wachmann 1 with component periods of the order of a day \citep{Meech1993}. However, recent modeling based on multiple disparate data sets suggest a much longer rotation period for that comet (e.g., \citealt {Miles2016, Schambeau2017, Schambeau2019}).

The fraction of periodic comets observed to be in NPA states is much smaller than what the theory suggests based on the timescale for rotation period change of a simple rotator, $\tau_\mathrm{P}$, due to outgassing torques \citep{Samarasinha1986, Jewitt1997}.
This timescale is indeed a lower limit to the timescale for rotational excitation; it will also depend on the relative component moments of inertia of the nucleus and the components of the torque. The rotational excitation of a nucleus is a complicated process and \citet{Gutierrez2003b} cite the following as a necessary (but not sufficient) condition for a nucleus to get rotationally excited and to remain that way.
    \begin{equation}
    \frac{N_\mathrm{l}}{N_\mathrm{s}} >
    \frac{(I_\mathrm{s}-I_\mathrm{i})}
    {(I_\mathrm{i}-I_\mathrm{l})}
    \frac{|\Omega_\mathrm{s}|}
    {|\Omega_\mathrm{l}|}
\end{equation}
where $N$, $I$, and $\Omega$ represent torque, moment of inertia, and angular velocity of the nucleus, respectively whereas 
$\mathrm{l}$, $\mathrm{i}$, and $\mathrm{s}$ represent respective components along long, intermediate, and short principal axes.
From this condition, it can be seen that as near-prolate nuclei have nearly equal $I_\mathrm{s}$ and $I_\mathrm{i}$, such nuclei are comparatively easy to excite. Also, slow rotators (i.e., small $\Omega_\mathrm{s}$) are easier to excite than fast rotators (i.e., large $\Omega_\mathrm{s}$) and the former have comparatively small rotational kinetic energies. An explanation is presented in Section~\ref{sec:rot_period_change} as to why small nuclei are comparatively easier to excite. The reader is referred to \citet{Gutierrez2003b} for an in depth discussion on the process of rotational excitation.

Even among the periodic comet nuclei, which experience recurrent torques due to outgassing, only a very small fraction of comets are observed to be in NPA states. In addition to the observational biases against discovering NPA states discussed earlier, a nucleus may be  prevented from being excited if the damping timescale is shorter than the excitation timescale. A flexible nucleus (e.g., rubble-pile like) will have shorter damping timescales and most comets are likely to be structurally flexible objects with low tensile strengths. The loss of mechanical energy of a non-rigid body in an excited spin state causes it to ultimately rotate around the maximum moment of inertia. The corresponding damping timescale, $\tau_\mathrm{damp}$, was derived by \citet {Burns1973} to be ${\tau}_\mathrm{damp}~{\propto}~r_\mathrm{n}^{-2}\Omega^{-3}$,  where $r_\mathrm{n}$ is the effective nuclear radius and $\Omega$ is the effective angular velocity. An immediate conclusion is large comets that are fast rotators have comparatively short damping timescales and are thus less likely to be in NPA spin states.
\citet {Samarasinha2008} discusses difficulties associated with deriving accurate damping timescales primarily due to our poor understanding of the values for structural parameters of comets. For example, for a 1~km radius nucleus with an effective rotation period of 1~day, ${\tau}_\mathrm{damp}$ can easily vary from ${\sim}10^4$ to ${\sim}10^{10}$ years depending on the structural flexibility of the nucleus. Therefore, ${\tau}_\mathrm{damp}$ could be of the same order as the orbital period for a few km sized fast rotating nucleus that is extremely flexible. On the other hand, ${\tau}_\mathrm{damp}$ could be as large as the age of the Solar System if the nucleus is comparatively more rigid. As most nuclei appear to be simple rotators (or close to that), it is likely that this is another pointer to the low rigidness of cometary nuclei.

%\vspace{-0.5mm}    
\subsubsection{Range of Measured Rotation Periods}
Among the rotational parameters that define the spin state of a comet, the rotation period of the nucleus is the parameter that has measurements for the largest number of comets, albeit a relatively small number when compared with the asteroids with known rotation periods. This paucity of data is primarily due to impediments caused by cometary comae  when deriving nuclear lightcuves of comets close to us (see Section~\ref{sec:how_we_know}). 

Rotation periods can be determined from rotational lightcurves, coma morphology, or radar. As the latter two techniques were described in Section~\ref{sec:how_we_know}, we only discuss lightcurves here. Ideally, the flux in the photometric aperture would be dominated by the nucleus, but  techniques that measure the coma flux have been successful for some comets. Common techniques for determining the correct rotation period from a lightcurve include phase dispersion minimization (commonly called PDM; \citealt{Stellingwerf1978}), Fourier transformations \citep[e.g.,][]{Belton1990}, and simply aligning the lightcurves by eye.  When the correct rotation period has been identified, data phased to this period will have their features (peaks and troughs) from different rotation cycles aligned with each other. Although rotational lightcurves are ideally acquired close together in time so that the viewing geometry is nearly constant, recent work by \citet{Kokotanekova2017} has shown that bare nucleus data acquired at widely separated epochs can reliably be combined in some circumstances.

The range of measured rotation periods of unambiguous comets varies from approximately four hours to multiple days (see Table~\ref{tab:physical} and Figure~\ref{fig:nucleus_histograms}).  Clearly this distribution is affected by observational selection effects. For example, to confidently derive a particular rotation period, typically an observational window longer than the rotation period is required, and therefore, there is a bias against deriving long rotational periods. Similarly, groundbased lightcurve observations have a bias against detecting rotation periods which are multiples of diurnal observing windows (e.g., periods near 12 and 24 hours) since nearly the same portion of the lightcurve is seen from night to night at the same location. Rotation periods based on the repetition of coma morphology have a distinct disadvantage against detecting short rotation periods because of the higher likelihood of coma structures getting smeared. The smearing occurs when the spatial separation of repeating features is less than the image resolution. Short rotation periods and low outflow speeds (e.g., of dust when compared with that of gas) cause closely spaced repetitive features whereas large geocentric distances cause low spatial resolution images.

In addition, there are a number of observational biases against measuring accurate  rotation periods. They include lightcurve observations with small amplitude variability, either because of the intrinsic nuclear shapes or due to the specific observing geometries (for a nucleus lightcurves) or due to the diurnal activity pattern (for a coma lightcurve), and difficulties associated with detecting periods of NPA rotations discussed in the previous subsection.
    
Finally, different observational techniques could yield different manifestations of nuclear rotation (e.g., a nuclear lightcurve indicating a double-peaked lightcurve vs. coma morphology showing a periodic coma structure originating from a particular location on the nucleus vs radar observations of the nucleus) and care must be made to accurately interpret them and provide a consistent picture.

\subsubsection{A Comet Spin Barrier?} 

\begin{figure}[t!]
\begin{center}
\includegraphics[width=3.3in]{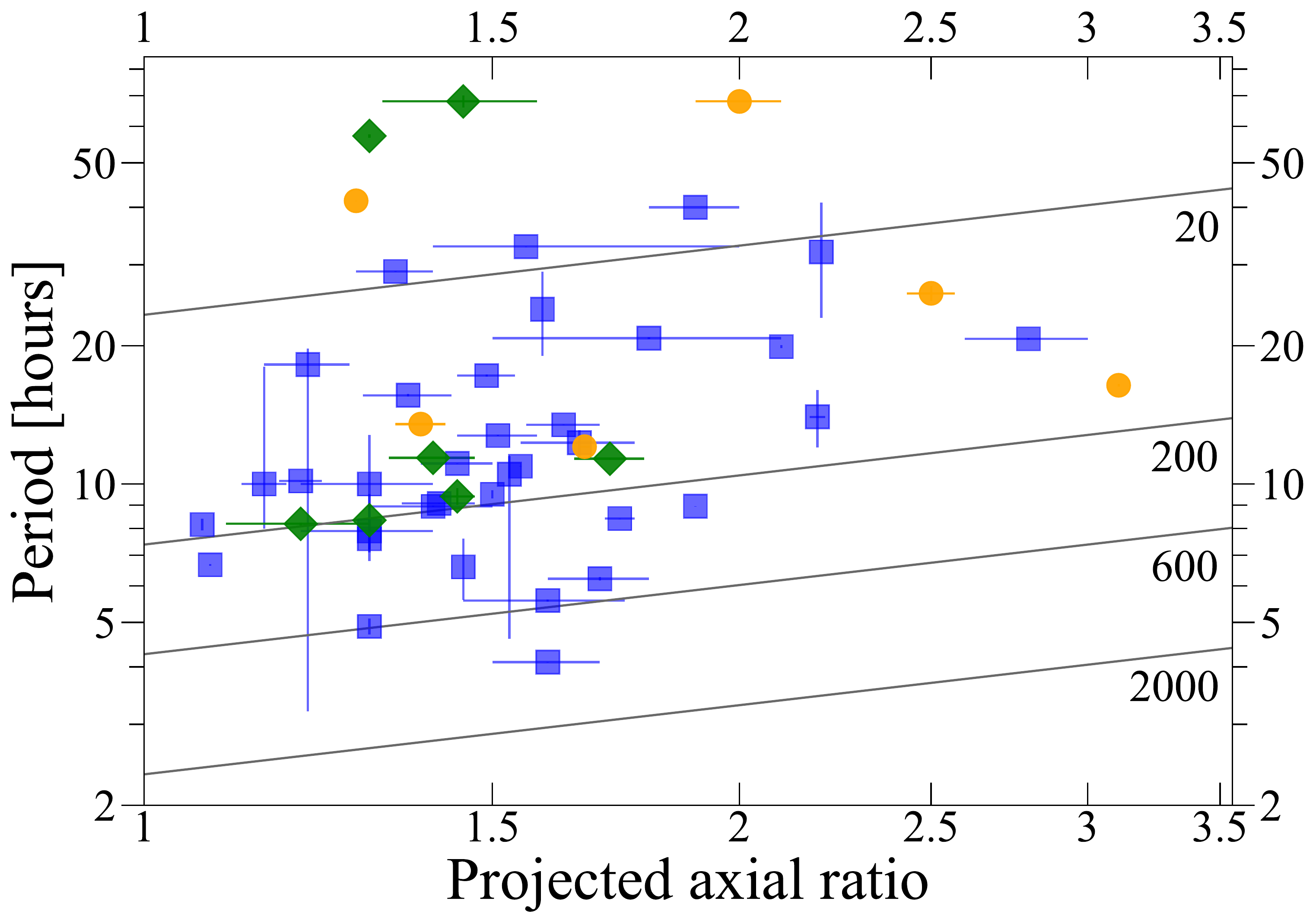}
\caption{Rotation period against projected axial ratio for comet nuclei, updated from \citet{Kokotanekova2017}. The blue squares are JFCs and the green diamonds are HTCs and LPCs. For these points, the axial ratio is a lower limit and the uncertainties are plotted when they were stated by the authors. When authors provided multiple equally likely periods, the shortest rotation period is plotted since this is most constraining.
Orange circles are comets having shape models derived from spacecraft data. The diagonal lines indicate the minimum density (denoted in kg/m$^{3}$ to the right), which a strength-less body of the given axial ratio and spin period requires to remain intact. Apart from the unusual case of 96P, which is undergoing continued fragmentation \citep{Eisner2019} potentially related to rotational spin-up (see Section~\ref{sec:causes_of_fragmentation}) no comet requires a density greater than $\sim$600 kg/m$^{3}$ to remain stable against rotational 
breakup.}
\label{fig:P_vs_elong}
%\vspace{-7mm}
\end{center}
\end{figure}

Figure~\ref{fig:P_vs_elong} plots rotation period versus axial ratio. This can be used to derive a lower limit to the bulk density of the nucleus under the strength-less body assumption.  \citet{Lowry2003a} derived a lower limit to the density of $\sim$600 kg/m$^3$ from such a plot -- corresponds to a minimum rotation period near 6 hours (also see \citealt {Weissman2004}, \citealt{Lamy2004}, and \citealt{Kokotanekova2017}) -- a value consistent with direct density measurements for comet 67P by the Rosetta mission \citep{Jorda2016}. A review by \citet{Groussin2019} found a mean bulk density of $480 \pm 220$ kg/m$^3$ from 20 published estimates using a variety of techniques (see also Guilbert-Lepoutre et al., in this volume). A lower limit to the density estimated from a plot such as that in Figure~\ref{fig:P_vs_elong} is derived for an ensemble and is extremely sensitive to a small number of comets spinning at periods near their rotational breakup periods. Thus this estimate should be considered only as a gross ensemble property but not as a density estimate for a particular comet.

In contrast to this ``minimum'' rotation period of $\sim$6~hours mentioned above for comets, the corresponding spin barrier for comparatively-sized asteroids occur at $\sim$2.2~hours (\citealt{Warner2009}, also see \citealt{Hu2021}). For comparatively-sized Jupiter trojans the barrier occurs at $\sim$4~hours (\citealt{Chang2021}) due to their higher densities than those of comets.

As noted earlier, there is a slight correlation between period and axial ratio. We suspect that this is an observational bias, due to the difficulty in detecting both low amplitude lightcurves and long rotation periods. However, fast rotators could also be missed, regardless of amplitude, if observations are not frequent enough. Future analyses with more robust observational data, e.g., from LSST, should investigate this further, but observational biases should be accounted for prior to interpreting the results.

\subsubsection{Evidence and Implications of Detected Rotation Period Changes}    
\label{sec:rot_period_change}
The idea that the rotation of comets could be altered by sublimation of gases dates back to the seminal work of \citet{Whipple1950} in which he introduced the ``icy conglomerate'' model of the nucleus. In \citet {Whipple1979}, this concept was further expanded to examine how the torque caused by the jet force due to sublimation can result in the precession of the rotation axis of a non-rigid body nucleus rotating around its maximum moment of inertia. However, the actual situation could be much more complex. The net torque due to outgassing could cause rotational excitation to an NPA spin state as well as changes to its period(s) of rotation, in addition to the forced precession of the direction of the rotational angular momentum vector (e.g., \citealt{Samarasinha2004} and references therein). As the rotation period has the largest spin parameter database including the most number of determinations on their temporal changes, we now concentrate on the current state of known rotation period changes.
    
Rotation period changes require precise measurements and typically (but not always) observations spanning multiple apparitions, making them challenging to determine. Just over a dozen have been measured or meaningfully constrained (see Table~\ref{tab:period_change}), most since {\it Comets~II}.

\citet{Samarasinha2013}, based on four periodic comets, pointed out that the rotation period changes seem to be independent of the active fraction of a nucleus where active fraction is the fractional surface area of a solid water ice nucleus needed to reproduce the observed water production of the comet. They introduced a parameter $X$ given by
    \begin{equation}
    X=\frac{|\delta P|r_\mathrm{n}^{2}}{P^{2}\zeta}
\end{equation}
where $P$ is the rotation period, $|\delta P|$ is the absolute value of the change of rotation period per orbit, $r_\mathrm{n}$ is the effective nuclear radius, and $\zeta$ is the total water production per unit surface area of the nucleus per orbit. It was found that $X$ was nearly constant for all comets despite a large scatter in the range of active fractions. They suggest highly active comets undergo more cancellation of torques. \citet{Mueller2018} added two more comets and found $X$ is still nearly constant. This near-constancy enables one to estimate $|\delta P|$ for various comets. \citet {Steckloff2018} proved that the sublimative torque model of \citet{Steckloff2016} is consistent with the $X$ parameter model.

In Table~\ref{tab:period_change}, $X$ is confined to approximately an order of magnitude and the geometrical mean of $\frac{X}{X_\mathrm{E}}$ of the 11 JFCs is $\sim$2 where $X_\mathrm{E}$ is the value of $X$ for comet 2P/Encke. The observations cover only portions of the active orbit instead of the entire active orbit for C/1990~K1 (Levy) \citep{Schleicher1991,Feldman1992} and C/2001~K5 \citep{Drahus2006}; for the latter, only the rate of spin change at the middle of the observing window is quoted (instead of the net change), so there is insufficient information for inclusion in Table~\ref{tab:period_change}. Due to these reasons, we cannot properly estimate the $X$-parameters for these two comets. Upper limits exist for rotation period changes of three more JFCs -- 14P/Wolf, 143P/Kowal-Mrkos, and 162P \citep{Kokotanekova2018}. However, as they do not meaningfully constrain the $X$-parameter, they are not listed in Table~\ref{tab:period_change}.

\citet{Jewitt2021} demonstrated that current observational data show that the active fraction, $f_\mathrm{A}$, and the effective dimensionless moment arm corresponding to sublimative torques,  $k_\mathrm{T}$, approximately follow   $f_\mathrm{A} \propto r_\mathrm{n}^{-2}$ and   $k_\mathrm{T} \propto r_\mathrm{n}^{2}$ and therefore $f_\mathrm{A} k_\mathrm{T}$ is nearly constant. They also showed $f_\mathrm{A} k_\mathrm{T}$ is proportional to the $X$ parameter of \citet{Samarasinha2013}. They argued these relationships for $f_\mathrm{A}$ and $k_\mathrm{T}$ are affected by significant observational biases and a debiased sample would make the $r_\mathrm{n}$ dependencies shallower for both $f_\mathrm{A}$ and $k_\mathrm{T}$. Therefore the net result would still be $f_\mathrm{A} k_\mathrm{T}$ (and therefore the $X$ parameter) is nearly constant. 

The most direct conclusion from this spin evolution analysis is comets that have large active fractions show similar rotational evolution to comets with small active fractions. Comet nuclei have a range of active fractions spanning nearly three orders of magnitude \citep[e.g.,][]{AHearn1995}; the near-constant $X$ spans a factor of $\sim$500 in active fraction \citep{Jewitt2021}. The exceptions appear to be comets with very low active fractions (${\ll}0.1\%$), where the sublimative torques are apparently too weak to cause any spin evolution.

The rotation period change during an orbit is not necessarily a monotonic function of time (e.g., 9P; \citealt{Belton2011, Chesley2013}, 67P; \citealt{Jorda2016, Kramer2019a}, 46P; \citealt{Farnham2021}). So far, 46P is the most extreme case where rotation period changes during one segment of the orbit get cancelled during the subsequent segment of the orbit (see \citealt{Farnham2021}) and this is likely due to the comet's high obliquity of 70$^\circ$ (\citealt{Knight2021}) as different torques dominate over different parts of the orbit. In fact, among the four comets whose intra-orbit spin changes are well-characterized, the percentage cancellation of $\delta P$ per orbit in increasing order are for comets 103P, 67P, 9P, and 46P. The obliquity (Table~\ref{tab:period_change}) increases in order: 9P, 103P, 67P, and 46P; however, as 103P is an NPA rotator, its obliquity-induced seasonal effects will be moderated and presumably the cancellation of torques too. This hints at percentage cancellation of spin change during an orbit correlating with obliquity. Could nuclear shape also play a part in this? Additional observations during the coming decades may help answer this.

\renewcommand{\thetable}{2}
\begin{table*}[t]
\begin{center}
\setlength{\tabcolsep}{0.04in}
\caption{Rotation Period Change per orbit ($\delta P$), effective nuclear radius ($r_n$), rotation period ($P$), total water production per unit surface area per orbit ($\zeta$), and $X$-parameter of comets (with respect to those of 2P/Encke, denoted by subscript~$\mathrm{E}$) followed by the spin axis direction rounded to the nearest degree in (RA, Dec) and (orbital longitude, obliquity). In the row corresponding to 2P, the absolute values for 2P are given within parentheses in units of minutes, km, hours, kg~m$^{-2}$, and m$^4$~kg$^{-1}$s$^{-1}$, respectively. The orbital longitude is the angle measured prograde from the sun-perihelion direction in the orbital plane. The sense of rotation is not known for some spin axis directions and thus could be the diametrically opposite directions. The references are listed below the Table as footnotes.}

\begin{tabular}{l c c c c c c c c c}
 \hline
 & & & & & \multicolumn{4}{c}{\hfill Spin Axis Direction \hfill} & \\
 \cmidrule(lr){7-10}
 Comet & $\frac{|\delta P|}{|\delta P_\mathrm{E}|}$ & $\frac{r_\mathrm{n}}{r_\mathrm{nE}}$ & $\frac{P}{P_\mathrm{E}}$ & $\frac{\zeta}{\zeta_\mathrm{E}}$ & $\frac{X}{X_\mathrm{E}}$ & RA ($^\circ$) & Dec ($^\circ$) & Orb Lon ($^\circ$) & Obliq ($^\circ$) \\ 
\noalign{\vskip 2 pt}
 \hline
 \noalign{\vskip 2 pt}
2P/Encke\textsuperscript{a,b} & 1 & 1 & 1 & 1 & 1 & 218 & 8 & 48 & 58 \\
 & (4) & (2.4) & (11) & (1.66$\times10^4$) & (5.3$\times10^{-5}$) & & & & \\ 
 9P/Tempel 1\textsuperscript{a,c} & 3.5 & 1.18 & 3.73 & 0.35 & 1.0 & 255 & 64 & 285 & 16 \\
 10P/Tempel 2\textsuperscript{a,d} & 0.07 & 2.50 & 0.82 & 0.37 & 1.7 & 162 & 58 & 193 & 49 \\
 19P/Borrelly\textsuperscript{a,e} & $>$ 10 & 1.04 & 2.55 & 0.35 & $>$ 4.8 & 214 & -5 & 146 & 103 \\
 21P/Giacobini-Zinner\textsuperscript{f} & 3.6, 6.3 & 0.42 & 0.86 & 0.43 & 1.9, 3.4 & 169 & 73 & 199 & 10 \\
 41P/Tuttle-Giacobini-Kres{\'a}k\textsuperscript{g} & 390 & 0.29 & 3.2 & 0.46 & 7.0 &  &  &  &  \\
 46P/Wirtanen\textsuperscript{h} & 0.6, 3.0 & 0.25 & 0.82 & 0.45 & 0.12, 0.62 & 319 & -5 & 240 & 70 \\
 49P/Arend-Rigaux\textsuperscript{i} & $<$0.06 & 1.92 & 1.2 & 0.34 & $<$0.45 &  &  &  &   \\
 67P/Churyumov-Gerasimenko\textsuperscript{a,j} & 5.3 & 0.69 & 1.09 & 0.38 & 5.5 & 70 & 64 & 21 & 52 \\
 96P/Machholz 1\textsuperscript{k} & $<$0.54 & 1.42 & 0.37 & 1.67 & $<$4.8 &  &  &  &  \\
 103P/Hartley 2\textsuperscript{a,l} & 37.5 & 0.24 & 1.64 & 0.43 & 1.9 & 8 & 54 & 340 & 48 \\
 C/1990 K1 (Levy)\textsuperscript{m} & 57 &  & 3.3 & 0.38 & &  &  &  & 
 \label{tab:period_change} \\ 
 \hline
 \hline
 % MMK: If I don't put the \label in a line in the table it either adds a vertical line (if placed on it's own line inside the tabular call) or changes the Table number to 3.4.3 when referenced in the text (if placed outside of tabular) 
\end{tabular}
\end{center}

\footnotesize
\textsuperscript{a}\citet{Mueller2018}; 
\textsuperscript{b}\citet{Woodney2007}; also \citet{Sekanina1988, Festou2000};
\textsuperscript{c}\citet{Thomas2013Tempel1};
\textsuperscript{d}\citet{Knight2012};
%\textsuperscript{e}\citet{Fernandez2013, Kokotanekova2018}};
\textsuperscript{e}\citet{Farnham2002, Schleicher2003};
\textsuperscript{f}There are two solutions for the spin period change; \citet{Goldberg2023};
\textsuperscript{g}As the change in rotation period is large, the mean value is considered for the calculations; The possibility of NPA spin is suggested for this comet (see text) and if that is the case, this $X$-parameter estimate is highly uncertain; \citet{Bodewits2018, Howell2018, Schleicher2019, Jewitt2021}; 
\textsuperscript{h}The first entry for $\frac{P}{P_\mathrm{E}}$ and $\frac{X}{X_\mathrm{E}}$ represent the net change per orbit and the second entry for those correspond to the difference between the maximum and minimum rotation periods during the orbit (see text); \citet{Farnham2021,Knight2021};
\textsuperscript{i}\citet{Eisner2017};
\textsuperscript{j}\citet{Preusker2015, Jorda2016};
\textsuperscript{k}\citet{Eisner2019};
\textsuperscript{l}This comet is in an evolving NPA spin state and a representative average based on the observations during the 2010 apparition is chosen as the rotation period; \citet{Belton2013}; \citet{Knight2015};
\textsuperscript{m}Observations cover only a segment of the active orbit ($\sim$1 month); \citet{Schleicher1991, Feldman1992}.

%\rone{Table 1: I appreciate the idea of presenting the values relative to 2P, but I suggest to give the absolute values for 2P (instead of all the 1).} 
%\rone{The caption of the table should contain the definition of the columns (the table should be readable by itself)."}
%\mmk{I think that the rotation period changes won't be in our main table, so they still need to be cited here. I suggest to include them in line and make this a full page wide, so we can also include obliquity. Given how long the footnote is, it won't appreciably lengthen the paper if we make it two columns but eliminate the footnote.} \nhs{Note that I didn't use the ``edit'' command to show new text in the Table and in the footnotes.}

\end{table*}

The timescale for rotation period change for a simple rotator, $\tau_\mathrm{P}$, can be expressed by (see \citealt{Samarasinha1986, Jewitt2021})
\begin{equation}
    \tau_\mathrm{P}= k \frac{\rho_\mathrm{n} r_\mathrm{n}^{2}}
    { k_\mathrm{T} f_\mathrm{A} P \zeta}
\end{equation}
\noindent where $k$ is approximately a constant. Since $f_\mathrm{A} k_\mathrm{T}$ is nearly constant, $\tau_\mathrm{P} \propto r_\mathrm{n}^{2}$. Therefore, smaller nuclei will change their rotation periods more rapidly. As can be seen in \citet{Kokotanekova2017} and \citet{Mueller2018}, shorter rotation periods are preferentially seen among the smaller nuclei. This is consistent with sublimation driven spin evolution being the primary cause of spin changes among comets. For short period comets, $\tau_\mathrm{P}$ is smaller than their dynamical lifetime and small nuclei in particular should have undergone extreme alterations to their spin sometimes causing potential splitting events. As a result, particularly for periodic comets, the spin we observe today is not primordial. However, for extremely large comets (e.g., $r_\mathrm{n} \gg$ 10 km) or for DNCs, the spin alterations during their lifetimes are much smaller and they may retain some residual evidence of the primordial distribution. This interpretation is also consistent with $\tau_\mathrm{P} \sim 10^2 r_\mathrm{n}^2$ suggested by \citet{Jewitt2021} for comets with perihelia around 1-2~au, where $r_\mathrm{n}$ is in km and $\tau_\mathrm{P}$ is in years.

 The changes to spin states due to outgassing torques can result in either a net spin-up or spin-down during an orbit. The nuclei which undergo spin-down may ultimately undergo spin-up, i.e., they will spin-down to extremely long periods and then reverse course and start spinning-up. During this epoch near extremely long rotation periods (with minimal rotational kinetic energies), if the conditions for rotational excitation are satisfied (see Section~\ref{sec:NPA_spin_changes} and \citealt{Gutierrez2003b}), the torques can put the nucleus into an excited spin state and it starts to spin-up. As $\tau_\mathrm{P} \propto r_\mathrm{n}^{2}$, the spin-down process too (similar to spin-up) is more efficient for smaller nuclei. Consequently, the rotational excitation is also more likely for smaller nuclei. The spin-up of nuclei is likely to be a cause of splitting events and is discussed in Section~\ref{sec:causes_of_fragmentation}.

As sublimation induced forces are the cause for the non-gravitational forces that change the orbits of comets, \citet{Rafikov2018} investigated the likely changes in the rotation period with corresponding non-gravitational forces and suggested they are correlated.  As sublimation is the primary driver for both the rotational and non-gravitational force induced orbital changes of comets, simultaneous modeling of both these motions may be used to derive better constrained bulk densities for comets than just using orbital changes. Rafikov further determined that the dimensionless lever arm responsible for torquing was small, with a typical value around 1\%, implying that most outgassing induced torques cancel themselves. \citet{Jewitt2021} arrived at a similar conclusion from order of magnitude estimates of spin-up timescales.

\subsubsection{Orientation of Spin Axes and Their Changes}
\label{sec:orientation_of_spin_axes}

Typically, the term spin axis refers to the axis defined by the angular velocity. For a simple rotator, the spin axis of the body aligns with the direction of the rotational angular momentum vector. For an NPA rotator, the instantaneous spin axis is not fixed either in the body frame or in the inertial frame. In this chapter, as most comet nuclei are simple rotators, the term ``spin axis'' is used for simplicity; however, in the case of an NPA rotator, the discussion implicitly refers to the direction of the rotational angular momentum vector despite the usage of the term ``spin axis''.

Other than by space missions to comets, the primary means of determining spin axes of comets is using (a) coma morphology observations taken preferentially at different observing geometries, and (b) multi-epoch observations of the bare-nucleus lightcurves corresponding to different observing geometries. Unlike for asteroids, such multi-epoch bare-nucleus lightcurves are difficult to carry out and expensive in terms of telescope time. Therefore, most spin axis orientations of comets determined to-date are via morphological observations; however, the lightcurve method may become practical for many more comets in the LSST era (see Section~\ref{sec:future}). 

Initial morphology-based spin axis determinations were based on broadband continuum features in the coma representative of scattering of sunlight by dust grains (e.g., \citealt{Sekanina1991morphology, Samarasinha2002, Farnham2002, Schleicher2003, Schleicher2003_Hyakutake}). With the relatively widespread use of narrowband filters within the last 2-3 decades, using primarily the morphology of CN features in the coma, many investigators were able to derive spin axes orientations of multiple comets or place useful constraints (e.g., \citealt{Farnham2007,Woodney2007, Samarasinha2011,Bair2018,Knight2021}). 

As mentioned earlier (Section~\ref{sec:rot_period_change}), \citet{Whipple1979} argued that the spin axis of a non-rigid oblate nucleus would ``precess'' due to non-gravitational torques. Numerical simulations by \citet{Samarasinha1997} point to evolution of the spin axis due to sublimative torques, irrespective of if the spin state is excited or not. The spin axis gradually evolves either towards the orbital direction of peak outgassing (typically near perihelion) or the one diametrically opposite to that. Such a configuration is comparatively stable since that will provide the least net torque over time in the inertial frame. Subsequent work by \citet{Neishtadt2002} using a highly idealized model obtained the same result as one of the main evolutionary paths for the spin axis. Therefore, statistically, more spin axis directions may tend to align closer to the major axis of the orbit. As the current database for spin axis directions of comets is sparse, only future observations (e.g., from the LSST era) could test this prediction.

\subsubsection{Properties and Processes Associated with Spin}
 
Spin studies provide insight into how spin is related to activity and other physical and structural parameters of the nucleus. The discussions above make it clear how activity drives the spin state changes; it is appropriate to note that the spin state, too, is responsible for determining the activity of a comet, the seasonal effects of outgassing caused by a specific spin state being a well-known example. As remarked earlier, since the timescale for rotation period change is larger for large nuclei \citep[e.g.,][]{Jewitt2021}, their spin changes are difficult to measure and there is an observational selection effect against detecting spin changes of large nuclei. 
    
\begin{figure}[t!]
\begin{center}
\includegraphics[width=3.3in]{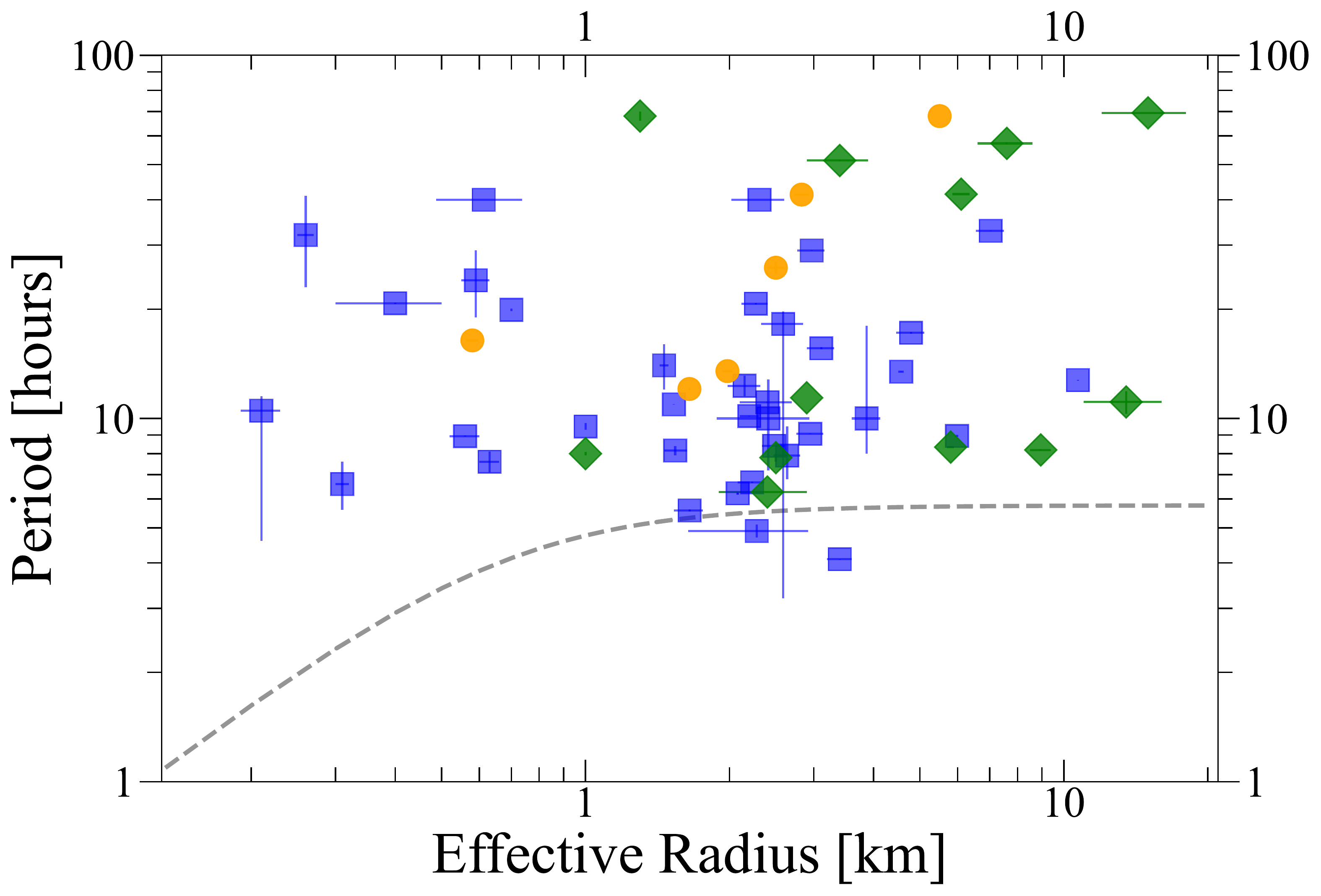}
\caption{Rotation period against effective radius of comet nuclei, updated from \citet{Kokotanekova2017}. The blue squares are JFCs, green diamonds are HTCs and LPCs, and orange circles are comets visited by spacecraft. The dashed line is the model for density 532~kg/m$^{3}$, axial ratio 2, and tensile strength 15~Pa, which corresponds to the parameters measured for 67P from Rosetta \citep{Groussin2015,Jorda2016}. The two comets below the line are 61P, which also has two plausible longer period solutions \citep{Lamy2011}, and 96P, which is undergoing continued fragmentation \citep{Eisner2019} potentially related to rotational spin-up (see Section~\ref{sec:causes_of_fragmentation}).}
\label{fig:P_vs_r}
\vspace{-1mm}
\end{center}
\end{figure}

Spin studies also provide insights into the structural properties of the nucleus (see Figure~\ref{fig:P_vs_r}). The existence of a potential spin barrier and the disruption of comet D/1993 F2 (Shoemaker-Levy 9) due to jovian tidal effects are consistent with low strengths for cometary nuclei (of the order of 10 Pa, about a million times weaker than water ice; \citealt{Weissman2004}). In addition, the paucity of NPA spin states is a pointer to  highly flexible nuclei with short damping timescales. However, both these structural traits should be considered as ensemble properties rather than specific to a particular comet. Here, the ``flexibility of the nucleus'' is meant to suggest that it is not a ``perfect'' rigid body but consists of a cohesive amalgamation of quasi sub-units held together weakly. This flexibility of the nucleus is also consistent with the frequent splitting events observed in comets (see next section). Although the activity and other activity\edit{-}caused effects (e.g., spin-up) could be the primary trigger responsible for most splitting events, the low tensile strengths of comets could facilitate effortless splitting. 

It is possible that the cometary surface layer could be somewhat more cohesive and stronger than the interior but it is still weak when compared to water ice or even some forms of snow (see Guilbert-Lepoutre et al., in this volume). Naturally, it is likely that the low strengths for comets mean that the surface constituents (e.g., boulders) as well as the interior could respond to the stresses and resultant strains caused by rotation, including that due to NPA spin, by even altering cometary shapes. This possibility of mass redistribution and also the initiation of localized activity on the nucleus facilitated by a combination of low strengths of the nucleus and the rotation is an investigation worth pursuing in the future.

Spin state changes can cause cometary outbursts in at least two ways: (a) the spin-up\edit{-}caused splitting can trigger outbursts due to exposure of new surface areas, and (b) the landslides triggered by rapid spin and/or spin changes can result in outbursts (\citet {Steckloff2018} and references therein). However, the majority of these outbursts are unlikely to be major contributors to changes in the spin state as the net torques due to outbursts are comparatively small (since most outbursts are of short duration).

\subsection{Evolution of Physical Properties} 
\label{sec:evolution_propertires}

Formation mechanisms and early evolution of comet nuclei are the focus of Simon et al.\ in this volume and were also the subject of excellent recent reviews by \citet{Dones2015} and \citet{Weissman2020}. Briefly, the two leading formation theories are hierarchical agglomeration whereby smaller cometesimals gradually merge into larger bodies through low-velocity collisions with minimal influence of gravity \citep[cf.][]{Weidenschilling1997,Davidsson2016}, and pebble accretion, in which mm- to cm-sized aggregates (``pebbles'') are formed out of dust and ice grains and then  concentrated by the streaming instability until coalescing in a gravitational collapse \citep[cf.][]{weidenschilling1977,Blum2017}. Formation presumably occurred between 15 and 30 au \citep[cf.][]{Gomes2004}, and the comets likely remained in this region for long enough that the population was significantly modified by collisional evolution \citep[cf.][]{Morbidelli2015,Jutzi2017} before being scattered by giant planet migration to the Oort cloud or perturbed outwards into the Kuiper belt \citep[e.g., the ``Nice Model''][]{Gomes2005,Tsiganis2005}.

Once in these reservoirs, comets were likely further modified. As reviewed by \citet{Weissman2020}, contributing processes are thought to include internal heating from short-lived radio-nuclides, heating by supernovae and passing stars, bombardment by high-energy particles, and accretion of and/or erosion by interstellar gas and dust grains. The first two processes would have heated nuclei at depth, affecting the volatile inventory and distribution. This has implications for the internal structure and composition (see chapters in this volume by Guilbert-Lepoutre et al.\ and Filacchione et al.). The last two processes are estimated to alter the top layers to perhaps 2~m (high-energy particles; see review by \citealt{Strazzulla1999}) or $\lesssim$0.1~m \citep[accretion/erosion; e.g.,][]{Odell1973,Stern1986}. The above processes likely cannot appreciably change the size or shape of comet nuclei, though recent work by \citet{Zhao2021} suggests that sublimation could have significantly altered the shape of Arrokoth with additional assistance due to its specific spin orientation.  The properties of the uppermost layers are undoubtedly altered by bombardment and accretion/erosion.
This may result in changes to the color, albedo, phase function, and polarization, among other things. However, these are hard to discern observationally because the outermost layer is lost as soon as activity begins and typically before the comet is observable. These macroscopic surface properties could theoretically be measured in incoming dynamically new comets before significant activity has begun, but such observations are extraordinarily challenging (see Section~\ref{sec:future}).

Subsequent perturbation from the Kuiper belt or Oort cloud into smaller perihelion distance orbits will result in the initiation of activity (see the chapter by Fraser et al.\ in this volume). Changes in activity are likely gradual as comets diffuse inwards from the Kuiper belt and Scattered Disk via the centaur region into JFCs, or rapidly for Oort cloud comets, which can have their perihelion distance change by many au in a single orbit \citep{Dybczynski2011}. The more rapid perihelion changes of Oort cloud comets and corresponding dramatic increase in insolation may trigger some of the fragmentation scenarios discussed below. Activity removes material to a depth of order 1 m per orbit for a typical comet, although this is not necessarily uniform across the surface due to seasons, topography, and fallback. This is small compared to the size of typical comet nuclei, so it will not change the size or shape appreciably in a single orbit. However, the cumulative effects of activity-driven mass loss may modify the sizes and shapes of comet nuclei over their active lifetimes, which is likely of order 10$^3$ to 10$^5$ yr \citep[e.g.,][]{Levison1994,Levison1997} and dependent on both size and number of close perihelion passages \citep{Nesvorny2017}. 

At the other end of their lifetimes, comets will be lost or destroyed by several mechanisms. \citet{Levison1997} estimate that the median dynamical lifetime of ``ecliptic comets'' (comets having $T_{J}>2$) is $4.5{\times}10^7$ years from the time of their first encounter with Neptune until they collide with the Sun or a planet, or are ejected from the solar system or into the Oort cloud. Thus, most comets will cease to be active long before they are ejected or destroyed by dynamical means. Inactivity may come about by loss of accessible volatiles or the formation of an inert crust; specific mechanisms and estimates of their timescales were reviewed by \citet{jewitt2004_comets2} in {\it Comets~II}. Such inactive comets could represent $\sim$10\% of the near-Earth asteroid population \citep{Fernandez2001}. See Jewitt and Hsieh, this volume, for additional discussion. Sub-km nuclei can be destroyed by mass loss due to sublimation alone  \citep[cf.][]{Samarasinha2007}. Fragmentation (or splitting) and disintegration are also common end states. The frequency of these events is not well constrained, but is estimated to be a few percent per orbit \citep[e.g.,][]{Weissman1980,Chen1994}. More than any other mechanisms just discussed, fragmentation has the potential to modify the nucleus properties discussed in the preceding subsections, so we explore it in more detail below.

\vspace{-0.5mm}
\subsubsection{Fragmentation Overview}
The modern history of comet fragmentation begins with 3D/Biela, which was observed as a pair of fragments in 1846 following regular observations as a single nucleus since its discovery in 1772. It was recovered again in 1852 as a single comet, but has never been recovered since, with heightened meteor activity in 1872, 1885, 1892, and 1899 calculated to have originated in Biela's orbit (see \citealt{Kronk1999,Kronk2003} and references therein). High profile split comets in recent decades include Shoemaker-Levy 9, which was discovered as a string of nuclei in 1993 and subsequently observed to impact Jupiter in 1994 \citep[cf.][]{Hammel1995,Weaver1995}, and 73P, which has had several apparently spontaneous fragmentation events since 1995 \citep[cf.][]{Sekanina2005} and whose fragments were observed in exquisite detail during a remarkable close approach to Earth in 2005 \citep[e.g.,][]{Weaver2006,Reach2009}. See \citet{Weissman1980}, \citet{Sekanina1982}, and \citet{Boehnhardt2004} for lists of other split comets and more detailed discussion of their splitting scenarios.

Despite the relatively high frequency of comet splitting, only a few groups of dynamically related comets are known. Most of the persistent groups are near-Sun comets, almost exclusively seen only by solar observatories. Best known is the Kreutz sungrazing group (perihelion distance ${\sim}$0.005 au) which includes a number of historically bright, ground-observed comets including C/1882 R1 (Great September Comet) and C/1965 S1 (Ikeya-Seki) \citep[cf.][]{Marsden1967,Marsden1989,Sekanina2004}, and thousands of minor fragments seen arriving in SOHO images every few days \citep[cf.][]{Biesecker2002,Knight2010}. The minor fragments are seen only near the Sun for hours to days before they are destroyed, but are thought to have been produced by cascading fragmentation from one of a few major fragments since the previous perihelion passage \citep[e.g.,][]{Sekanina2002}. The Marsden and Kracht groups have ${\sim}10{\times}$ larger perihelion distances ($q~{\sim}~0.05$ au) and are thought to be descended from 96P/Machholz 1 as part of the ``Machholz Complex'' \citep{Ohtsuka2003,Sekanina2005_machholz}. Little is known about the Meyer group, which also has much larger perihelion distances than the Kreutz group ($q~{\sim}~0.04$ au); it has not been linked to any other known objects but its members are thought to be on long period orbits. See \citet{Marsden2005}, \citet{Battams2017}, and \citet{Jones2018} for reviews of these near-Sun groups. 

Few other groups of persistent families are known. \citet{Boehnhardt2004} lists several likely split pairs/families, but uncertainties in orbital integrations, worsened by non-gravitational accelerations, make firm linkages difficult unless they are recently split; \citet{Tancredi1995} estimated that even well known comet orbits cannot be integrated reliably for more than $\sim$1000 years. The only strong linkage of comets on typical JFC orbits seems to be 42P/Neujmin 3 and 53P/Van Biesbroeck \citep{Carusi1985}, while several likely split pairs of Oort cloud comets have been identified.

Most fragmentation events result in an initial  brightening of the inner coma. As the new material expands, the inner coma brightness gradually returns to pre-outburst levels and one or more fragments may be discernible. There is generally a primary fragment having brightness approximately unchanged from before the event  and fainter fragments which diminish in brightness or simply disappear on timescales of days to weeks. Frequently, but not always, the fainter fragments do not persist long enough to gain individual designations, and it is rare that they survive to be re-observed on subsequent apparitions. It is speculated \citep[e.g.,][]{Jewitt2021} that these short lifetimes are due to rapid spin-up (see below), though other mechanisms such as rapid depletion of volatiles could also be responsible.

\vspace{-0.5mm}
\subsubsection{Outburst, Fragmentation, and Disintegration}
\begin{figure*}[t!]
\begin{center}
\includegraphics[width=6.5in]{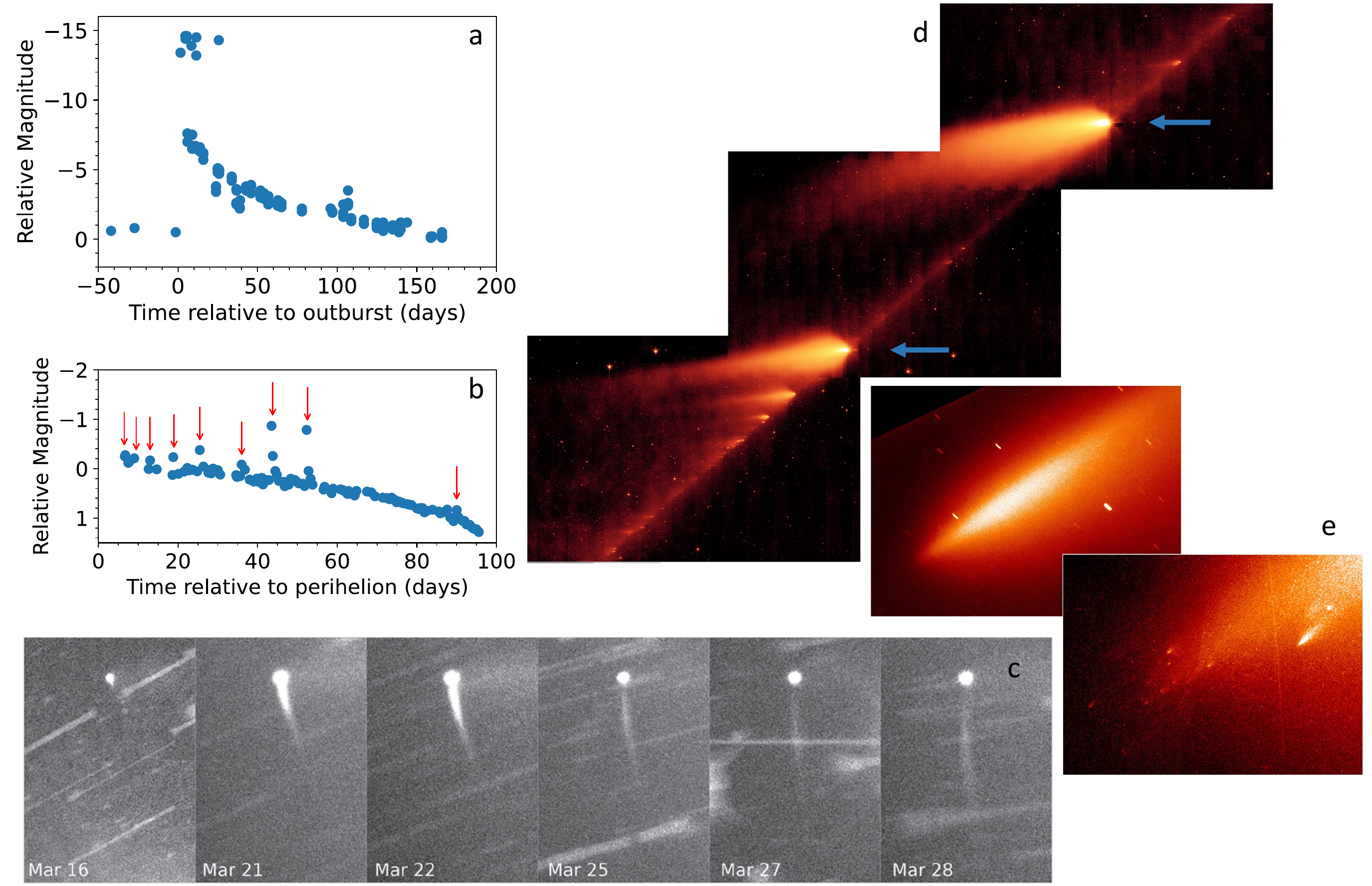}
\caption{Examples of outbursts/fragmentation/disintegration. {\bf Panel a:} Massive outburst of 17P/Holmes using selected ``total'' coma magnitudes archived with the Minor Planet Center. Time is measured from 2007 October 23.0. {\bf Panel b:} Recurring minor outbursts of 7P/Pons-Winnecke adapted from \citet{Lister2022}. Time is measured from perihelion (2021 May 27.1), and outbursts are indicated with vertical arrows. {\bf Panel c:} Evolution of minor outburst morphology of 49P/Arend-Rigaux in 2012 from \citet{Eisner2017}. {\bf Panel d:} Extensive fragmentation of 73P/Schwassman-Wachmann~3 imaged 2006 May 4-6 by Spitzer. Persistent fragments are indicated with horizontal arrows. (Credit: NASA/JPL-Caltech/W. Reach (SSC/Caltech)) {\bf Panel e:} Disintegration of C/1999 S4 (LINEAR) as seen on 2000 August 5. The lower right image by HST shows numerous discrete fragments that were not visible in the near-simultaneous upper left image acquired from the ground. (Credit: H. Weaver (the Johns Hopkins University), and the HST Comet LINEAR Investigation Team, University of Hawaii and NASA/ESA)}
\label{fig:outburst}
\vspace{-3mm}
\end{center}
\end{figure*}

Fragmentation may be part of a continuum of phenomena, including outbursts and disruption/disintegration (see Figure~\ref{fig:outburst}), that can modify the size, shape, and spin of a comet nucleus by progressively larger amounts. Most benign are outbursts, where a comet suddenly brightens due to a rapid, and usually short-lived, increase in activity. Historically, outbursts needed to be large to be detected \citep[e.g.,][]{Ishiguro2016,Miles2016b} unless a comet was already under regular observation. However, with the improved cadences and depth of modern surveys, smaller outbursts are being discovered with much higher frequency than previously known. For instance, M.~Kelley (U.~of Maryland) and collaborators have reported outbursts of more than a dozen comets in Zwicky Transient Facility images since 2019 \citep[e.g.,][]{Kelley2019,Lister2022}. Deep Impact and Rosetta showed that outbursts too small to be detected from the ground happen frequently  \citep{Farnham2007_deepimpact, Vincent2016_fireworks}, so it is reasonable to assume that the behavior is widespread. \citet{Belton2010} proposed a taxonomy to explain a range of activity/outbursts of different strengths observed in spacecraft target comets.

The largest outburst recorded is that of comet 17P/Holmes in 2007 where it brightened by nearly a million times just within two days \citep[e.g.,][]{Montalto2008, Lin2009, Reach2010}. However, except in the largest outbursts, the mass of material involved is likely very small compared to the whole nucleus. Estimates of the masses of the large outbursts reported in \citet{Ishiguro2016} are $10^{8} - 10^{11}$~kg, considerably less than the ${\sim}10^{16}$~kg mass of 67P \citep{Patzold2019}. For an assumed density of 600~kg/m$^3$, a $10^{8}$~kg outburst corresponds to a sphere of equivalent radius $\sim$34~m. Thus, the volume of material involved in most outbursts is likely comparable to the sizes of the pits and boulders seen on 67P \citep[e.g.,][]{Vincent2015,Pajola2016} and is insufficient, individually, to make appreciable changes in a comet's size, shape, or rotation period. 

Outbursts are occasionally accompanied by discrete brightness condensations in the coma which are identified as fragments. Fragments need not be associated with outbursts, and many are only discovered a considerable time after they were presumably produced. Identification of fragments and measurements of their sizes are limited by viewing geometry as well as the activity level of the coma. \citet{Graykowski2021} identified fragments as small as 1~m in radius in HST observations of the debris field around the B and G fragments of 73P at 0.23~au from Earth in 2006, while the smallest discrete fragments identified by \citet{Weaver2001} of C/1994 S4 (LINEAR) at $\sim$0.7~au from Earth were $\sim$25 m in radius. We surmise that  many outbursts which appear devoid of discrete fragments likely contain fragments either too small or too heavily enshrouded in coma to be detected.

Fragmentations can result in significant alteration of the original nucleus, particularly when the resulting pieces are large enough to persist. The primary surviving fragment of 73P is fragment C, which was estimated to have an equivalent radius of $\sim$0.5 km in 2006 \citep{Toth2005,Graykowski2019} as compared to the pre-breakup radius of $\sim$1.1 km for the full comet \citep{Boehnhardt1999}. Sizes have not been published for the other fragments of 73P, but presuming they make up the bulk of the lost material, such mass loss could carry enough angular momentum and changes to the moments of inertia to significantly affect the rotation period and rotation state, in addition to the obvious changes to the size and shape. 

The most extreme phenomenon in this continuum is disintegration, when the nucleus completely disappears. This may be accompanied by discrete fragments (e.g., C/1999 S4) or simply a diffuse cloud of material devoid of brightness condensations like C/2010 X1 (Elenin), in which no fragments were detected down to $\sim$40~m in radius \citep{Li2015}. Spontaneous disintegration seems to be far more common among dynamically new comets, an idea which we will revisit in the next subsection.

\subsubsection{Causes of Fragmentation} 
\label{sec:causes_of_fragmentation}
\begin{figure}[t!]
\begin{center}
\includegraphics[width=2.0in]{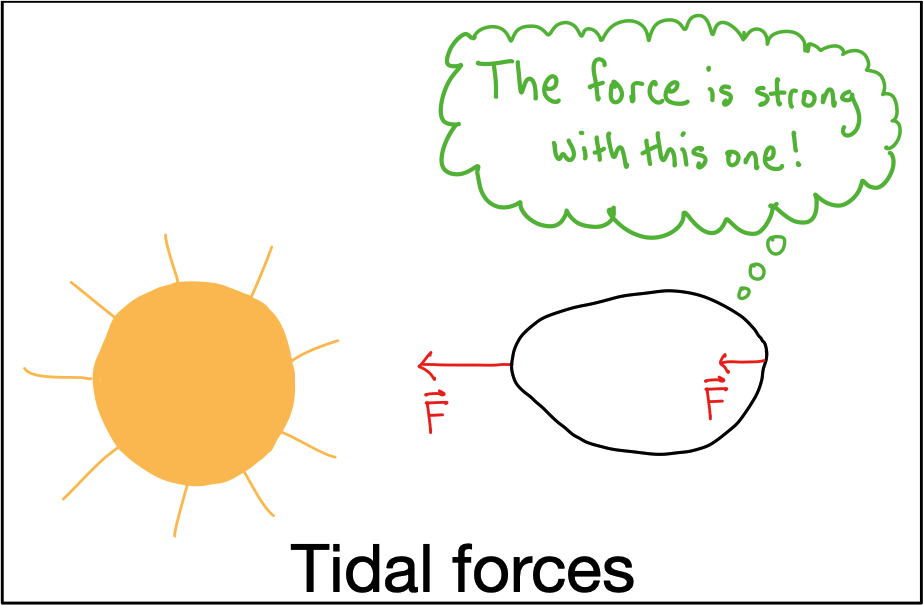}
\includegraphics[width=2.0in]{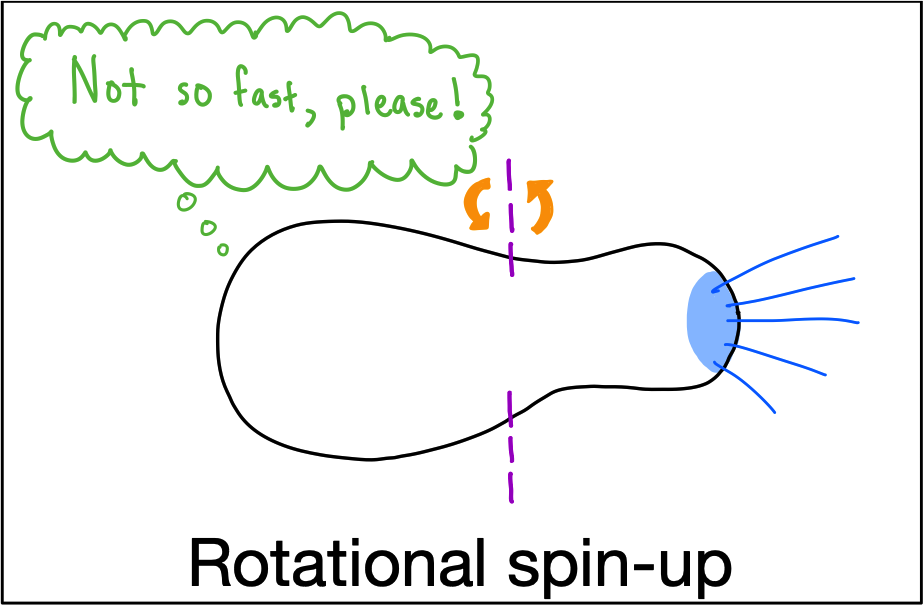}
\includegraphics[width=2.0in]{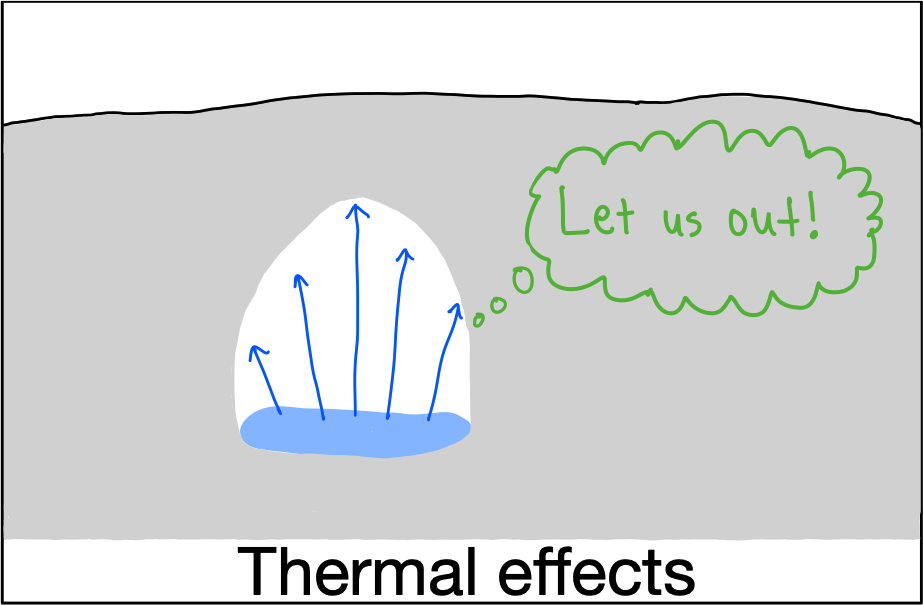}
\includegraphics[width=2.0in]{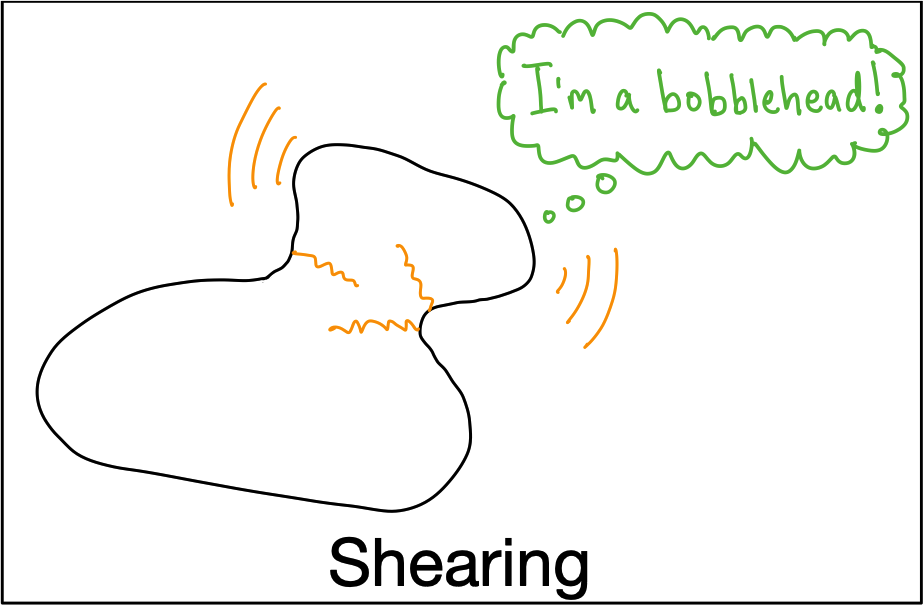}
\includegraphics[width=2.0in]{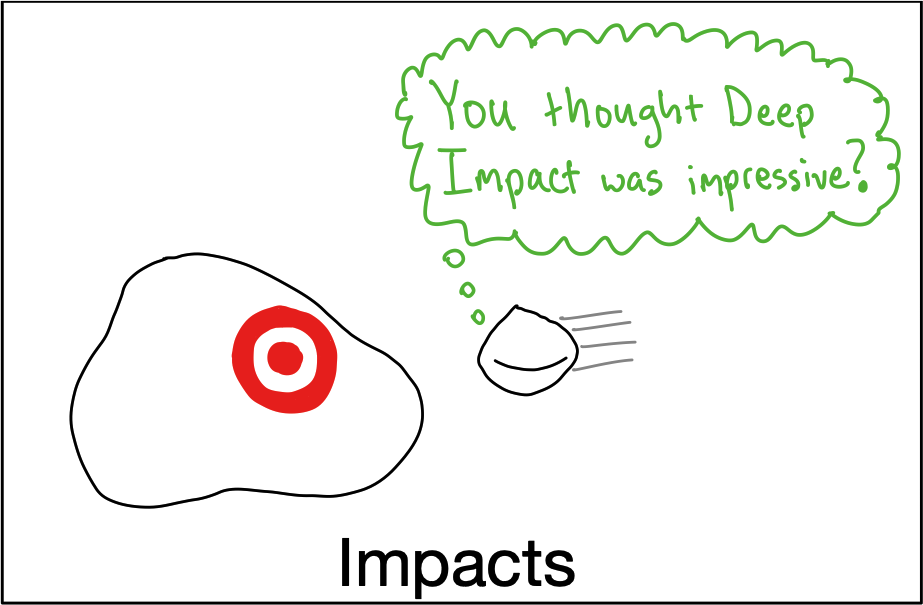}
\caption{Causes of fragmentation (see text for additional details).}
\label{fig:frag_cartoon}
\vspace{-4mm}
\end{center}
\end{figure}

A variety of mechanisms have been proposed to explain why comets split. In a few cases, the cause can be determined with some confidence, but most are conjecture. The most commonly accepted explanations for comet fragmentation are discussed below and illustrated in Figure~\ref{fig:frag_cartoon}.

\smallskip
\noindent {\it \ul {Tidal Forces}} --- The differential gravitational force from a large mass felt on opposite sides of a less massive body can cause disruption of the latter if the differential gravitational force exceeds the forces acting to hold the body together. The splitting of sungrazing comets C/1882 R1 and Ikeya-Seki when near perihelion inspired \citet{Whipple1963} and \citet{Opik1966} to think of comet nuclei as (nearly) strength-less bodies, prone to split due to tidal forces based on their radius and density. The dramatic discovery of the string of fragments in Shoemaker-Levy 9 motivated a host of studies, which lead to the robust conclusion that it had been disrupted during a close approach to Jupiter in 1992 \citep[e.g., ][]{Sekanina1994, Asphaug1996}. Subsequent work has shown that the distance at which tidal splitting occurs can be modified to some extent by other properties of the nucleus such as the axial ratio, sense of rotation, and rotation period (see \citealt{Knight2013} and references therein). However, such modifications are relatively small, and to first order, tidal splitting can be assumed to occur at the Roche limit. 

Using the classical definition of the Roche limit as ${\approx}2.45~R_\mathrm{p}(\frac{{\rho}_\mathrm{p}}{{\rho}_\mathrm{c}})^{\frac{1}{3}}$, where $R_\mathrm{p}$ is the radius of the primary body, and ${\rho}_\mathrm{p}$ and ${\rho}_\mathrm{c}$ are the bulk densities of the primary and comet, a comet with a density of 600~kg/m$^3$ is in danger of tidal splitting if it has perihelion within $\sim$3.2 solar radii (the common upper limit for orbits deemed to be ``sungrazing''; see review by \citealt[][]{Jones2018}), perijove within $\sim$3.2 Jovian radii, or perigee within $\sim$5.1 Earth radii. Encounters sufficiently close to induce tidal splitting are infrequent, with the only clear examples being several bright Kreutz sungrazing comets, Shoemaker-Levy~9, and 16P/Brooks~2 \citep{Sekanina1985}.

\smallskip
\noindent {\it \ul {Rotational Spin-Up}} --- A comet nucleus can split if the centrifugal forces due to rotation exceed the forces holding it together. As previously discussed, torques caused by the asymmetric outgassing of the nucleus can lead to changes in the rotation period and spin state. Although non-principal axis rotation states are known to exist, for simplicity, we consider the case of simple rotation here. Torques can act to either decrease (``spin-up'') or increase (``spin-down'') the rotation period. Spin-down decreases the centrifugal force and, thus, does not lead to splitting, while spin-up can lead to destruction, as has been discussed extensively in the literature, \citep[e.g.,][]{Sekanina1982,Jewitt1997,Jorda2002}. 

As discussed in Section~\ref{sec:rot_period_change}, smaller nuclei can spin up more rapidly than larger nuclei. Order of magnitude calculations \citep[e.g.,][]{Jewitt2021} suggest that this can lead to catastrophic breakup of very small comet nuclei on surprisingly short timescales consistent with observations: 20--200 days at 1.6 au for 20~m radius fragments of 332P/Ikeya-Murakami \citep{Jewitt2016} and $\sim$1.5 days at $\sim$0.06 au for 50~m radius sungrazing comets \citep{Knight2010}. \citet{Jewitt2021} further argues that, for typical short-period comets of up to 10 km in radius, the timescale for destruction by spin-up is of order $10^4$ yr or less, and is shorter than the devolatilization timescale. Due, presumably, to the difficulty of making rotation period measurements for LPCs as well as no possibility of measuring a change in rotation period on subsequent orbits, only two Oort cloud comets have had a change in period detected, so a similar analysis cannot be conducted on the LPC population.

Even if these timescales are underestimated,
observational evidence suggests spin-up is a frequent mechanism in the apparently spontaneous demise of comets. For example, size distributions of JFCs reveal an under-abundance of sub-km nuclei \citep[e.g.,][]{Fernandez2013,Bauer2017} that can be explained by spin-up, while \citet{Jewitt2022_LPCs} used a sample of 27 LPCs to argue that rotational breakup can explain the destruction of near-Sun long-period comets. However, note that spin-up is not necessarily the controlling factor in the destruction of small nuclei. As demonstrated by \citet[][]{Jewitt2021} and consistent with earlier work by \citet{Samarasinha2007}, either mass loss or spin-up can be the main driver of fragmentation. 

A possibly related phenomenon is the disappearance of faint Oort cloud comets near the Sun. \citet{Bortle1991} found that about 70\% of Oort cloud comets for whom the absolute magnitude $H_{10}$ in the \citet{Vsekhsviatskii1964} system was greater than $7.0 + 6.0q$, where $q$ is the perihelion distance in au, did not survive their perihelion passage. This equation is often referred to as the ``Bortle limit.'' \citet{Sekanina2019} found even less frequent survival using a high quality set of modern orbits. On the assumption that fainter comets have smaller nuclei, this behavior can naturally be explained by sublimation driven spin-up, with increasingly larger nuclei being destroyed as perihelion distance decreases. In the most extreme case of a sungrazing comet, simulations by \citet{Samarasinha2013} predicted significant changes to the rotational state for C/2012 S1 (ISON) due to sublimation.

\smallskip
\noindent {\it \ul {Thermal Effects}} ---
We use ``thermal effects'' to broadly encompass several processes related to the heating of the nucleus. The most commonly proposed mechanism is the buildup of subsurface gas pressure which is eventually released catastrophically as an outburst or fragmentation. A variety of processes have been proposed including the transition of amorphous to crystalline H$_2$O ice \citep[e.g.,][]{Patashnick1974,Prialnik1999} and exothermic dissolution of gases like CO and CO$_2$ \citep{Miles2016b}. Such processes are likely to be most important at large heliocentric distances; see reviews by \citealt{Hughes1991} and \citealt{Gronkowski2016} for additional details. Thermal stress has often been cited as a likely cause of breakups \citep[e.g.,][]{Boehnhardt2004}. \citet{Fernandez2009} pointed out that the results from Deep Impact suggest that the thermal pulse from diurnal heating can only penetrate a few centimeters, so is unlikely to create large amounts of thermal stress. 

Another thermal effect is the complete loss of the volatiles holding a comet together. As envisioned by \citet[][]{Sekanina2018}, a comet experiencing this might remain intact as a fluffy aggregate, but would be subject to extreme non-gravitational accelerations and would be expected to dissipate rapidly as the grains crumble. This loss of the volatile ``glue'' would be more likely to occur at small heliocentric distances and might, therefore, help explain the ``Bortle limit'' discussed above. In this scenario, there would be a survivor bias in which comets not prone to such behavior would survive, thus decreasing the frequency of disintegration by returning comets.

\smallskip
\noindent {\it \ul {Shearing}} --- The Rosetta team identified fractures that were centered in the neck region of 67P and propagated  during the mission. \citet{Matonti2019} concluded that these expanding fractures were due to shearing deformation caused by stresses from 67P's two lobes and speculated that this could cause outbursts at large distances or ultimately lead to 67P breaking up. Given the apparent prevalence of highly elongated/bilobate nuclei, this could be an important evolutionary mechanism across the comet population.

\smallskip
\noindent {\it \ul {Impacts}} ---
Although collisional evolution of the comet population undoubtedly occurred in the early stages of the solar system, spatial densities throughout the solar system are too low for impacts onto comets to trigger fragmentation with any appreciable regularity today. However, small impacts undoubtedly occur on occasion, and may be capable of triggering increased activity by exposing fresh, subsurface ice.

\subsubsection{Potential Insight Gained}
Splitting provides the unique opportunity to study the freshly exposed interior of a nucleus. If each fragment is sufficiently separated spatially to allow its chemical composition to be measured, the compositions of fragments can be compared to investigate whether or not the parent nucleus was heterogeneous. To date, this has only been accomplished for 73P, whose fragments had similar compositions \citep[e.g.,][]{DelloRusso2007,Schleicher2011}. Measuring compositions of more  split comets is highly desirable, as the results should help discriminate between nucleus formation scenarios, notably how bilobate nuclei come to be. 

Rotationally resolved observations of recently split nuclei while they are weakly active or inactive would permit searches for color, albedo, and/or polarimetric variations. Since no comet has been shown to have appreciable macroscopic variations in any of these properties, the detection of variations in a recently split comet would suggest that fresh materials have different properties from older surface material. 

If comets split along paths through their interiors with less structural integrity, the resulting more consolidated fragments may be the cometesimals out of which the original comet formed. It is interesting to note that the sizes of the fragments sampled in occasional high-resolution images of recent fragmentations -- e.g., 73P, C/1999 S4, 332P (all discussed above) --  are of similar sizes to the inferred sizes of the SOHO-observed Kreutz comets \citep{Knight2010}. Furthermore, all of these are consistent with \citet{Weidenschilling1997}'s preferential sizes of cometesimals and roughly comparable in size to the boulders and pits on 67P \citep[cf.][]{Vincent2015,Pajola2016}. If the fragments produced in splitting events are the fundamental units of comet formation, then measuring their size distribution will be valuable for constraining formation models. At present, the measured sizes of fragmented comets are thought to be limited by detector sensitivity. Future observations with greater survey depth should be capable of testing the small-end of this size distribution.

\section{FUTURE ADVANCES}
\label{sec:future}

We conclude this chapter by considering what lies ahead for the study of comet nuclei. We limit this discussion to the discoveries that will advance the topics discussed in the preceding sections via remote observations. Future advances are obviously difficult to predict, but we shall make educated guesses based on the technological advances already underway or planned as well as well-established long term goals of the community. We supplement this with some wishful thinking if cost and/or luck were no impediment.  

\subsection{Near-term Telescope Advances}

The easiest future insights to foresee are those that result from existing technology and soon-to-be completed facilities. The two major facilities expected to revolutionize planetary science in the near future are JWST and the Vera Rubin Observatory's Legacy Survey of Space and Time (LSST).

\subsubsection{JWST}
JWST, a 6.5-m space-based telescope located at the Earth-Sun L2 Lagrange point, began operations in mid-2022 and is capable of acquiring images and spectra from 0.6 to 28.3 $\mu$m. Its unprecedented thermal IR capabilities provide fascinating possibilities for nucleus studies. Most known JFC nuclei should be detectable at aphelion, meaning that a dedicated survey could theoretically measure the nucleus sizes of essentially the entire JFC population. In reality, many are likely to be active at such distances \citep{Kelley2013}, making nucleus detections more challenging. 

A more compelling advance would come from JWST detections of the nuclei of long period comets at large heliocentric distances post-perihelion, when activity has died down or is sufficiently low to allow the coma-nucleus separation techniques of \citet{Fernandez2013}. The efficiency is likely to be dependent on the dust-to-nucleus ratio, which would not be known in advance, but given the paucity of direct detections of the nuclei of comets coming from the Oort Cloud, would be well worth exploring. A study comparable in scope to SEPPCoN \citep{Fernandez2013}  would have enormous value for shaping our understanding of the properties of LPC nuclei. Furthermore, JWST will enable surface spectroscopy of many more nuclei than have ever been studied in this manner. This may lead to the first remote detection of exposed ices and will make possible studies of spectral changes with rotation. 

\subsubsection{LSST}

LSST is expected to begin full operations in early 2025. Surveying the entire sky visible from its southern hemisphere site in Chile every few nights to a limiting magnitude of $r{\approx}24.7$, LSST is predicted to increase the number of known comets by about an order of magnitude \citep{Silsbee2016,Ivezic2019}. Specific operational details are still being worked out and will have some effect on the number and nature of discoveries --- a proposed twilight survey would find many new comets at small solar elongations \citep{Seaman2018}, while the proposed northern ecliptic survey would detect more distantly active bodies in the main asteroid belt, Kuiper belt, and (possibly) inner Oort cloud \citep{Schwamb2018} --- but it is clear that LSST will discover fainter and more distant objects than current surveys. The distant discoveries are especially enticing for the possibility of detecting bare nuclei of DNCs, allowing us to monitor the initiation of activity for the first time.

Most LSST discoveries will be too faint for detailed follow up, but the sheer number should vastly improve our understanding of the population of small and/or faint comets with large perihelion distances. The improved statistics should allow meaningful tests of population models \citep[e.g.,][]{Silsbee2016}. Frequent observations of individual comets may permit measurement of the physical properties of the nuclei of inactive or weakly active comets such as phase function, rotation period and axial ratios from lightcurves, color, and size for an assumed albedo. Regular monitoring will also allow the detection of unexpected activity or activity changes, both at individual epochs and by stacking multiple epochs. In addition to discovering many new weakly active comets, this will likely reveal many more fragmentation events than are currently known. The rapid identification of activity changes \citep[e.g.,][]{Kelley2019} will enable the triggering of follow-up observations on specialized instruments that should yield even better data, such as size distributions of newly formed fragments.

\subsection{Longer Term Telescope Advances}
NEO Surveyor, the planned follow up to WISE/NEO-WISE, is slated to launch in 2026 and operate for at least 5 years \citep{Sonnett2021}. As proposed, it will have roughly 50\% more collective power than WISE/NEOWISE and contain two passively cooled channels at mid-IR wavelengths, 4-5 $\mu$m and 6-10 $\mu$m. The longer wavelength channel should be virtually free of reflected sunlight, making it more effective than the post-cryogen phase NEOWISE at obtaining thermal measurements of comet nuclei. However, it will likely be less effective than the cryogen phase of the WISE mission, since the 4-5 $\mu$m region will suffer CO and CO$_2$ gas contamination, meaning the results will depend only on the longer wavelength channel (WISE had two gas-free bandpasses during its cryogen phase). On the whole, we expect NEO Surveyor to be capable of detecting the nuclei of smaller and more distant comets than WISE/NEOWISE, but to produce results that are more model dependent. NEO Surveyor's much longer planned duration than the cryogen phase of WISE means it will detect many nuclei at multiple epochs; the repeat sampling should increase confidence in marginal results and may yield insight into the axial ratio of higher S/N targets.

A rebuilt Arecibo would reinstate our ability to directly measure nucleus properties of close-approaching comets. If it could be made more powerful, the volume of space in which such observations can be made greatly increases.

One or more 30-m class telescopes are expected to become operational by the late 2020s: the Extremely Large Telescope (39-m diameter in Chile), the Thirty Meter Telescope (30-m; Hawaii), and Giant Magellan Telescope (24.5-m; Chile). After JWST, the next major space-based observatory is expected to be the 2.4-m Nancy Grace Roman telescope with a launch date in the mid-2020s. Primarily planned to study dark energy and hunt for exoplanets, its primary value for comet studies is likely via serendipitous observations at large distances \citep{Holler2021}. The true successor to HST as a flexible, multi-wavelength space-based observatory is likely to be the Large Ultraviolet Optical Infrared Surveyor (LUVOIR). The LUVOIR concept is still being developed, but it is likely to have a 10-m class mirror and support imaging and spectroscopic capabilities from UV to IR wavelengths.
Thirty meter class telescopes and next generation space telescopes will possess collecting power roughly an order of magnitude greater than the most powerful existing telescopes, thereby permitting detection and characterization of smaller and/or more distant nuclei. Those that overlap with the LSST era could enable detailed studies that are impossible with current technology. The high angular resolution of 30-m class telescopes could resolve nuclei of moderately sized comets without requiring extreme close approaches, might allow jets to be traced down to the active areas on the nucleus, or might detect sub-meter-scale fragments.

\vspace{-0.5mm}
\subsection{Serendipitous Advances}
Advances that come from clever use of new technologies or simply luck are harder to predict. Entirely beyond our control are comets on orbits that bring them close enough to Earth for their nucleus properties to be measured. In general, encounters within 0.1 au yield outstanding data via numerous techniques. With diverse and often flexible missions in operation throughout the solar system, there are now more opportunities for close approaches of some kind than ever before. Indeed, the closest known approach of a comet to any planet was not a flyby of Earth, but C/2013 A1 (Siding Spring)'s 0.0009 au pass by Mars in October 2014 which yielded the first spatially resolved images of a dynamically new comet \citep{Farnham2017}. 

Occultations are another opportunity that depend, to a great extent, on luck. To date, no comet nucleus has conclusively been detected by an occultation, though \citet{Fernandez1999} constrained Hale-Bopp to ${\leq}$30 or ${\leq}48$~km depending on model parameters. The technique depends on measuring a shadow on Earth that is only the size of the occulting body, and whose position on Earth is uncertain in proportion to the positional uncertainty of the orbit and its distance from Earth. Comet occultations are more challenging than asteroids and even Kuiper belt objects due to their combination of small sizes, vastly larger positional uncertainty owing to their non-gravitational forces from outgassing, and optical depth effects. However, there is no technical reason preventing the observation of comet nucleus occultations. The highly successful campaign for New Horizons extended mission target Arrokoth \citep{Buie2020}, which correctly identified it as a contact binary and provided accurate estimates of the size of each lobe as well as the albedo, engenders optimism that a similar campaign could be mounted for the right comet. In principle, a relatively low cost program could be developed to attempt campaigns for promising targets. The logistics are daunting -- tens to hundreds of observers spaced evenly across tens of km -- but with persistence, chords across the nucleus of multiple long period comets might be obtained, unequivocally constraining their sizes at a fraction of the cost of a dedicated space mission.

Unexpected results from new technology provide another possible advance. A prime example of relevance to this chapter is the discovery in SOlar Heliospheric Observatory (SOHO) images that there exists a nearly continuous stream of small comets approaching the Sun. Small sungrazing comets belonging to the Kreutz family were known prior to SOHO \citep{Marsden1989}, but the discovery of thousands of such comets \citep[cf.][]{Battams2017} was unexpected, as was the existence of several other groups of dynamically related comets \citep{Marsden2005}. There is no facility currently planned that will rival SOHO's surveying power, but the recently launched Parker Solar Probe \citep{Fox2016} and Solar Orbiter \citep{Muller2013}, which are on trajectories with perihelion distance decreasing to 0.05~au and 0.28~au, respectively, might discover previously unknown groups of small comets near the Sun by fortuitously passing close to their orbits.

Moving further from luck to clever design, is the opportunity to use instrumentation in untraditional ways. JWST will allow parallel observations for many of its instruments. Parallel observations will have no control over telescope orientation, but careful planning for serendipitous pointing that include known comets or simply data mining to search for unknown objects might prove a low-cost way to explore comet nucleus properties. Data mining for serendipitous observations of known solar system objects has become relatively common (e.g., asteroids in Gaia data by \citealt{Carry2021}); similar work with JWST data could be particularly compelling if it can detect the nuclei of distant comets. \citet{Schlichting2012} constrained the size distribution of sub-km Kuiper belt objects by searching for occultations of stars in HST images; perhaps future work might be capable of probing the size distribution of the Oort cloud \citep[e.g.,][]{Nihei2007,Ofek2010}?

Finally, computationally intensive data mining of ongoing and future surveys will likely prove fruitful. Analyses that can combine many epochs of data at a multitude of possible rates of motion and then efficiently search the results may be capable of discovering much smaller and/or more distant comets than currently known. These will not only yield new discoveries, but could be used to determine many of the properties discussed in this chapter. The proliferation of stable, space-based datasets -- Kepler/K2, TESS, GAIA -- as well as deep ground-based surveys from larger telescopes like DECam on the Blanco 4-m at CTIO and Hypersuprime Cam on the 8-m Subaru telescope and, of course, LSST should provide ample input for such analyses over the next decade.

\subsection{Open Questions}

\subsubsection{What is the range of comet nucleus properties?}

As discussed previously, the size, axial ratio, albedo, color, phase slope, and rotation period have been measured for enough comet nuclei that we have a sense of their likely ranges. However, there are still few enough data points that individual measurements continue to have value in fleshing out the group statistics. A very reachable objective of the next decade or so is to measure each of these properties for enough comets to draw statistically significant conclusions, with the goal of making assessments akin to the revelation of the ``spin barrier'' among near-Earth asteroids \citep{Pravec2000}. Hints of this are already evident in the inference that comets had low densities prior to a direct measurement being made by Rosetta \citep{Lowry2003a}. However, might a clear limit be established for all comets, by dynamical class, or for a subset which share some common feature? 

With sufficient data, exploration of the bounds of different properties might be especially insightful. Some questions which might be answerable in the relatively near future include: Is there a minimum size for comet nuclei? Is there a minimum and/or maximum albedo? What is the distribution of rotation periods? How common are highly-elongated/bilobate nuclei and what is their long-term stability if they outgas continuously? Are there binary nuclei similar to main-belt comet 288P (300163) \citep{Agarwal2017}? Can phase function be used to discriminate between traditional asteroids and inactive comet nuclei on near-Earth object orbits?

\subsubsection{What Are the Bulk Properties of the Nuclei of LPCs?} 

Tied closely to understanding the range of nucleus properties is the desire to understand the bulk properties of the comets originating from the Oort cloud. As discussed previously, few nucleus properties of LPCs are known, but a dedicated program to detect the nuclei of LPCs outbound at large heliocentric distances in the thermal IR with JWST may be capable of detecting sufficient numbers of LPC nuclei to determine bulk properties of this class of comets.

Another possibility is distant detections of inbound LPCs prior to the onset of activity. The recent discovery of comet C/2014 UN271 illustrates how challenging this will be. C/2014 UN271 was discovered in archival DECam images at $\sim$29~au \citep{Bernardinelli2022}.  It is apparently exceptionally large and/or bright, so detection of a meaningful number of comets inbound requires being sensitive to smaller sizes. Since C/2014 UN271 was discovered at $V~{\approx}~22.5$, surveys would need to go to $V~{\approx}~27$, to detect 10 km objects at comparable distances. C/2014 UN271 exhibited clear activity in TESS data by a heliocentric distance of 23.8~au \citep[][]{Farnham2021b} and likely activity at $\sim$25~au in DECam images \citep{Bernardinelli2021}, so discoveries must be made at even larger distances if the nucleus is to be detected outright. 
Alternatively, discovering that only a small fraction of LPCs activate at large heliocentric distances would support the recently proposed lack of hypervolatile rich comets in the Oort cloud \citep{Lisse2022}.

Such observations might be feasible with one of the proposed 30-m class telescopes, though they would likely require a large field of view and dedicated time to survey a sufficiently large area of sky to get meaningful statistics about the population.  This could be approached in a manner similar to the New Horizons team's search for an extended mission target. Many epochs of deep imaging of the same fields eventually yielded the discovery of Arrokoth and compelling insights into physical properties of the faintest KBOs \citep{Benecchi2015}. Perhaps a 30-m class telescope ``deep field'' akin to the famous Hubble Deep field could achieve such a feat.

\section{SUMMARY}
\label{sec:summary}

Since the publication of {\it Comets~II}, our knowledge of comet nucleus properties has grown considerably and a few hundred comet nuclei, dominated by JFCs but spanning across all dynamical types, have one or more physical and surface properties measured. The vast majority of this growth has come via remote observations. Several dedicated missions have provided ground-truth measurements and confirmed that remote observations can yield accurate measurements of bulk properties. Thus, it is becoming possible to use the cumulative nucleus properties to meaningfully constrain models of solar system formation and evolution. Highlights include:
\begin{itemize}[leftmargin=*]
    \item The effective radius has been measured or constrained for a few hundred comets. Most of these ($>$200) were included in two large-scale IR surveys whose data permit a debiased assessment of the size distribution.
    \item Ever improving observational capabilities and, in some instances, multi-orbit observations have yielded more than 50 axial ratio measurements and more than 60 rotation period measurements. In situ imaging from missions and radar observations have demonstrated that highly elongated and/or bilobate nuclei are common, motivating new investigations into how this might come to be.
    \item The number of comets with changes in rotation period measured or meaningfully constrained has increased from just one to about a dozen, leading to the surprising finding that rotation period changes seem to be nearly independent of the active fraction of the nucleus.
    \item The first nucleus polarization and delay-Doppler radar measurements have been made.
    \item Albedo and nucleus phase coefficients have each been measured for more than 20 comets, while nucleus color indices have been measured for more then 50 comets. A sufficient number of comets have had multiple nucleus surface properties measured to allow investigations into correlations between properties as well as comparisons with other small bodies in the solar system. When remote observations are combined with detailed mission data, studies into the ``age'' of comet surfaces are becoming viable.
    \item Detailed observations of several comets with highly favorable apparitions as well as greatly expanded surveying capabilities have led to advances in our  understanding of nucleus spin states, outbursts, and disintegration/fragmentation.
\end{itemize}

We are on the cusp of the JWST and LSST eras that are expected to provide incredible new sources of data, potentially making every currently known JFC accessible and yielding direct detections of fainter and more distant comets than previously possible. These telescopes, along with several 30-m class ground-based facilities and eventual successors to HST and NEOWISE should enable the first widespread studies of Oort cloud comet nuclei and, undoubtedly, result in paradigm-shifting breakthroughs in our understanding of small bodies in the solar system.

%--------------------------------- 
\vskip .2in
\noindent \textbf{Acknowledgments.} \\

MMK acknowledges support from NASA Solar System Observations program grant NNX17AK15G / 80HQTR20T-0060. RK acknowledges support through the ESO Fellowship Programme. Funding via past NASA Planetary Science Division grants were instrumental in NHS's contributions and this support is gratefully acknowledged.
We thank Ellen Howell, Yan Fernandez, Michael S.P. Kelley, Abbie Donaldson, and Jian-Yang Li for helpful discussions.

%---------------------------------
%\bibliographystyle{ppvi_lim3.bst}
%\bibliographystyle{sss-full.bst}
\bibliographystyle{sss-three.bst}
%\bibliography{refs.bib}

\end{document}